\documentclass[times, twocolumn, twocolappendix, tighten]{aastex631}

\usepackage{amsmath, upgreek, mathcomp}
\usepackage[T1]{fontenc}

\begin{document}

\title{Oxygen Isotopic Compositions of Chondrules as Probes of Solar Protoplanetary Disk Formation}

\correspondingauthor{Sota Arakawa}
\email{arakawas@jamstec.go.jp}

\author[0000-0003-0947-9962]{Sota Arakawa}
\affiliation{Center for Mathematical Science and Advanced Technology, Japan Agency for Marine-Earth Science and Technology, 3173-25 Showa-machi, Kanazawa-ku, Yokohama, 236-0001, Japan}

\author[0000-0002-5934-5076]{Takayuki Ushikubo}
\affiliation{Kochi Institute for Core Sample Research, EMS, Japan Agency for Marine-Earth Science and Technology, 200 Monobe-otsu, Nankoku, Kochi, 783-8502, Japan}
\affiliation{Marine Core Research Institute, Kochi University, 200 Monobe-otsu, Nankoku, Kochi 783-8502, Japan}

\author[0000-0002-8596-3505]{Ryosuke T. Tominaga}
\affiliation{Department of Earth and Planetary Sciences, Institute of Science Tokyo, 2-12-1 Ookayama, Meguro, Tokyo, 152-8550, Japan}



\begin{abstract}

Chondrules are thought to have formed during transient flash-heating events in dust-enriched regions of the solar protoplanetary disk.
Although laboratory studies have characterized the oxygen isotopic compositions of chondritic materials, quantitative interpretations based on simulations of disk formation and evolution remain limited.
Here, we perform one-dimensional simulations of disk formation and evolution by solving a diffusion--advection equation with mass infall from the parental cloud core.
We compute the temporal evolution of oxygen isotopic compositions using an experimentally derived isotope-exchange model.
We examine how the oxygen isotopic signatures of the disk depend on the radial distribution of infalling material and the composition of the parental cloud core.
We find that the oxygen isotopic compositions of carbonaceous-chondrite chondrules can be reproduced if either (i) the radial extent of mass infall onto the disk is moderate ($\sim 10~{\rm au}$), or (ii) it is large ($> 10~{\rm au}$) and the parental cloud core was ice-depleted and/or experienced weaker CO self-shielding than is commonly assumed.
We further suggest the scenario that the observed bimodal trends in oxygen isotopic composition and redox state reflect the partial escape of H$_{2}$O vapor from chondrule-forming regions during heating.
In contrast, if ordinary-chondrite chondrules formed inside the snow line under background temperatures of $\lesssim 500~{\rm K}$, their oxygen isotopic compositions may be difficult to explain within the present disk-evolution model, because oxygen isotopic exchange between silicates and vapor species proceeds efficiently only in the inner disk at $T \gtrsim 500$--$600~{\rm K}$.

\end{abstract}



\section{Introduction} \label{sec:intro}

Meteorites and their components preserve a record of physicochemical processes in the solar system.
In particular, their oxygen isotopic compositions exhibit a large diversity \citep[e.g.,][]{2017GeCoA.201...83K, 2018E&PSL.481..264F, 2018E&PSL.496..132M, 2018crpd.book..196T, 2021GeCoA.293..328U, 2025SSRv..221...85T}, which would be a clue to unveiling the solar system's formation history.
As oxygen is a major constituent of both dust and gas, its isotopic variations reflect both dynamical and thermal processes.

Oxygen isotopic variations are commonly interpreted as the result of mixing between isotopically distinct reservoirs in the solar protoplanetary disk \citep[e.g.,][]{1977E&PSL..34..209C, 1998Sci...282..452Y, 2012GeCoA..90..242U}.
\citet{2004Sci...305.1763Y} proposed an origin for these isotopically distinct reservoirs: they were formed through CO self-shielding in the parental molecular cloud \citep[see also][]{2002Natur.415..860C, 2005Natur.435..317L, 2020SciA....6.2724K}.
Minor CO isotopologues (i.e., C$^{17}$O and C$^{18}$O) are selectively photodissociated by ultraviolet irradiation \citep[e.g.,][]{1988ApJ...334..771V}, producing $^{16}$O-poor oxygen atoms that are subsequently converted into $^{16}$O-poor H$_{2}$O ice via grain-surface reactions.
By mass balance, the residual CO gas becomes enriched in $^{16}$O; consequently, the three major oxygen reservoirs (H$_{2}$O, CO, and silicates) acquire distinct isotopic compositions.
Evidence for a $^{16}$O-poor H$_{2}$O reservoir in the solar protoplanetary disk is recorded in cometary H$_{2}$O ice \citep[e.g.,][]{2019ARA&A..57..113A} and in $^{16}$O-poor magnetite (``cosmic symplectite'') found in the primitive carbonaceous chondrite Acfer 094 \citep[e.g.,][]{2007Sci...317..231S}.

In a protoplanetary disk, dust particles undergo radial transport driven by gas drag and turbulent diffusion \citep[e.g.,][]{1977MNRAS.180...57W}, and H$_{2}$O ice sublimates around the H$_{2}$O snow line.
Such differential transport between solids and gas can locally enrich or deplete $^{16}$O-poor H$_{2}$O relative to $^{16}$O-rich CO interior to the snow line \citep[e.g.,][]{2004ApJ...614..490C}.
In the hot inner disk, both H$_{2}$O and CO vapors can exchange oxygen isotopes with silicate grains \citep[e.g.,][]{2018ApJ...865...98Y, 2024GeCoA.374...93Y}, potentially driving silicates toward $^{16}$O-poor or $^{16}$O-rich compositions.

Protoplanetary disks form through the gravitational collapse of molecular cloud cores \citep[e.g.,][]{1981Icar...48..353C, 1994ApJ...421..640N}.
A fraction of the material originating from the parental cloud core falls into the hot inner disk \citep[e.g.,][]{2012M&PS...47...99Y, 2024A&A...687A..65W}, where isotopically distinct reservoirs (H$_{2}$O, CO, and silicates) can exchange oxygen isotopes.
These processed materials are subsequently transported outward by radial expansion during the disk-formation stage \citep[e.g.,][]{2006ApJ...640L..67D, 2024A&A...687A.158B, 2024A&A...691A.147M}.
As a result, the oxygen isotopic compositions of these reservoirs after disk formation may not retain their original compositions in the parental cloud core.

Chondrules are thought to have formed through transient flash-heating events in dust-enriched regions of the solar protoplanetary disk \citep[e.g.,][]{2008Sci...320.1617A, 2015E&PSL.430..308M, 2018crpd.book..151E, 2018crpd.book...57J}.
Their oxygen isotopic compositions are therefore expected to reflect the mean isotopic composition of the disk's solid component \citep[e.g.,][]{2018crpd.book..196T}.
Petrographic and isotopic systematics largely support local formation of chondrules after disk formation, rather than disk-wide transport following chondrule formation \citep[e.g.,][]{2024SSRv..220...69M}.
Chondrules in carbonaceous chondrites are thought to have formed in the cold outer disk, where H$_{2}$O existed as ice \citep[e.g.,][]{2020SSRv..216...55K}.
In contrast, chondrules in ordinary chondrites are thought to have formed in the hot inner region, where H$_{2}$O existed as vapor \citep[e.g.,][]{2018ApJ...854..164S}.
The oxygen isotopic compositions of chondrules may thus allow us to constrain the oxygen isotopic compositions of individual reservoirs in the solar protoplanetary disk, as well as their relative abundances \citep[e.g.,][]{2010GeCoA..74.6610K}.
They may also provide insights into the properties of the parental cloud core, and into the characteristic sizes of dust aggregates in the disk and their spatial variations.

To date, numerous laboratory analyses have been conducted of the oxygen isotopic compositions of meteoritic materials, including chondrules, and various scenarios have been proposed to explain the observed variations.
However, quantitative discussions based on numerical simulations of protoplanetary disk formation and evolution have remained limited.
In this study, we construct a basic model to investigate the spatiotemporal evolution of the oxygen isotopic compositions of solids and gases during the disk formation phase (Section \ref{sec:model}).
We perform one-dimensional numerical simulations of disk formation and evolution by solving a diffusion--advection equation with mass infall from the parental cloud core.
The temporal evolution of the oxygen isotopic composition of each reservoir is computed using an experimentally derived isotope-exchange reaction model \citep[e.g.,][]{2024GeCoA.374...93Y}.
We investigate how the oxygen isotopic signatures of the solar protoplanetary disk depends on the radial distribution of mass infall onto the disk, the aggregate radius, and the compositions of the parental cloud core (Section \ref{sec:results}).
By comparing our numerical results with the observed oxygen isotopic characteristics of chondrules in chondrites, we further discuss plausible parameters for the solar protoplanetary disk, as well as possible chondrule-forming conditions both beyond and interior to the snow line (Section \ref{sec:discussion}).

\section{Oxygen Isotopic Compositions of Solar System Materials}
\label{sec:solar}

\begin{figure*}[]
\centering
\includegraphics[width = 0.8\textwidth]{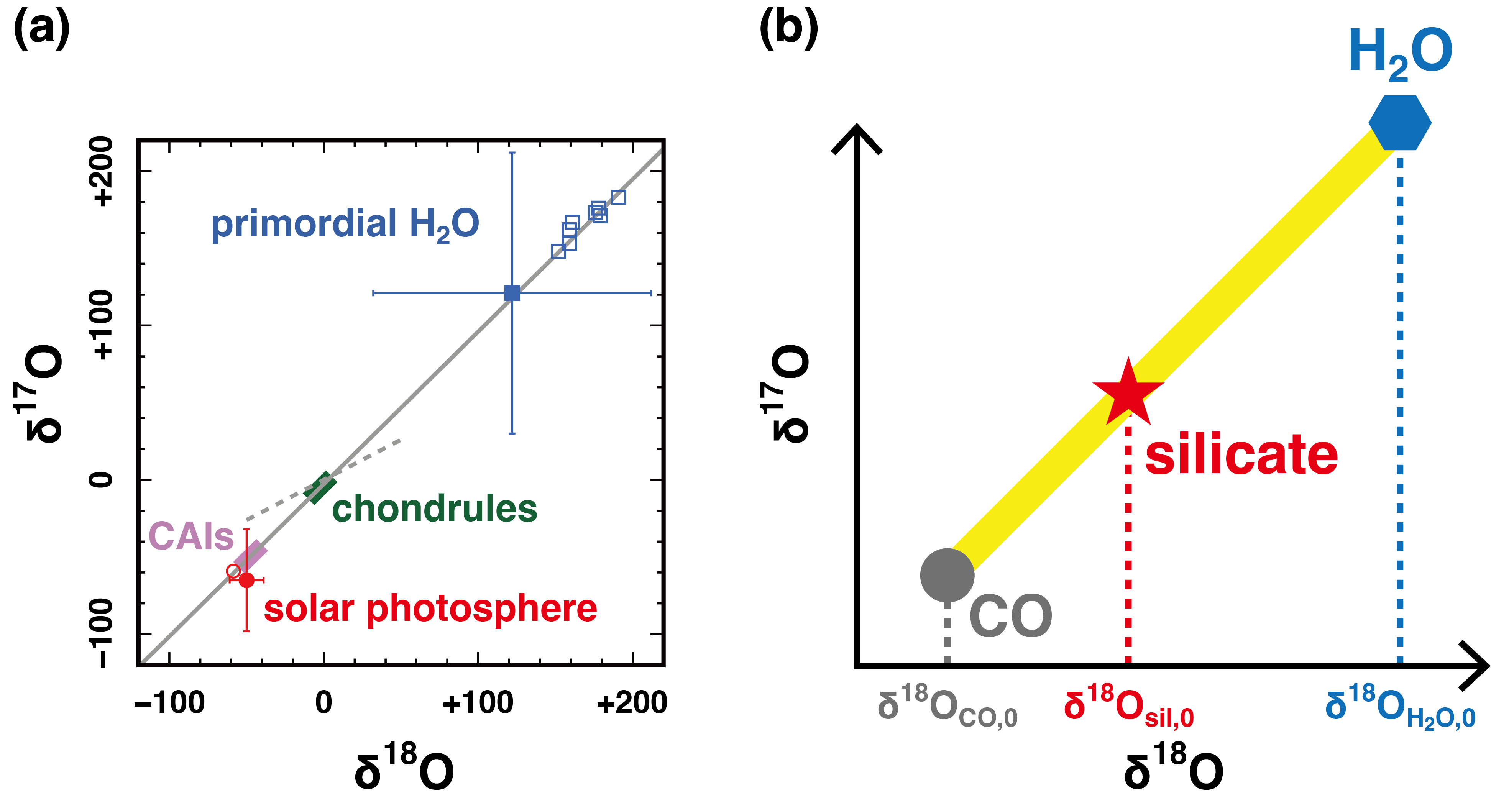}
\caption{
(a) Oxygen three-isotope diagram of solar system materials.
The gray solid line indicates the Primitive Chondrule Materials (PCM) line \citep{2012GeCoA..90..242U}, with a slope of approximately 1, whereas the gray dashed line indicates the terrestrial fractionation (TF) line, with a slope of $0.52$.
Red open circle: solar photosphere inferred from solar-wind samples collected by Genesis \citep{2011Sci...332.1528M}.
Red filled circle: solar photosphere derived from CO absorption lines \citep{2018NatCo...9..908L}.
Blue open squares: cosmic symplectites found in Acfer 094 \citep{2007Sci...317..231S}.
Blue filled square: H$_{2}$O in comet 67P \citep[][and references therein]{2023SSRv..219...32H}.
Violet thick line: primitive (least altered) CAIs in unmetamorphosed chondrites \citep[e.g.,][]{2017GeCoA.201..103U}.
Green thick line: chondrules \citep[e.g.,][]{2018crpd.book..196T}.
(b) Oxygen isotopic compositions of each reservoir in the parental molecular cloud core.
In this study, we consider three reservoirs: silicates (red star), H$_{2}$O (blue hexagon), and CO (gray circle).
We assume that CO self-shielding in the molecular cloud leads to an oxygen-isotope trend along the PCM line \citep[yellow line;][]{2012GeCoA..90..242U}.
Silicates have the solar-system average composition, $\Delta^{17}{\rm O}_{\rm sil, 0} = -30.7\tcperthousand$ ($\updelta^{18}{\rm O}_{\rm sil, 0} = -60\tcperthousand$).
In our fiducial case, we set $\Delta^{17}{\rm O}_{\rm H_{2}O, 0} = +81.4\tcperthousand$ ($\updelta^{18}{\rm O}_{\rm H_{2}O, 0} = +180\tcperthousand$) and $\Delta^{17}{\rm O}_{\rm CO, 0} = -105.4\tcperthousand$ ($\updelta^{18}{\rm O}_{\rm CO, 0} = -220\tcperthousand$).
Variations in oxygen isotopic composition driven by isotope exchange in the protoplanetary disk occur along the mixing line.
}
\label{fig:1}
\end{figure*}

In Section \ref{sec:solar}, we briefly review the oxygen isotopic compositions of extraterrestrial materials considered in this study.
Solar system materials exhibit a diverse range of oxygen isotopic composition which we summarize in Figure \ref{fig:1}(a).
Their oxygen isotopic compositions are generally expressed using the $\updelta$ notation: $\updelta^{A} {\rm O} (\tcperthousand) = {( {R^{A/16}} / {R^{A/16}_{\rm ref}} - 1 )} \times 10^{3}$, where $A = 17$ or $18$ denotes the mass number of the minor isotopes $^{17}$O or $^{18}$O, $R^{A/16}$ is the oxygen isotope ratio of the sample (defined as the number ratio of $^{A}$O to $^{16}$O), and $R^{A/16}_{\rm ref}$ is that of the terrestrial reference material called Vienna Standard Mean Ocean Water \citep{1976E&PSL..31..341B}.

The oxygen isotopic variations of solar system materials are distributed largely along a mass-independent trend with a slope of $\approx 1$ \citep[e.g.,][]{1977E&PSL..34..209C, 1998Sci...282..452Y}.
\citet{2012GeCoA..90..242U} proposed a mixing line called the Primitive Chondrule Materials (PCM) line, defined as follows (gray solid line):
\begin{equation}
\label{eq:PCM}
\updelta^{17}{\rm O} = 0.987 \times \updelta^{18}{\rm O} - 2.7\tcperthousand.
\end{equation}
In contrast, the oxygen isotopic variations of terrestrial materials are distributed largely along a mass-dependent trend called the terrestrial fractionation (TF) line, with a slope of $0.52$ (gray dashed line).
In meteoritics, the deviation from the TF line, $\Delta^{17}{\rm O}$, is widely used as a measure of mass-independent fractionation.
Here, $\Delta^{17}{\rm O}$ is defined as
\begin{equation}
\label{eq:Delta}
\Delta^{17}{\rm O} \equiv \updelta^{17}{\rm O} - 0.52 \times \updelta^{18}{\rm O}.
\end{equation}

\subsection{Solar photosphere}

The bulk average oxygen isotopic composition of the solar system has been estimated from that of the solar photosphere.
The NASA Genesis mission measured the oxygen isotopic composition of solar wind samples, and \citet{2011Sci...332.1528M} reported that the solar wind has a composition of $\updelta^{17}{\rm O} = {( -80.8 \pm 5.0 )}\tcperthousand$ and $\updelta^{18}{\rm O} = {( -102.3 \pm 3.3 )}\tcperthousand$, implying photospheric values of $\updelta^{17}{\rm O} = -59.1\tcperthousand$ and $\updelta^{18}{\rm O} = -58.5\tcperthousand$ (red open circle).
The corresponding $\Delta^{17}{\rm O}$ value is $-28.4\tcperthousand$.
\citet{2018NatCo...9..908L} analyzed the strengths of CO absorption lines in the solar photosphere and reported photospheric values of $\updelta^{17}{\rm O} = {( -65 \pm 33 )}\tcperthousand$ and $\updelta^{18}{\rm O} = {( -50 \pm 11 )}\tcperthousand$ (red filled circle).
These two estimates agree with each other within analytical uncertainty, and a solar-system average value of $\Delta^{17}{\rm O} \sim -30\tcperthousand$ has commonly been assumed in the community.

\subsection{Primordial H$_{2}$O}

It is widely accepted that $^{16}$O-poor primordial H$_{2}$O existed in the solar protoplanetary disk.
\citet{2007Sci...317..231S} found extremely $^{16}$O-poor magnetite grains with isotopic compositions around $\updelta^{17}{\rm O} \sim \updelta^{18}{\rm O} \sim +180\tcperthousand$, known as cosmic symplectites, in the primitive carbonaceous chondrite Acfer 094 (blue open squares).
As cosmic symplectites are thought to have formed via oxidation of Fe--Ni metal and sulfides by water, their oxygen isotopic compositions are considered to reflect that of the reacting H$_{2}$O.
Evidence for $^{16}$O-poor H$_{2}$O is also recorded in cometary H$_{2}$O ice.
The isotopic composition of H$_{2}$O in comet 67P/Churyumov--Gerasimenko has been reported as $\updelta^{17}{\rm O} = {( +121 \pm 91 )}\tcperthousand$ and $\updelta^{18}{\rm O} = {( +122 \pm 90 )}\tcperthousand$ \citep[blue filled square;][and references therein]{2023SSRv..219...32H}.
Although the analytical uncertainty is large, this value is broadly consistent with that of cosmic symplectites.

\subsection{Calcium--aluminum-rich inclusions}

Calcium--aluminum-rich inclusions (CAIs) in chondrites are the oldest known refractory condensates in the solar system \citep[4567 Ma; e.g.,][]{2010E&PSL.300..343A, 2012Sci...338..651C}.
Assuming a spatially homogeneous initial abundance of $^{26}$Al, most CAIs are inferred to have formed within 0.2 Myr \citep[e.g.,][]{2020GeCoA.279....1K}.
The oxygen isotopic compositions of primitive (least altered) CAIs in unmetamorphosed (type 3.0) chondrites commonly show a concentrated distribution around $\Delta^{17}{\rm O} \approx -20\tcperthousand$ to $-25\tcperthousand$ \citep[violet thick line;][]{2014E&PSL.401..327B, 2017GeCoA.201..103U, 2019M&PS...54.1647K}.
Moreover, primitive amoeboid olivine aggregates (AOAs), which are also regarded as high-temperature condensates, exhibit similarly $^{16}$O-rich oxygen isotopic compositions around $\Delta^{17}{\rm O} \approx -25\tcperthousand$ \citep[e.g.,][]{2019PNAS..11623461M, 2021GeCoA.293..544F}.

\subsection{Chondrules}

Chondrules mainly consist of ferromagnesian silicates, with minor phases including high-Ca pyroxene, plagioclase, glass, Fe--Ni metal, and sulfides.
Chondrules in both carbonaceous and non-carbonaceous chondrites have oxygen isotopic compositions of $\Delta^{17}{\rm O} \sim 0\tcperthousand$ \citep[green thick line;][and references therein]{2018crpd.book..196T}, which are clearly distinct from those of the solar photosphere, primordial H$_{2}$O, and CAIs.

\citet{2012GeCoA..90..242U} measured the oxygen isotopic compositions of chondrules in the least altered carbonaceous chondrite Acfer 094.
They found that chondrule phenocrysts and mesostasis have similar oxygen isotopic compositions, and that most chondrules could be classified into two groups with $\Delta^{17}{\rm O} \approx -2\tcperthousand$ and $-5\tcperthousand$.
Their oxygen isotopic compositions are correlated with their Mg\# (defined as the mole percent of ${\rm MgO}/[{\rm MgO}+{\rm FeO}]$ in chondrule ferromagnesian silicates): the group with $\Delta^{17}{\rm O} \approx -5\tcperthousand$ has a narrow Mg\# distribution around $99 \pm 1$, whereas the group with $\Delta^{17}{\rm O} \approx -2\tcperthousand$ has a broad Mg\# distribution ranging from $\approx 98$ to $40$ (see Figure \ref{fig:8}(h)).
Similar trends in the $\Delta^{17}{\rm O}$--Mg\# plane have been identified in other carbonaceous chondrites, including Ornans-type (CO) and Mighei-type (CM) chondrites, as well as an ungrouped carbonaceous chondrite Tagish Lake \citep[e.g.,][]{2013GeCoA.102..226T, 2015GeCoA.148..228T, 2021GeCoA.293..328U}.

The $\Delta^{17}{\rm O}$--Mg\# systematics of chondrules in non-carbonaceous chondrites differ from those in carbonaceous chondrites.
In ordinary chondrites, the $\Delta^{17}{\rm O}$ values of most chondrules range from $0\tcperthousand$ to $+2\tcperthousand$, and no correlation has been observed between $\Delta^{17}{\rm O}$ and Mg\# \citep[e.g.,][]{2010GeCoA..74.6610K, 2020GeCoA.282..133S}.
Similar trends have also been reported for enstatite, Kakangari, and Rumuruti chondrites \citep[e.g.,][]{2011GeCoA..75.6556W, 2017GeCoA.209...24M}.

The formation age distribution of chondrules is still under debate \citep[e.g.,][]{2009Sci...325..985V, 2012M&PS...47.1108K, 2019GeCoA.244..416P}.
Assuming that the short-lived nuclide $^{26}{\rm Al}$ was spatially homogeneous throughout the solar protoplanetary disk, with an initial abundance of $^{26}{\rm Al} / ^{27}{\rm Al} \approx 5 \times 10^{-5}$ \citep[e.g.,][]{2023Icar..40215607D, 2023Icar..39415427P}, the majority of chondrules in carbonaceous chondrites formed $\gtrsim 2.2~{\rm Myr}$ after CAI formation \citep[e.g.,][]{2019GeCoA.253..111H, 2019GeCoA.260..133T, 2022GeCoA.322..194F}, whereas those in ordinary chondrites formed $\approx 1.8$--$2.2~{\rm Myr}$ after CAI formation \citep[e.g.,][]{2021GeCoA.293..103S}.
In contrast, if $^{26}{\rm Al}$ was heterogeneously distributed, as suggested by \citet{2025ApJ...979L..29I}, the inferred chondrule formation ages for carbonaceous and ordinary chondrites shift to $\gtrsim 1.6~{\rm Myr}$ and $\approx 0.4$--$0.8~{\rm Myr}$ after CAI formation, respectively.
Although these estimates remain highly uncertain, chondrule formation in the solar protoplanetary disk may have begun $\sim 1$--$2~{\rm Myr}$ after CAI formation.
Additionally, recent studies on the dichotomy in nucleosynthetic isotopic anomalies have suggested that (proto-)Jupiter may have formed within $\lesssim 1~{\rm Myr}$ of CAI formation \citep[e.g.,][]{2017PNAS..114.6712K, 2020NatAs...4...32K}, potentially modifying the disk structure by opening a gap that inhibited the radial migration of dust particles.
Therefore, in this study, we focus on the spatiotemporal evolution of the solar protoplanetary disk during its first $1~{\rm Myr}$, that is, the interval preceding the likely onset of widespread chondrule formation, and leave a detailed discussion of the effects of (proto-)Jupiter formation to future work.

\section{Models}
\label{sec:model}

In Section \ref{sec:model}, we introduce our numerical model.
We simulate the formation of the solar protoplanetary disk via the gravitational collapse of the parental cloud core and compute the spatiotemporal evolution of oxygen isotopic compositions.
We summarize the oxygen isotopic composition of the parental cloud core in Section \ref{sec:mcc} and outline its chemical components in Section \ref{sec:solid_gas_mcc}.
The mass-accretion dynamics during the collapse are presented in Section \ref{sec:accretion}, and the disk-evolution model is described in Section \ref{sec:disk_evolution}.

\subsection{Oxygen isotopic composition in the parental molecular cloud core}
\label{sec:mcc}

We assume that each reservoir in the molecular cloud core has a distinct oxygen isotopic composition as a result of CO self-shielding.
We further assume that the isotopic composition of each reservoir is spatially uniform within the cloud core for simplicity.
If the oxygen-isotope distribution is produced by CO self-shielding, the isotopic compositions of the reservoirs lie along a mass-independent fractionation trend with a slope of approximately unity \citep{2004Sci...305.1763Y, 2009A&A...503..323V}.
In this study, we assume that CO self-shielding in the molecular cloud leads to an oxygen-isotope trend along the PCM line \citep{2012GeCoA..90..242U}.
By combining Equations \eqref{eq:PCM} and \eqref{eq:Delta}, we obtain the linear relationship between $\Delta^{17}{\rm O}$ and $\updelta^{18}{\rm O}$:
\begin{equation}
\Delta^{17}{\rm O} = 0.467 \times \updelta^{18}{\rm O} - 2.7\tcperthousand.
\end{equation}

We define $\chi_{\rm H_{2}O}$ and $\chi_{\rm CO}$ as the relative oxygen atomic abundances of H$_{2}$O and CO with respect to silicates in the cloud core.
\citet{2004Sci...305.1763Y} assumed $\chi_{\rm H_{2}O} = 2$ and $\chi_{\rm CO} = 3$ based on the solar chemical composition models of \citet{1998A&A...330..375G}, and we adopt these values as our fiducial case.
We also discuss the impacts of lower $\chi_{\rm H_{2}O}$ and $\chi_{\rm CO}$ on the oxygen isotopic evolution in Section \ref{sec:chi_and_C}.

Figure \ref{fig:1}(b) shows the oxygen isotopic compositions of each reservoir in the parental molecular cloud core.
As the oxygen isotope ratios of silicates in the cloud core are not altered by CO self-shielding, they are expected to have an isotopic composition close to the solar-system average.
We set the oxygen isotopic composition of silicates in the cloud core to $\updelta^{18}{\rm O}_{\rm sil, 0} = - 60\tcperthousand$ ($\Delta^{17}{\rm O}_{\rm sil, 0} = - 30.7\tcperthousand$).
From oxygen mass balance, the oxygen isotopic compositions of H$_{2}$O and CO in the cloud core, $\updelta^{18}{\rm O}_{\rm H_{2}O, 0}$ and $\updelta^{18}{\rm O}_{\rm CO, 0}$, satisfy the following equation:
\begin{equation}
\frac{ \updelta^{18}{\rm O}_{\rm sil, 0} - \updelta^{18}{\rm O}_{\rm CO, 0} }{ \updelta^{18}{\rm O}_{\rm H_{2}O, 0} - \updelta^{18}{\rm O}_{\rm sil, 0} } = \frac{\chi_{\rm H_{2}O}}{\chi_{\rm CO}}.
\label{eq:balance}
\end{equation}
In our fiducial case, we adopt $\updelta^{18}{\rm O}_{\rm H_{2}O, 0} = + 180\tcperthousand$ \citep[$\Delta^{17}{\rm O}_{\rm H_{2}O, 0} = + 81.4\tcperthousand$;][]{2007Sci...317..231S} and $\updelta^{18}{\rm O}_{\rm CO, 0} = - 220\tcperthousand$ ($\Delta^{17}{\rm O}_{\rm CO, 0} = - 105.4\tcperthousand$).
We also discuss the impacts of lower $\updelta^{18}{\rm O}_{\rm H_{2}O, 0}$ and higher $\updelta^{18}{\rm O}_{\rm CO, 0}$ on the oxygen isotopic evolution in Section \ref{sec:chi_and_C}.

\subsection{Solid and gas components}
\label{sec:solid_gas_mcc}

\begin{figure}[]
\centering
\includegraphics[width = 0.9\columnwidth]{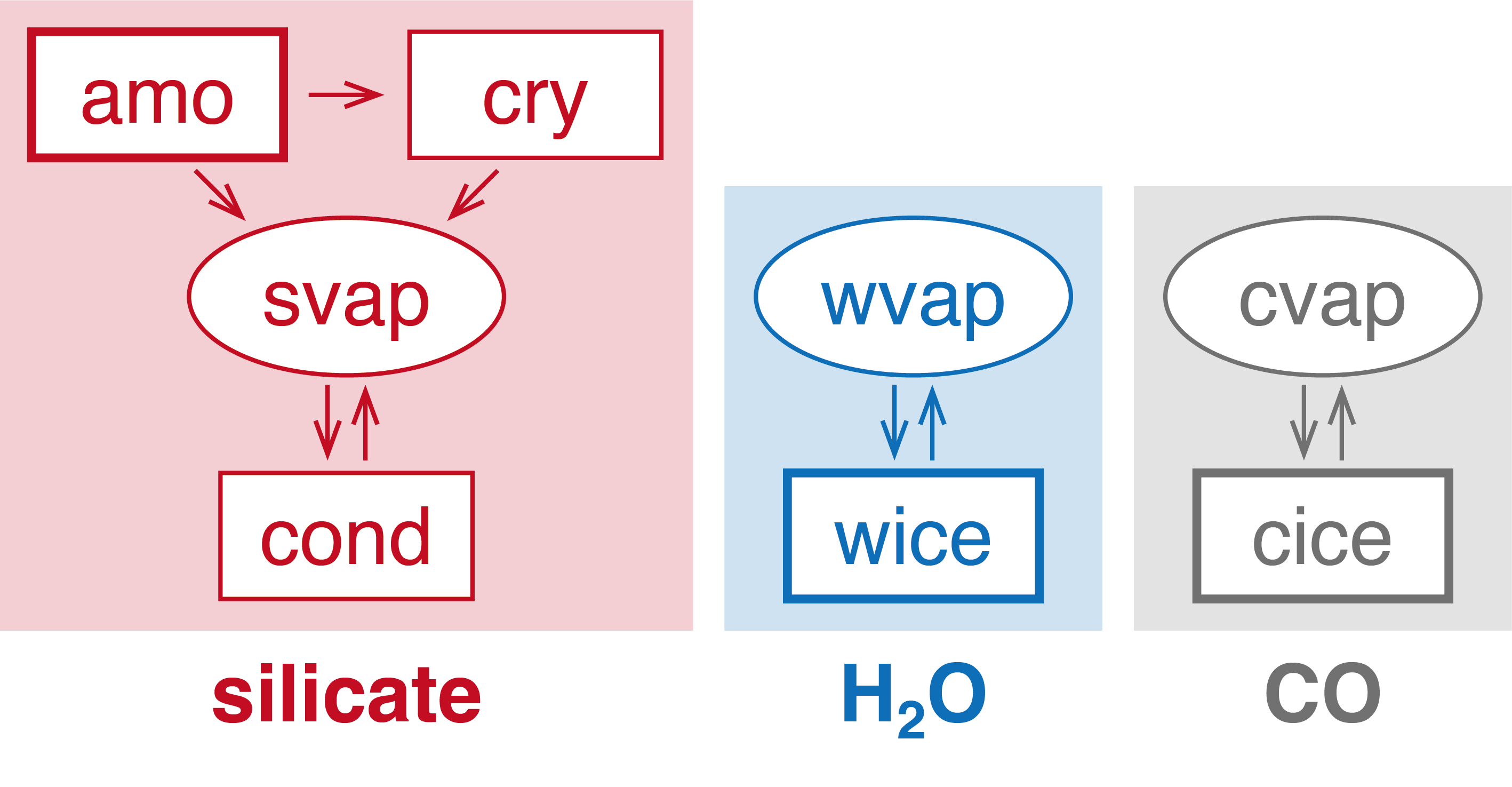}
\caption{
Phase-transition diagram of the oxygen reservoirs considered in this study.
For silicates, we consider four phases: amorphous ({\sf amo}), crystalline ({\sf cry}), condensate ({\sf cond}), and vapor ({\sf svap}).
For H$_{2}$O, we consider two phases: ice ({\sf wice}) and vapor ({\sf wvap}), and for CO we also consider two phases: ice ({\sf cice}) and vapor ({\sf cvap}).
In the molecular cloud core, we assume that silicates exist as {\sf amo}, H$_{2}$O as {\sf wice}, and CO as {\sf cice}.
We assume that evaporation and recondensation occur at the sublimation temperature of each species.
In contrast, crystallization from {\sf amo} to {\sf cry} is treated as an irreversible and temperature-dependent process.
}
\label{fig:2}
\end{figure}

We consider four chemical components within the system: (i) nebular gas dominated by ${\rm H}_{2}$ and ${\rm He}$, (ii) silicates represented by forsterite (Mg$_{2}$SiO$_{4}$), (iii) H$_{2}$O, and (iv) CO.
We do not consider the presence of other molecules (e.g., CO$_{2}$) because their isotope-exchange kinetics is poorly understood.
We also consider phase transitions for silicates, H$_{2}$O, and CO (Figure \ref{fig:2}).

Astronomical observations have revealed that the majority of silicate grains in the interstellar medium are in the amorphous phase \citep[e.g.,][]{2004ApJ...609..826K, 2010ARA&A..48...21H}, and we assume that all silicate grains are amorphous in the parental cloud core.
We also assume that H$_{2}$O and CO in the cloud core exist as solids because the temperature of the cloud core is set to be lower than their sublimation temperatures (see Section \ref{sec:accretion}).

Both H$_{2}$O and CO sublimate into the gas phase in the inner part of the protoplanetary disk where the temperature is higher than their sublimation temperatures, which we set to $170~{\rm K}$ \citep[e.g.,][]{1981PThPS..70...35H} and $20~{\rm K}$ \citep[e.g.,][]{2016ApJ...821...82O}, respectively.
We also consider their recondensation when the temperature decreases below their respective sublimation temperatures.

For the silicate component, we consider the crystallization of amorphous grains as an irreversible process (see Section \ref{sec:crystallization}).
Both amorphous and crystalline grains sublimate into the gas phase when the disk temperature reaches the silicate sublimation point, which is set to $2000~{\rm K}$ \citep[e.g.,][]{2010ARA&A..48..205D}.
As in the cases of H$_{2}$O and CO, recondensation of silicate vapor occurs at the sublimation point, and condensate grains form.
We assume that the condensates are crystalline silicates with a forsterite composition, but we distinguish them from crystalline dust formed via crystallization of amorphous dust.

\subsection{Mass accretion from collapsing molecular cloud core}
\label{sec:accretion}

Protoplanetary disks form through the gravitational collapse of prestellar molecular cloud cores.
The spatiotemporal evolution of the disk gas surface density, $\Sigma_{\rm g} = \Sigma_{\rm g} ( r, t )$, is governed by the following advection equation \citep[e.g.,][]{1974MNRAS.168..603L}:
\begin{equation}
\frac{\partial \Sigma_{\rm g}}{\partial t} = \frac{1}{r} \frac{\partial}{\partial r} {\left( r \Sigma_{\rm g} v_{\rm g} \right)} + S_{\rm g},
\label{eq:a--d_g}
\end{equation}
where $r$ is the heliocentric distance, $v_{\rm g}$ is the radial advection velocity of gas (see Appendix \ref{app:nu}), and $S_{\rm g}$ is the radial distribution of mass accretion to the disk.
In Section \ref{sec:accretion}, we derive $S_{\rm g}$ using the molecular cloud core model of \citet{2013ApJ...770...71T}.

Following earlier studies on disk formation \citep[e.g.,][]{2013ApJ...770...71T, 2025ApJ...985..106K}, we assume that the cloud core is a Bonnor--Ebert sphere and adopt its radial density profile \citep[e.g.,][]{1955ZA.....37..217E, 1956MNRAS.116..351B}.
The radial density profile of the cloud core, $\rho_{\rm cd} = \rho_{\rm cd} (R)$, satisfies the following differential equation: ${\rm d}{(\ln{\rho_{\rm cd}} )} / {\rm d}R = - M (R) / {\left( 4 \pi \rho_{\rm cd, 0} {R_{0}}^{2} R^{2} \right)}$.
The density at $R = 0$ is $\rho_{\rm cd, 0} = 1 \times 10^{-18}~{\rm g}~{\rm cm}^{-3}$, $M (R) = \int_{0}^{R}~4 \pi {R'}^{2} \rho_{\rm cd}~{\rm d}R'$ is the mass within a radius $R$, and $R_{0} = c_{\rm s, cd} / \sqrt{ 4 \pi \mathcal{G} \rho_{\rm cd, 0} / f_{\rm cd}}$ is the characteristic length of the cloud core \citep{2018ApJ...865..102T}.
Here $\mathcal{G}$ denotes the gravitational constant.
The factor $f_{\rm cd}$ represents the ratio between the gravitational potential and the thermal energy density in the initial state, and we set $f_{\rm cd} = 1.4$ following \citet{2013ApJ...770...71T}.
The isothermal sound speed in the cloud core is $c_{\rm s, cd} = \sqrt{k_{\rm B} T_{\rm cd} / m_{\rm g}}$, where $k_{\rm B}$ is the Boltzmann constant, $T_{\rm cd} = 10~{\rm K}$ is the temperature of the cloud core, and $m_{\rm g} = 3.9 \times 10^{-24}~{\rm g}$ is the mean molecular mass of nebular gas \citep[e.g.,][]{2014prpl.conf...27A}.

\begin{figure}[]
\centering
\includegraphics[width = 0.9\columnwidth]{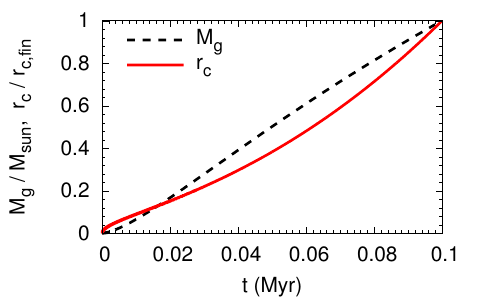}
\caption{
Temporal evolution of the cumulative mass of accreted gas, $M_{\rm g}$ (gray dashed line), and the centrifugal radius, $r_{\rm c}$ (red solid line).
In our setting, the duration of mass accretion is $t_{\rm acc} = 1.0 \times 10^{5}~{\rm yr}$.
}
\label{fig:3}
\end{figure}

In this study, we define $t = 0$ as the onset of mass accretion to the disk.
There is a finite time lag between the onset of the collapse of the cloud core and the onset of mass accretion to the disk, given by $t_{0} = C_{\rm cd} \sqrt{3 / {( 8 \pi \mathcal{G} \rho_{\rm cd, 0} )}}$, where $C_{\rm cd} = 2.567$.\footnote{
\citet{2013ApJ...770...71T} divided the cloud core into spherical shells and modeled the collapse of each shell.
Since the collapsing shells spend most of their time at larger radii, they assumed a constant pressure gradient force during collapse and obtained $C_{\rm cd} = \int_{0}^{1}~{( {f_{\rm cd}}^{-1} \ln{ x } + x^{-1} - 1 )}^{-1/2}~{\rm d}x = 2.567$ (see their Equations~(3)--(5) for details).
}
The mass accretion rate at $t$ is given by
\begin{equation}
\dot{M}_{\rm g} = 4 \pi \rho_{\rm cd} {R_{\rm in}}^{2} \frac{{\rm d}R_{\rm in}}{{\rm d}t},
\end{equation}
where $R_{\rm in} = {\left[ { 2 \mathcal{G} M_{\rm g} {\left( t + t_{0} \right)}^{2} } / {C_{\rm cd}}^{2} \right]}^{1/3}$ denotes the initial radius in the cloud core of the material accreting onto the disk at time $t$.
The cumulative mass of accreted gas at $t$, $M_{\rm g}$, is given by $M_{\rm g} (t) = \int_{0}^{t}~\dot{M}_{\rm g} (t')~{\rm d}t'$.
Mass accretion to the disk ceases when $M_{\rm g}$ reaches $1 M_{\odot}$, and the duration of mass accretion is $t_{\rm acc} = 1.0 \times 10^{5}~{\rm yr}$ in our setting.
The gray dashed line in Figure \ref{fig:3} represents the temporal evolution of $M_{\rm g}$ in our cloud-core model.

Following \citet{2005A&A...442..703H}, we assume that the cloud core gas initially located at $R$ collapses onto the disk within a centrifugal radius $r_{\rm c}$, which is determined by conservation of angular momentum during the collapse.
Assuming that the cloud core rotates rigidly with a constant angular velocity ${\omega}_{\rm cd}$, $r_{\rm c}$ is given by
\begin{equation}
r_{\rm c} = \frac{{{\omega}_{\rm cd}}^{2} {R_{\rm in}}^{4}}{\mathcal{G} M_{\rm g}}.
\end{equation}
The centrifugal radius at the end of accretion, $r_{\rm c, fin}$, is given by
\begin{equation}
r_{\rm c, fin} = 10.6 {\left( \frac{\omega_{\rm cd}}{1 \times 10^{-14}~{\rm s}^{-1}} \right)}^{2}~{\rm au}.
\end{equation}
The radial distribution of mass accretion to the disk, $S_{\rm g}$, is a function of $t$ and the heliocentric distance $r$.
\citet{2005A&A...442..703H} showed that $S_{\rm g}$ is given by
\begin{equation}
S_{\rm g} = \frac{\dot{M}_{\rm g}}{8 \pi {r_{\rm c}}^{2}} {\left( \frac{r}{r_{\rm c}} \right)}^{- 3/2} {\left( 1 - \sqrt{ \frac{r}{r_{\rm c}} } \right)}^{- 1/2}.
\end{equation}
The red solid line in Figure \ref{fig:3} represents the temporal evolution of $r_{\rm c}$ in our cloud-core model.

In this study, we adopt the classical picture in which mass from the molecular cloud core accretes onto the disk interior to the centrifugal radius $r_{\rm c}$.
However, magnetohydrodynamic simulations of collapsing molecular cloud cores have shown that angular-momentum removal by magnetic braking can strongly affect the radial extent over which material falls onto the disk \citep[e.g.,][]{2008A&A...477....9H, 2009ApJ...698..922M, 2011PASJ...63..555M, 2016MNRAS.457.1037W, 2021A&A...648A.101L}.
This effect can be mimicked by adopting a smaller value of $r_{\rm c}$.
For example, \citet{2022NatAs...6...72M} modeled the impact of magnetic braking by setting $r_{\rm c} = 0.35 {( M_{\rm g} / 1 M_{\odot} )}^{- 0.5}~{\rm au}$.
In this study, we treat ${\omega}_{\rm cd}$ as a parameter and vary $r_{\rm c, fin}$ accordingly, to examine how the resulting disk oxygen isotopic composition depends on the radial extent over which material falls onto the disk.

We assume that the masses of H$_{2}$/He gas and amorphous silicates in the parental cloud core are $M_{\rm g, cd} = 1 M_{\odot}$ and $M_{\rm amo, cd} = 5 \times 10^{-3} M_{\odot}$, respectively \citep[e.g.,][]{2021A&A...653A.141A}.
As the formula mass of Mg$_{2}$SiO$_{4}$ is $140$ and it contains four oxygen atoms, the silicate mass per oxygen atom is $m_{1, {\rm sil}} = 35 {N_{\rm A}}^{-1}~{\rm g}$, where $N_{\rm A}$ is the Avogadro number.
Similarly, the H$_{2}$O mass per oxygen atom is $m_{1, {\rm H_{2}O}} = 18 {N_{\rm A}}^{-1}~{\rm g}$, and the CO mass per oxygen atom is $m_{1, {\rm CO}} = 28 {N_{\rm A}}^{-1}~{\rm g}$.
From the relative oxygen atomic abundances of H$_{2}$O and CO (see Section \ref{sec:mcc}), the masses of H$_{2}$O and CO ices in the cloud core are given by $M_{\rm wice, cd} = \chi_{\rm H_{2}O} {( m_{1, {\rm H_{2}O}}  / m_{1, {\rm sil}} )} M_{\rm amo, cd}$ and $M_{\rm cice, cd} = \chi_{\rm CO} {( m_{1, {\rm CO}}  / m_{1, {\rm sil}} )} M_{\rm amo, cd}$, respectively.
The infall mass fluxes of species $i$ ($i = $ {\sf amo}, {\sf wice}, or {\sf cice}) are then given by
\begin{equation}
S_{i} = \frac{M_{i, {\rm cd}}}{M_{\rm g, cd}} S_{\rm g}.
\end{equation}

\subsection{Temporal evolution of the solar protoplanetary disk}
\label{sec:disk_evolution}

We model the temporal evolution of the radial distribution of materials in the solar protoplanetary disk using a one-dimensional disk-evolution code \citep{2021ApJ...920...27A}.
The spatiotemporal evolution of the surface density of species $i$ ($i$ denotes {\sf amo}, {\sf cry}, {\sf cond}, {\sf svap}, {\sf wice}, {\sf wvap}, {\sf cice}, or {\sf cvap}), $\Sigma_{i} = \Sigma_{i} ( r, t )$, is governed by the following advection‐-diffusion equation \citep[e.g.,][]{2006ApJ...640L..67D, 2012M&PS...47...99Y, 2019ApJ...884...32J}:
\begin{align}
\frac{\partial \Sigma_{i}}{\partial t} = & \frac{1}{r} \frac{\partial}{\partial r} {\left[ r \Sigma_{i} v_{i} + r D_{i} \Sigma_{\rm g} \frac{\partial}{\partial r} {\left( \frac{\Sigma_{i}}{\Sigma_{\rm g}} \right)} \right]} \nonumber \\
                                         & + S_{i} + \dot{\Sigma}_{i, {\rm cnv}},
\label{eq:a--d}
\end{align}
where
\begin{equation}
v_{i} = v_{\rm g} + \frac{{\rm St}_{i}}{1 + {{\rm St}_{i}}^{2}} {\left( \eta v_{\rm K} - {\rm St}_{i} v_{\rm g} \right)}
\end{equation}
and $D_{i} = \nu / {( 1 + {{\rm St}_{i}}^{2} )}$ are the radial advection velocity and the diffusion coefficient of species $i$, respectively \citep[e.g.,][]{1986Icar...67..375N, 2007Icar..192..588Y}.
Here, $v_{\rm K}$ is the Keplerian velocity, $\eta = - r {( {\rm d}P_{\rm g}/{\rm d}r )} / {( \rho_{\rm g} {v_{\rm K}}^{2} )}$, $P_{\rm g}$ is the midplane gas pressure, and $\nu$ is the effective gas viscosity (see Appendix \ref{app:nu}).
The term $\dot{\Sigma}_{i, {\rm cnv}}$ in Equation \eqref{eq:a--d} represents conversion among species via crystallization, sublimation, and recondensation (see Section \ref{sec:crystallization}).

We set ${\rm St}_{i} = 0$ when species $i$ is in the gas phase.
In contrast, when species $i$ is in the solid phase, the Stokes number ${\rm St}_{i}$ is given by \citep[e.g.,][]{1977MNRAS.180...57W}
\begin{equation}
{\rm St}_{i} = \frac{\rho_{\rm agg} a_{\rm agg}}{\sqrt{8 / \pi} \rho_{\rm g} c_{\rm s}} \max{\left( 1, \frac{4 a_{\rm agg}}{9 \lambda_{\rm mfp}} \right)},
\end{equation}
where $a_{\rm agg}$ and $\rho_{\rm agg}$ denote the radius and density of dust aggregates, $\rho_{\rm g}$ and $c_{\rm s}$ denote the midplane gas density and the isothermal sound speed, and $\lambda_{\rm mfp}$ is the mean-free path of the gas molecules, respectively.
We assume that $a_{\rm agg} = a_{\rm agg, in}$ for the inner disk region with $T \ge 170~{\rm K}$, and $a_{\rm agg} = a_{\rm agg, out}$ for the outer disk region with $T < 170~{\rm K}$.
Both $a_{\rm agg, in}$ and $a_{\rm agg, out}$ are treated as parameters and are set to either $0.1~{\rm mm}$ or $10~{\rm mm}$ \citep[e.g.,][]{2014prpl.conf..339T}.
The aggregate density is set to $\rho_{\rm agg} = 0.9~{\rm g}~{\rm cm}^{-3}$ for $T \ge 170~{\rm K}$ and to $0.45~{\rm g}~{\rm cm}^{-3}$ for $T < 170~{\rm K}$.
These assumptions correspond to a volume packing fraction of $\sim 30\%$ \citep[e.g.,][]{2018SSRv..214...52B}.

The disk midplane temperature, $T$, is given by
\begin{equation}
T = \min{\left[ {\left( {T_{\rm acc}}^{4} + {T_{\rm irr}}^{4} + {T_{\rm cd}}^{4} \right)}^{1/4}, 2000~{\rm K} \right]},
\label{eq:T}
\end{equation}
where $T_{\rm acc}$ and $T_{\rm irr}$ correspond to the accretion and irradiation heating terms, respectively (see Appendix \ref{app:T}).
We use a classical accretion heating model that assumes energy release at the disk midplane to compute $T_{\rm acc}$ \citep[e.g.,][]{1994ApJ...421..640N}.
In addition, we set the sublimation temperature to $2000~{\rm K}$ and cap $T$ at $2000~{\rm K}$ to mimic dust sublimation.
Such a sublimation-regulated upper limit on $T$ is also seen in detailed thermo-chemical models that explicitly account for mineral evaporation \citep[e.g.,][]{2024A&A...687A..65W}.

The evolutionary equation for the isotopic composition of species $i$, $\Delta^{17}{\rm O}_{i}$, was derived by \citet{2024PASJ...76..881H}, and we follow their formulation \citep[see also][]{2017A&A...600A.140S}.
The column density of oxygen atoms in species $i$ is given by $\mathcal{N}_{i} = \Sigma_{i} / m_{1, i}$, where $m_{\rm 1, amo} = m_{\rm 1, cry} = m_{\rm 1, cond} = m_{\rm 1, svap} = m_{\rm 1, sil}$, $m_{\rm 1, wice} = m_{\rm 1, wvap} = m_{\rm 1, H_{2}O}$, and $m_{\rm 1, cice} = m_{\rm 1, cvap} = m_{\rm 1, CO}$ (see Section \ref{sec:accretion}).
Here we define the product of $\Delta^{17}{\rm O}_{i}$ and $\mathcal{N}_{i}$ as $U_{i}$:
\begin{equation}
U_{i} \equiv \Delta^{17}{\rm O}_{i} \times \mathcal{N}_{i}.
\label{eq:U_i}
\end{equation}
The spatiotemporal evolution of $\Delta^{17}{\rm O}_{i}$ is then obtained by solving the advection--diffusion equation for $U_{i}$:
\begin{align}
\frac{\partial U_{i}}{\partial t} = & \frac{1}{r} \frac{\partial}{\partial r} {\left[ r U_{i} v_{i} + r D_{i} \Sigma_{\rm g} \frac{\partial}{\partial r} {\left( \frac{U_{i}}{\Sigma_{\rm g}} \right)} \right]} \nonumber \\
                                    & + \Delta^{17}{\rm O}_{i, 0} \frac{S_{i}}{m_{1, i}} + \dot{U}_{i, {\rm cnv}} + \dot{U}_{i, {\rm exc}}.
\label{eq:a--d_U}
\end{align}
The terms $\dot{U}_{i, {\rm cnv}}$ and $\dot{U}_{i, {\rm exc}}$ in Equation \eqref{eq:a--d_U} represent conversion (see Section \ref{sec:crystallization}) and isotope exchange (see Section \ref{sec:exchange}) among species, respectively.
The material accreted from the molecular cloud core consists of {\sf amo}, {\sf wice}, and {\sf cice}, with $\Delta^{17}{\rm O}_{{\rm amo}, 0} = \Delta^{17}{\rm O}_{{\rm sil}, 0}$, $\Delta^{17}{\rm O}_{{\rm wice}, 0} = \Delta^{17}{\rm O}_{{\rm H_2O}, 0}$, and $\Delta^{17}{\rm O}_{{\rm cice}, 0} = \Delta^{17}{\rm O}_{{\rm CO}, 0}$.
The isotopic composition of species $i$ is then given by $\Delta^{17}{\rm O}_{i} = U_{i} / \mathcal{N}_{i}$ (Equation \eqref{eq:U_i}).

\subsubsection{Crystallization, sublimation, and recondensation}
\label{sec:crystallization}

We regard the crystallization of amorphous grains as an irreversible process.
The crystallization timescale of amorphous silicate, $t_{\rm amo \to cry}$, is a function of $T$ \citep[e.g.,][]{2000ApJ...535..247H, 2018ESC.....2..778Y}:
\begin{equation}
t_{\rm amo \to cry} = t_{\rm cry} \exp{\left( \frac{E_{\rm cry}}{R_{\rm gas} T} \right)},
\end{equation}
where $t_{\rm cry} = 3.48 \times 10^{-18}~{\rm s}$ is a prefactor, $E_{\rm cry} = 414~{\rm kJ}~{\rm mol}^{-1}$ is the activation energy for crystallization, and $R_{\rm gas}$ is the gas constant.
In protoplanetary disks, amorphous forsterite grains crystallize at $\sim 800~{\rm K}$ \citep[e.g.,][]{2023ApJ...957...47I}.
Then, $\dot{\Sigma}_{\rm cry, cnv}$ and $\dot{U}_{\rm cry, cnv}$ are given as follows:
\begin{align}
\dot{\Sigma}_{\rm cry, cnv} = - \dot{\Sigma}_{\rm amo, cnv} = \frac{\Sigma_{\rm amo}}{t_{\rm amo \to cry}},  \\
\dot{U}_{\rm cry, cnv}      = - \dot{U}_{\rm amo, cnv}      = \frac{U_{\rm amo}}{t_{\rm amo \to cry}}.
\end{align}

In contrast, we regard sublimation and recondensation as instantaneous processes occurring at the sublimation temperature.
The sublimation temperatures of silicate, H$_{2}$O, and CO are set to $2000~{\rm K}$, $170~{\rm K}$, and $20~{\rm K}$, respectively.
We assume that $\Sigma_{i}$ and $U_{i}$ are transferred between the solid and gas phases immediately when the temperature crosses the sublimation temperature.

\subsubsection{Isotope exchange reaction}
\label{sec:exchange}

When two species $i$ and $j$ exchange oxygen isotopes, the temporal evolution of $U_{i}$ due to isotope exchange with $j$, $\dot{U}_{i \leftrightarrow j}$, is given by
\begin{equation}
\dot{U}_{i \leftrightarrow j} = \frac{U_{j} \mathcal{N}_{i} - U_{i} \mathcal{N}_{j}}{\mathcal{N}_{i} + \mathcal{N}_{j}} \frac{1}{t_{i \leftrightarrow j}},
\end{equation}
where $t_{i \leftrightarrow j}$ is the isotope exchange timescale between $i$ and $j$.
We note that $t_{i \leftrightarrow j} = t_{j \leftrightarrow i}$ holds by definition.

We consider isotope exchange between solid silicates ({\sf amo}, {\sf cry}, {\sf cond}) and H$_{2}$O/CO vapors ({\sf wvap}, {\sf cvap}).
We also take into account isotope exchange between H$_{2}$O and CO vapors \citep[e.g.,][]{2004GeCoA..68.3943A, 2009GeCoA..73.4998L}.
The exchange timescales for these reactions are summarized in Appendix \ref{app:exchange}.
The term $\dot{U}_{i, {\rm exc}}$ in Equation \eqref{eq:a--d_U} is given by the sum of $\dot{U}_{i \leftrightarrow j}$ over $j$; for example, $\dot{U}_{{\rm amo}, {\rm exc}} = \dot{U}_{{\rm amo} \leftrightarrow {\rm wvap}} + \dot{U}_{{\rm amo} \leftrightarrow {\rm cvap}}$.
Furthermore, in the innermost region where $T = 2000~{\rm K}$ and all components (silicates, H$_{2}$O, and CO) exist in the gas phase, we assume instantaneous isotopic equilibrium among them.

\section{Results}
\label{sec:results}

We present our numerical results in Section \ref{sec:results}.
In this study, we focus on the spatiotemporal evolution of the solar protoplanetary disk during its first $1~{\rm Myr}$, that is, the interval preceding the likely onset of widespread chondrule formation \citep[e.g.,][]{2022GeCoA.322..194F}.
A fraction of the material originating from the parental cloud core falls into the hot inner disk and is subsequently transported outward by radial expansion \citep[e.g.,][]{2012M&PS...47...99Y, 2024A&A...691A.147M}.
The radial extent of mass infall depends on the core angular velocity, $\omega_{\rm cd}$, and we show how the resulting oxygen isotopic signatures varies with $\omega_{\rm cd}$ in Section \ref{sec:omega}.
Differential transport between $^{16}$O-poor H$_{2}$O ice and $^{16}$O-rich CO vapor beyond the snow line has been proposed as a possible cause of $^{16}$O-rich isotopic compositions of chondritic materials relative to the solar-system average \citep[e.g.,][]{2004Sci...305.1763Y}.
As the strength of dynamical coupling between solids and gas depends on the aggregate radius, $a_{\rm agg}$, we demonstrate how the resulting oxygen isotopic signature varies with $a_{\rm agg}$ in Section \ref{sec:a_agg}.
The relative abundances of H$_{2}$O and CO, as well as their initial isotopic compositions in the parental cloud core, remain under debate \citep[e.g.,][]{2010ApJ...710.1009W, 2013ChRv..113.9043V, 2015ARA&A..53..541B, 2025NatAs...9..883S}.
As both the abundances and the initial compositions affect the subsequent evolution in the disk, we examine these impacts in Section \ref{sec:chi_and_C}.

\subsection{Dependence on $\omega_{\rm cd}$}
\label{sec:omega}

\begin{figure*}[]
\centering
\includegraphics[width = \textwidth]{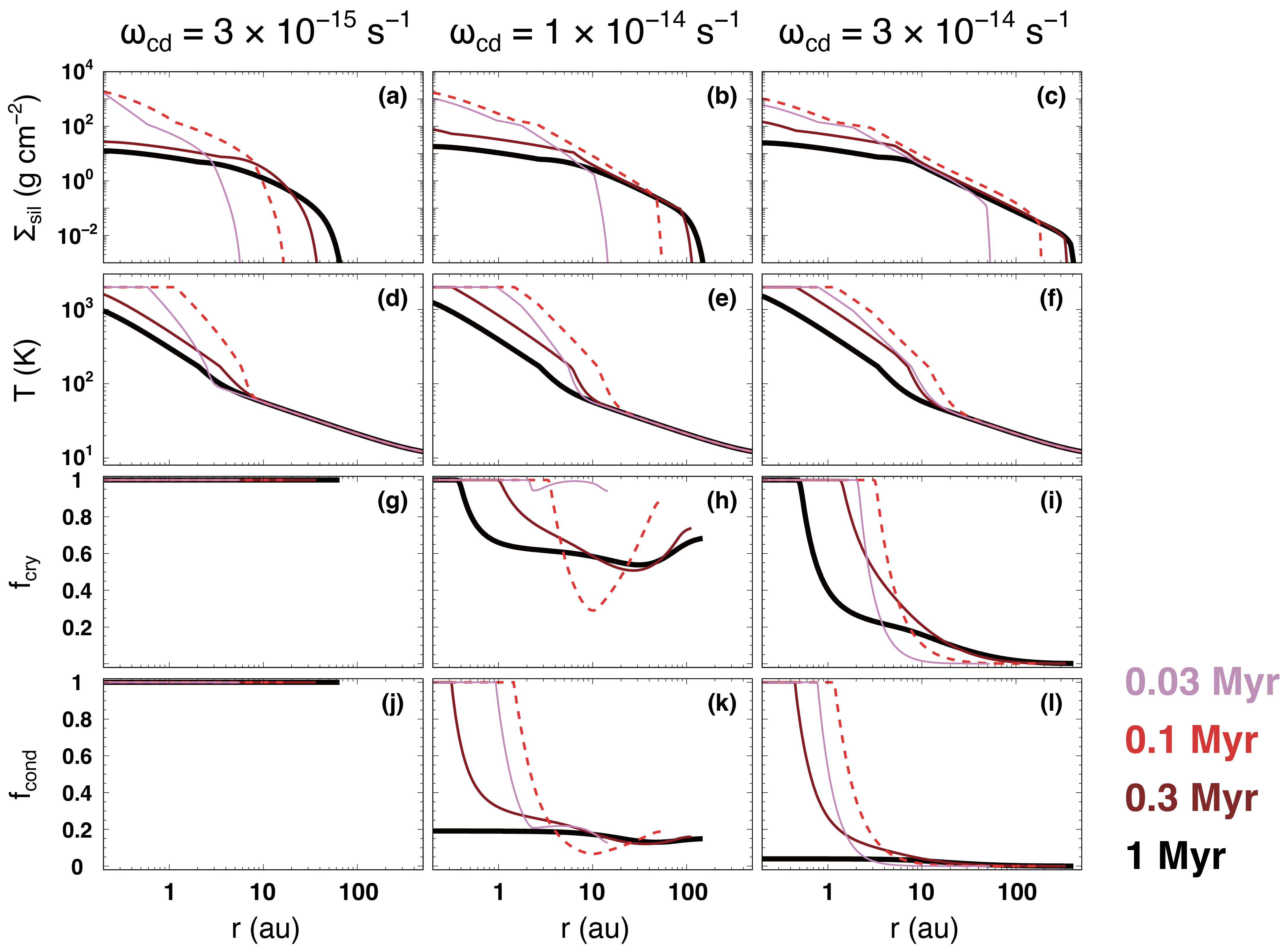}
\caption{
Temporal evolution of the radial distribution of the protoplanetary disks.
The left, middle, and right columns correspond to $\omega_{\rm cd} = 3 \times 10^{-15}~{\rm s}^{-1}$ (i.e., $r_{\rm c, fin} \approx 1~{\rm au}$), $\omega_{\rm cd} = 1 \times 10^{-14}~{\rm s}^{-1}$ (i.e., $r_{\rm c, fin} \approx 10~{\rm au}$), and $\omega_{\rm cd} = 3 \times 10^{-14}~{\rm s}^{-1}$ (i.e., $r_{\rm c, fin} \approx 100~{\rm au}$), respectively.
We set $a_{\rm agg, in} = a_{\rm agg, out} = 0.1~{\rm mm}$ in these calculations.
Different lines represent snapshots at different times, and red dashed lines ($t = 0.1~{\rm Myr}$) correspond to the time when infall from the molecular cloud core terminates ($t_{\rm acc}$).
Panels (a)--(c) show the silicate surface density, $\Sigma_{\rm sil}$.
Panels (d)--(f) show the disk midplane temperature, $T$.
Panels (g)--(i) show the crystallinity, $f_{\rm cry}$, and panels (j)--(l) shows the fraction of condensate silicates, $f_{\rm cond}$.
}
\label{fig:4}
\end{figure*}

First, we discuss disk formation through mass accretion from the molecular cloud core and its dependence on the core angular velocity, $\omega_{\rm cd}$.
Figures \ref{fig:4}(a), \ref{fig:4}(b), and \ref{fig:4}(c) show the time evolution of the silicate surface density,
\begin{equation}
\Sigma_{\rm sil} \equiv \Sigma_{\rm amo} + \Sigma_{\rm cry} + \Sigma_{\rm cond} + \Sigma_{\rm svap},
\end{equation}
for $\omega_{\rm cd} = 3 \times 10^{-15}~{\rm s^{-1}}$, $1 \times 10^{-15}~{\rm s^{-1}}$, and $3 \times 10^{-14}~{\rm s^{-1}}$, respectively.
Different lines represent snapshots at different times, and the red dashed line ($t = 0.1~{\rm Myr}$) corresponds to the time when infall from the molecular cloud core terminates ($t_{\rm acc}$).
In these calculations, we set $a_{\rm agg, in} = a_{\rm agg, out} = 0.1~{\rm mm}$, so that dust and gas are tightly coupled dynamically.

Mass accretion from the molecular cloud core occurs in the region interior to the centrifugal radius $r_{\rm c}$.
At the end of the accretion phase, $t = t_{\rm acc}$, the centrifugal radius is $r_{\rm c, fin} \approx 1~{\rm au}$ for $\omega_{\rm cd} = 3 \times 10^{-15}~{\rm s^{-1}}$, $r_{\rm c, fin} \approx 10~{\rm au}$ for $\omega_{\rm cd} = 1 \times 10^{-14}~{\rm s^{-1}}$, and $r_{\rm c, fin} \approx 100~{\rm au}$ for $\omega_{\rm cd} = 3 \times 10^{-14}~{\rm s^{-1}}$.
Figures \ref{fig:4}(a)--\ref{fig:4}(c) show that, for $t < t_{\rm acc}$, $\Sigma_{\rm sil}$ increases within $r_{\rm c}$ due to infall, and the disk radius expands with time.
In contrast, for $t > t_{\rm acc}$, viscous evolution causes $\Sigma_{\rm sil}$ to decrease in the inner region, while the disk radius continues to expand with time.

Figures \ref{fig:4}(d)--\ref{fig:4}(f) show the time evolution of the midplane temperature, $T$.
In all cases, at $t = t_{\rm acc}$ (red dashed lines), the temperature reaches $T = 2000~{\rm K}$ at $r \lesssim 1~{\rm au}$.
The radial extent of the sublimation region, $\sim 1~{\rm au}$, is broadly consistent with that observed in an extrasolar disk around a Class I protostar \citep{2025Natur.643..649M}.
For $t > t_{\rm acc}$, $T$ decreases as $\Sigma_{\rm sil}$ decreases.
At $t = 1~{\rm Myr}$ (black solid lines), $T \approx T_{\rm irr}$ in the outer region at $r \gtrsim 10~{\rm au}$.
The evolutionary picture of $\Sigma_{\rm sil}$ and $T$ presented above is consistent with that reported in earlier studies \citep[e.g.,][]{2012M&PS...47...99Y, 2022NatAs...6...72M}.

Figures \ref{fig:4}(g)--\ref{fig:4}(i) show the time evolution of the crystallinity, $f_{\rm cry}$:
\begin{equation}
f_{\rm cry} \equiv 1 - \frac{\Sigma_{\rm amo}}{\Sigma_{\rm sil}}.
\end{equation}
Since amorphous forsterite crystallizes at $\approx 800~{\rm K}$ \citep[e.g.,][]{2023ApJ...957...47I}, $f_{\rm cry}$ reflects the fraction of silicates that have experienced temperatures above $800~{\rm K}$.
Similarly, Figures \ref{fig:4}(j)--\ref{fig:4}(l) show the time evolution of the fraction of condensate silicates, $f_{\rm cond} \equiv 1 - {( \Sigma_{\rm amo} + \Sigma_{\rm cry} )} / \Sigma_{\rm sil}$.
We assume that evaporation and recondensation occur at $T = 2000~{\rm K}$; therefore, $f_{\rm cond}$ traces the fraction of silicates that have experienced temperatures above $2000~{\rm K}$.

For $\omega_{\rm cd} = 3 \times 10^{-15}~{\rm s^{-1}}$, mass infall from the molecular cloud core occurs inside $\approx 1~{\rm au}$.
As the inner region reaches $T = 2000~{\rm K}$, all silicates undergo evaporation and recondensation, becoming condensate silicates.
These condensate silicates are subsequently transported to the outer disk through viscous evolution, resulting in $f_{\rm cry} = f_{\rm cond} = 1$ throughout the disk (Figures \ref{fig:4}(g) and \ref{fig:4}(j)).
In contrast, for $\omega_{\rm cd} = 3 \times 10^{-14}~{\rm s^{-1}}$, mass infall from the molecular cloud core occurs over a broad region of the disk at $r \lesssim 100~{\rm au}$.
As a result, most of the infalling material is transported to the outer disk after experiencing only a low-temperature environment with $T \ll 800~{\rm K}$.
Consequently, except for the hot inner region at $r \lesssim 1~{\rm au}$, we find $f_{\rm cry} \ll 1$ and $f_{\rm cond} \ll 1$ (Figures \ref{fig:4}(i) and \ref{fig:4}(l)).

\begin{figure*}[]
\centering
\includegraphics[width = \textwidth]{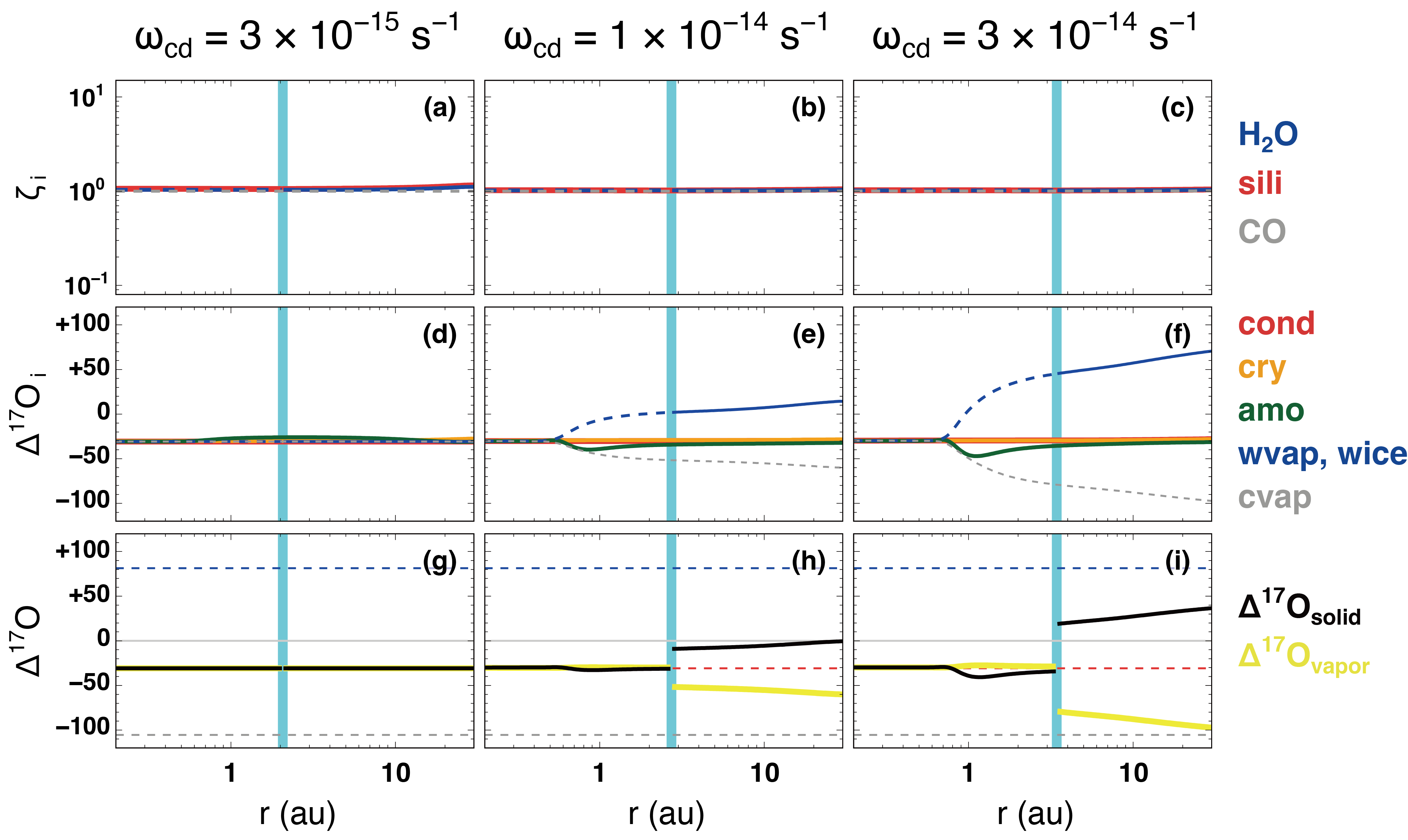}
\caption{
Radial distribution of each oxygen reservoir and their oxygen isotopic composition at $t = 1~{\rm Myr}$.
The left, middle, and right columns correspond to $\omega_{\rm cd} = 3 \times 10^{-15}~{\rm s}^{-1}$ (i.e., $r_{\rm c, fin} \approx 1~{\rm au}$), $\omega_{\rm cd} = 1 \times 10^{-14}~{\rm s}^{-1}$ (i.e., $r_{\rm c, fin} \approx 10~{\rm au}$), and $\omega_{\rm cd} = 3 \times 10^{-14}~{\rm s}^{-1}$ (i.e., $r_{\rm c, fin} \approx 100~{\rm au}$), respectively.
We set $a_{\rm agg, in} = a_{\rm agg, out} = 0.1~{\rm mm}$ in these calculations.
The vertical cyan lines represent the location of H$_{2}$O snow line.
Panels (a)--(c) show the enrichment factor of each oxygen reservoir, $\zeta_{i}$ ($i$ denotes {\sf sil}, {\sf H$_{2}$O}, or {\sf CO}).
Panels (d)--(f) show the oxygen isotopic composition of each species, $\Delta^{17}{\rm O}_{i}$ ($i$ denotes {\sf amo}, {\sf cry}, {\sf cond}, {\sf wice}, {\sf wvap}, or {\sf cvap}).
Panels (g)--(i) show the mean oxygen isotopic compositions of the solid and vapor phases, $\Delta^{17}{\rm O}_{\rm solid}$ and $\Delta^{17}{\rm O}_{\rm vapor}$.
The red, blue, and gray horizontal dashed lines in panels (g)--(i) indicate the oxygen isotopic compositions of the corresponding reservoirs in the parental cloud core.
}
\label{fig:5}
\end{figure*}

Oxygen isotope exchange reactions between amorphous silicates and CO/H$_{2}$O vapors proceed in regions with $T \gtrsim 500~{\rm K}$ \citep{2024GeCoA.374...93Y}.
This temperature is lower than the crystallization temperature of amorphous silicates, but it is still higher than the evaporation temperature of H$_{2}$O ice ($170~{\rm K}$).
In Figure \ref{fig:4}, we have shown how the spatial distribution of $f_{\rm cry}$ depends on $\omega_{\rm cd}$.
By analogy, we expect that the material transported to the outer disk preserves the oxygen isotopic signature inherited from the molecular cloud core only for large $\omega_{\rm cd}$.
We confirm this expectation in Figure \ref{fig:5}.

We define the enrichment factor of each oxygen reservoir, $\zeta_{i}$ ($i$ denotes {\sf sil}, {\sf H$_{2}$O}, or {\sf CO}), as
\begin{equation}
\zeta_{i} \equiv \frac{\Sigma_{i} / \Sigma_{\rm g}}{M_{i, {\rm cd}} / M_{\rm g, cd}},
\end{equation}
where $\Sigma_{\rm H_{2}O} \equiv \Sigma_{\rm wice} + \Sigma_{\rm wvap}$, $\Sigma_{\rm CO} \equiv \Sigma_{\rm cice} + \Sigma_{\rm cvap}$, $M_{\rm sil, cd} = M_{\rm amo, cd}$, $M_{\rm H_{2}O, cd} = M_{\rm wice, cd}$, and $M_{\rm CO, cd} = M_{\rm cice, cd}$ (see Section \ref{sec:accretion}).
Figures \ref{fig:5}(a)--\ref{fig:5}(c) show the spatial distribution of $\zeta_{i}$ at $t = 1~{\rm Myr}$ for $\omega_{\rm cd} = 3 \times 10^{-15}~{\rm s^{-1}}$, $1 \times 10^{-14}~{\rm s^{-1}}$, and $3 \times 10^{-14}~{\rm s^{-1}}$, respectively.
In these calculations, we set $a_{\rm agg, in} = a_{\rm agg, out} = 0.1~{\rm mm}$, as in the cases shown in Figure \ref{fig:4}.
In all cases, $\zeta_{i} \approx 1$ and is nearly constant with $r$, indicating that dust aggregates are dynamically coupled to the gas.

Figures \ref{fig:5}(d)--\ref{fig:5}(f) show the radial profiles of the oxygen isotopic composition at $t = 1~{\rm Myr}$ for each component (i.e., silicate, H$_{2}$O, or CO) and phase (e.g., solid or gas).
Solid lines represent the solid-phase components, whereas dashed lines represent the gas-phase components.
As discussed above, when $\omega_{\rm cd}$ is small (i.e., $r_{\rm c}$ is small), material accreted from the molecular cloud core experiences high temperatures with $T \gg 500~{\rm K}$ before being transported to the outer disk; consequently, the material in the outer disk does not retain the oxygen isotopic composition inherited from the cloud core (Figure \ref{fig:5}(d)).
In contrast, when $\omega_{\rm cd}$ is large (i.e., $r_{\rm c}$ is large), the material in the outer disk largely preserves the oxygen isotopic composition of the molecular cloud core (Figure \ref{fig:5}(f)).
For reference, the radial profiles of $\zeta_{i}$ and $\Delta^{17}{\rm O}_{i}$ at the end of mass infall from the molecular cloud core ($t = 0.1~{\rm Myr}$) are also shown in Appendix \ref{app:zeta_delta} (see Figure \ref{fig:A1}).

Figures \ref{fig:5}(g)--\ref{fig:5}(i) show the mean oxygen isotopic compositions of the solid and vapor phases at $t = 1~{\rm Myr}$, denoted by $\Delta^{17}{\rm O}_{\rm solid}$ and $\Delta^{17}{\rm O}_{\rm vapor}$, respectively.
These quantities are defined as follows:
\begin{align}
\Delta^{17}{\rm O}_{\rm solid} & = \frac{U_{\rm amo} + U_{\rm cry} + U_{\rm cond} + U_{\rm wice} + U_{\rm cice}}{ \mathcal{N}_{\rm amo} + \mathcal{N}_{\rm cry} + \mathcal{N}_{\rm cond} + \mathcal{N}_{\rm wice} + \mathcal{N}_{\rm cice} }, \\
\Delta^{17}{\rm O}_{\rm vapor} & = \frac{U_{\rm svap} + U_{\rm wvap} + U_{\rm cvap}}{ \mathcal{N}_{\rm svap} + \mathcal{N}_{\rm wvap} + \mathcal{N}_{\rm cvap} }.
\end{align}

Beyond the H$_{2}$O snow line, both silicates and H$_{2}$O exist in the solid phase.
If each reservoir satisfies $\zeta_{i} \approx 1$ and largely preserves the molecular-cloud-core composition, $\Delta^{17}{\rm O}_{\rm solid}$ in the outer disk beyond the snow line is $\Delta^{17}{\rm O}_{\rm solid} = {( \chi_{\rm H_{2}O} \Delta^{17}{\rm O}_{\rm H_{2}O, 0} + \Delta^{17}{\rm O}_{\rm sil, 0} )} / {( \chi_{\rm H_{2}O} + 1 )}$.
In our fiducial case \citep[$\chi_{\rm H_{2}O} = 2$;][]{2004Sci...305.1763Y}, this gives $\Delta^{17}{\rm O}_{\rm solid} \sim + 40\tcperthousand$ (Figure \ref{fig:5}(i)).
In contrast, if each reservoir undergoes sufficient isotope exchange in the hot region during disk formation, both H$_{2}$O and silicates approach the solar-system average value ($- 30.7\tcperthousand$), yielding $\Delta^{17}{\rm O}_{\rm solid} \approx - 30\tcperthousand$ (Figure \ref{fig:5}(g)).

Cosmochemical analyses of the oxygen isotopic compositions of chondrules in carbonaceous chondrites suggest that $\Delta^{17}{\rm O}_{\rm solid}$ beyond the snow line was approximately $\Delta^{17}{\rm O}_{\rm solid} \sim 0\tcperthousand$ \citep[e.g.,][]{2018crpd.book..196T}.
This condition can be achieved when the angular velocity of the parental molecular cloud core is moderate, which corresponds to $\omega_{\rm cd} \approx 1 \times 10^{-14}~{\rm s^{-1}}$ ($r_{\rm c, fin} \approx 10~{\rm au}$) in our model setup (Figure \ref{fig:5}(h)).

\subsection{Dependence on $a_{\rm agg}$}
\label{sec:a_agg}

In Section \ref{sec:omega}, we showed how the spatial distribution of the oxygen isotopic composition varies with $\omega_{\rm cd}$, particularly beyond the H$_{2}$O snow line.
In contrast, we find that $\Delta^{17}{\rm O}_{\rm solid} \approx - 30\tcperthousand$ in the region interior to the snow line, with little dependence on $\omega_{\rm cd}$, as long as we assume $a_{\rm agg, in} = a_{\rm agg, out} = 0.1~{\rm mm}$.
\citet{2004Sci...305.1763Y} proposed a scenario in which, if dust aggregates grow to large sizes beyond the snow line, differential radial drift makes the region interior to the snow line more H$_{2}$O-rich than the molecular-cloud-core composition; subsequent oxygen isotope exchange between H$_{2}$O and silicates drives silicates to become more $^{16}$O-poor than the solar-system average of $- 30\tcperthousand$.
In Section \ref{sec:a_agg}, we discuss how the oxygen isotopic composition in the region interior to the snow line varies with the dust aggregate radius, $a_{\rm agg}$.

In this section, we consider three cases for the dust aggregate radius: (i) $a_{\rm agg, in} = 10~{\rm mm}$ and $a_{\rm agg, out} = 0.1~{\rm mm}$, (ii) $a_{\rm agg, in} = 0.1~{\rm mm}$ and $a_{\rm agg, out} = 10~{\rm mm}$, and (iii) $a_{\rm agg, in} = 10~{\rm mm}$ and $a_{\rm agg, out} = 10~{\rm mm}$.
Here, we set $\omega_{\rm cd} = 1 \times 10^{-14}~{\rm s^{-1}}$ for all three cases.
Whether icy aggregates or non-icy aggregates grow more efficiently in disks remains an open question \citep[e.g.,][]{2024A&A...682A.144D, 2024NatAs...8.1148U}.
Astronomical observations suggest that aggregate radii are typically in the range $0.1$--$10~{\rm mm}$ \citep[e.g.,][]{2022A&A...664A.137G, 2023ApJ...957...11D}; we adopt the lower and upper bounds of this range.

\begin{figure*}[]
\centering
\includegraphics[width = \textwidth]{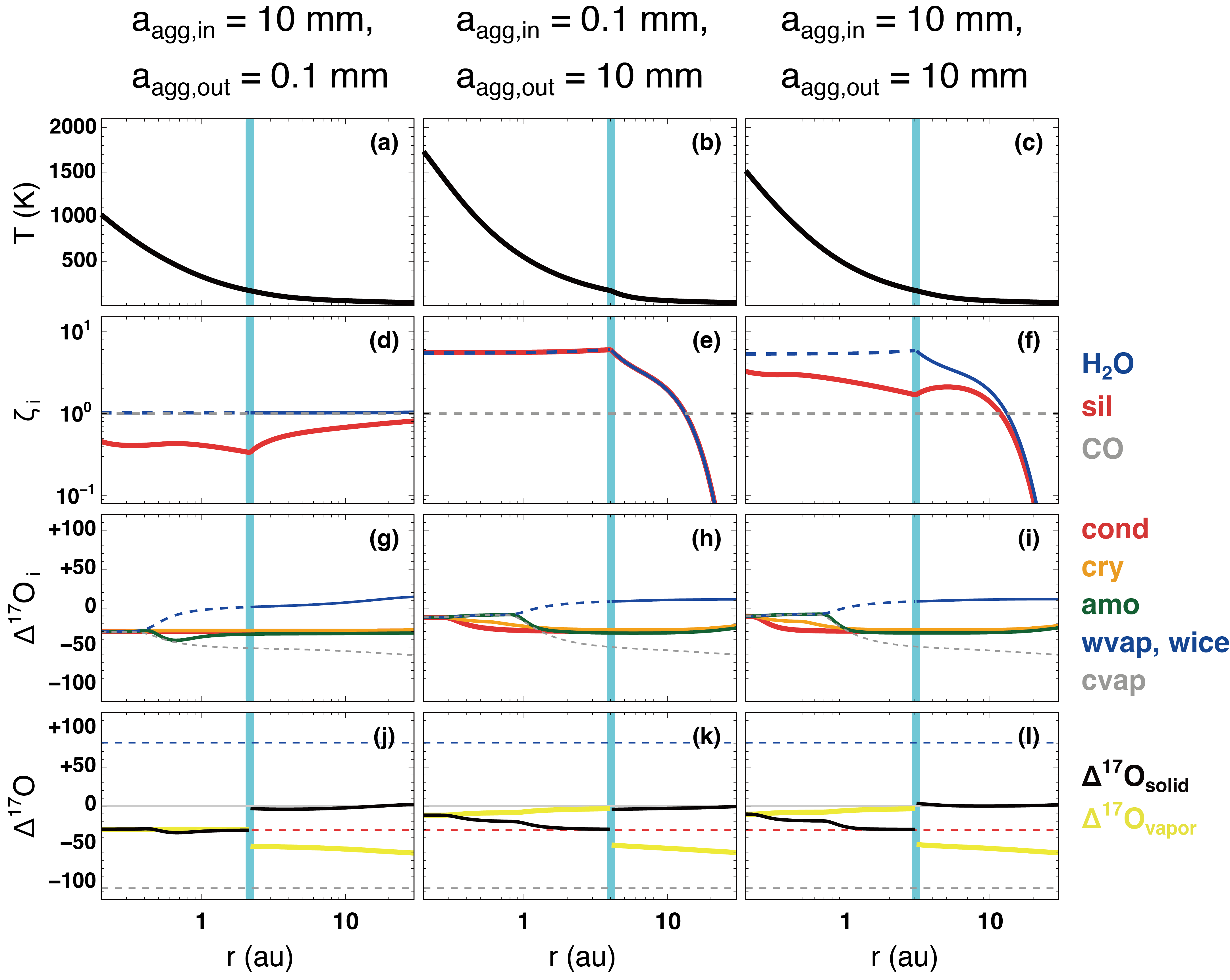}
\caption{
Radial distribution of the disk midplane temperature, each oxygen reservoir, and their oxygen isotopic composition at $t = 1~{\rm Myr}$.
The left, middle, and right columns correspond to case (i) ($a_{\rm agg, in} = 10~{\rm mm}$ and $a_{\rm agg, out} = 0.1~{\rm mm}$), case (ii) ($a_{\rm agg, in} = 0.1~{\rm mm}$ and $a_{\rm agg, out} = 10~{\rm mm}$), and case (iii) ($a_{\rm agg, in} = 10~{\rm mm}$ and $a_{\rm agg, out} = 10~{\rm mm}$), respectively.
We set $\omega_{\rm cd} = 1 \times 10^{-14}~{\rm s^{-1}}$ (i.e., $r_{\rm c, fin} \approx 10~{\rm au}$) in these calculations.
The vertical cyan lines represent the location of H$_{2}$O snow line.
Panels (a)--(c) show the disk midplane temperature, $T$.
Panels (d)--(f) show the enrichment factor of each oxygen reservoir, $\zeta_{i}$ ($i$ denotes {\sf sil}, {\sf H$_{2}$O}, or {\sf CO}).
Panels (g)--(i) show the oxygen isotopic composition of each species, $\Delta^{17}{\rm O}_{i}$ ($i$ denotes {\sf amo}, {\sf cry}, {\sf cond}, {\sf wice}, {\sf wvap}, or {\sf cvap}).
Panels (j)--(l) show the mean oxygen isotopic compositions of the solid and vapor phases, $\Delta^{17}{\rm O}_{\rm solid}$ and $\Delta^{17}{\rm O}_{\rm vapor}$.
The red, blue, and gray horizontal dashed lines in panels (j)--(l) indicate the oxygen isotopic compositions of the corresponding reservoirs in the parental cloud core.
}
\label{fig:6}
\end{figure*}

Figures \ref{fig:6}(a)--\ref{fig:6}(c) show the disk midplane temperature, $T$, at $t = 1~{\rm Myr}$, and Figures \ref{fig:6}(d)--\ref{fig:6}(f) show the enrichment factors of each oxygen reservoir, $\zeta_{i}$, at the same time.
For case (i) with $a_{\rm agg, in} = 10~{\rm mm}$ and $a_{\rm agg, out} = 0.1~{\rm mm}$, we find $\zeta_{\rm sil} < 1$ in the region interior to the snow line (red line in Figure \ref{fig:6}(d)).
In contrast, in cases (ii) and (iii) with $a_{\rm agg, out} = 10~{\rm mm}$, $\zeta_{\rm sil} > 1$ interior to the snow line (Figures \ref{fig:6}(e) and \ref{fig:6}(f)).
Inside the snow line, $T$ is regulated primarily by accretion heating and increases with $\Sigma_{\rm sil}$.
Therefore, in the case with $a_{\rm agg, in} = 10~{\rm mm}$ and $a_{\rm agg, out} = 0.1~{\rm mm}$ (Figure \ref{fig:6}(a)), the temperature interior to the snow line is lower than in the other two cases with $a_{\rm agg, out} = 10~{\rm mm}$ (Figures \ref{fig:6}(b) and \ref{fig:6}(c)).

The blue and gray lines in Figures \ref{fig:6}(d)--\ref{fig:6}(f) represent the enrichment factors of H$_{2}$O and CO, respectively.
Since the condensation temperature of CO is $20~{\rm K}$, CO remains in the gas phase over nearly the entire disk and is dynamically coupled to the disk gas (H$_{2}$ and He); thus, in all cases, we obtain $\zeta_{\rm CO} \approx 1$ throughout the disk.
In contrast, H$_{2}$O exists in the solid phase in regions with $T < 170~{\rm K}$.
When $a_{\rm agg, out} = 0.1~{\rm mm}$, icy aggregates are dynamically well coupled to the disk gas, yielding $\zeta_{\rm H_{2}O} \approx 1$ across the disk (Figure \ref{fig:6}(d)).
In contrast, when $a_{\rm agg, out} = 10~{\rm mm}$, differential radial drift leads to $\zeta_{\rm H_{2}O} > 1$ interior to the snow line (Figures \ref{fig:6}(e) and \ref{fig:6}(f)).
Moreover, interior to the snow line, H$_{2}$O is present in the gas phase, and $\zeta_{\rm H_{2}O}$ is nearly uniform with radius in this inner region at $t = 1~{\rm Myr}$.

Figures \ref{fig:6}(g)--\ref{fig:6}(i) show the radial profiles of the oxygen isotopic composition at $t = 1~{\rm Myr}$ for each component (i.e., silicate, H$_{2}$O, or CO) and phase (e.g., solid or gas).
For $a_{\rm agg, in} = 10~{\rm mm}$ and $a_{\rm agg, out} = 0.1~{\rm mm}$ (Figure \ref{fig:6}(g)), the oxygen isotopic compositions of crystalline (yellow) and condensate (red) silicates are spatially almost uniform at $\approx -30\tcperthousand$.
In contrast, the composition of amorphous silicates (green) varies with radius interior to the snow line.
At $r \lesssim 0.6~{\rm au}$ ($T \gtrsim 500~{\rm K}$), amorphous silicates equilibrate with CO, and at $r \lesssim 0.4~{\rm au}$ ($T \gtrsim 700~{\rm K}$), they further equilibrate with H$_{2}$O.
These equilibration fronts are consistent with experimentally reported temperatures for isotope exchange reactions \citep{2024GeCoA.374...93Y}.
Importantly, even for amorphous silicates, isotope exchange with H$_{2}$O and CO does not proceed at $T \sim 170~{\rm K}$, but only in hot regions with $T \gtrsim 500~{\rm K}$.

For $a_{\rm agg, in} = 0.1~{\rm mm}$ and $a_{\rm agg, out} = 10~{\rm mm}$ (Figure \ref{fig:6}(h)), amorphous silicates equilibrate with CO and H$_{2}$O at $r \lesssim 0.9~{\rm au}$ ($T \gtrsim 600~{\rm K}$).
In contrast, crystalline and condensate silicates equilibrate with CO and H$_{2}$O only at $r \lesssim 0.3~{\rm au}$ ($T \gtrsim 1300~{\rm K}$).
This oxygen isotope exchange temperature of $\sim 1300~{\rm K}$ for crystalline forsterite is consistent with estimates in previous studies \citep[e.g.,][]{2024GeCoA.374...93Y} and is significantly higher than the isotope exchange temperature for amorphous forsterite ($\sim 500~{\rm K}$).

In Section \ref{sec:a_agg}, we consider a disk formed from a molecular cloud core with a moderate angular velocity of $\omega_{\rm cd} = 1 \times 10^{-14}~{\rm s^{-1}}$, for which the disk crystallinity reaches $\sim 50\%$ (Figure \ref{fig:4}).
In this situation, interior to the snow line, $\Delta^{17}{\rm O}_{\rm solid}$ exhibits a two-step radial transition (Figure \ref{fig:6}(k)).
We find that $\Delta^{17}{\rm O}_{\rm solid} \sim -30\tcperthousand$, close to the solar-system average, just inside the snow line.
In regions with $T \gtrsim 600~{\rm K}$, amorphous silicates become $^{16}$O-poor, resulting in an increase in $\Delta^{17}{\rm O}_{\rm solid}$.
Moreover, in regions with $T \gtrsim 1300~{\rm K}$, crystalline silicates also become $^{16}$O-poor, leading to a further increase in $\Delta^{17}{\rm O}_{\rm solid}$.
A similar trend is also observed for the case with $a_{\rm agg, in} = 10~{\rm mm}$ and $a_{\rm agg, out} = 10~{\rm mm}$ (Figures \ref{fig:6}(i) and \ref{fig:6}(l)).

For reference, the radial profiles of $\zeta_{i}$ and $\Delta^{17}{\rm O}_{i}$ at the end of mass infall from the molecular cloud core ($t = 0.1~{\rm Myr}$) are also shown in Appendix \ref{app:zeta_delta} (see Figure \ref{fig:A2}).
At this time, $\zeta_{i} \sim 1$ throughout the disk for all three species, indicating that the radial drift timescales of dust aggregates are longer than $0.1~{\rm Myr}$ and that the aggregates remain dynamically coupled to the gas during the early stage of disk formation, when viscous expansion drives the outward transport of disk material (see Section \ref{sec:cai}).

\subsection{Dependence on the initial relative abundance and isotopic composition in the cloud core}
\label{sec:chi_and_C}

In this study, we adopt $\chi_{\rm H_2O} = 2$ and $\chi_{\rm CO} = 3$ as fiducial values for the relative abundances of H$_{2}$O and CO in the molecular cloud core.
These values were assumed in \citet{2004Sci...305.1763Y} and are based on the solar chemical composition model of \citet{1998A&A...330..375G}.
However, observations of ices in molecular clouds often suggest $\chi_{\rm H_2O} \lesssim 1$ and $\chi_{\rm CO} \lesssim 1$ \citep[e.g.,][]{2010ApJ...710.1009W, 2013ChRv..113.9043V}.
Although $\chi_{\rm H_{2}O}$ and $\chi_{\rm CO}$ in the solar system's parental molecular cloud core do not necessarily have to match these typical observational values, the assumptions adopted by \citet{2004Sci...305.1763Y} should be regarded as highly uncertain.
In addition, there is no strong basis for the absolute oxygen isotopic compositions assumed for each reservoir in the cloud core, $\updelta^{18}{\rm O}_{\rm H_{2}O, 0} = + 180\tcperthousand$ and $\updelta^{18}{\rm O}_{\rm CO, 0} = - 220\tcperthousand$, which are often treated as canonical reference values \citep[e.g.,][]{2017M&PS...52.1797A, 2022Eleme..18..175V}.
In Section \ref{sec:chi_and_C}, we examine how the spatial distribution of oxygen isotopic compositions changes when we vary $\chi_{\rm H_{2}O}$ and $\updelta^{18}{\rm O}_{\rm H_{2}O, 0}$.

Here, we assume $\omega_{\rm cd} = 3 \times 10^{-14}~{\rm s^{-1}}$ (i.e., $r_{\rm c, fin} \approx 100~{\rm au}$), $a_{\rm agg, in} = 0.1~{\rm mm}$, and $a_{\rm agg, out} = 10~{\rm mm}$.
With this setup, most of the material accreted from the molecular cloud core is transported to the cold outer disk without being heated above $\sim 500~{\rm K}$ (Section \ref{sec:omega}) and, except for the innermost region of the disk where $T \gtrsim 800~{\rm K}$, the crystallinity remains low, with $f_{\rm cry} \ll 1$.
In addition, differential radial drift of icy aggregates yields $\zeta_{\rm H_{2}O} > 1$ interior to the snow line at $t = 1~{\rm Myr}$, and in hot regions with $T \gtrsim 500~{\rm K}$, silicate dust becomes more $^{16}$O-poor than the solar-system average (Section \ref{sec:a_agg}).

\begin{figure*}[]
\centering
\includegraphics[width = \textwidth]{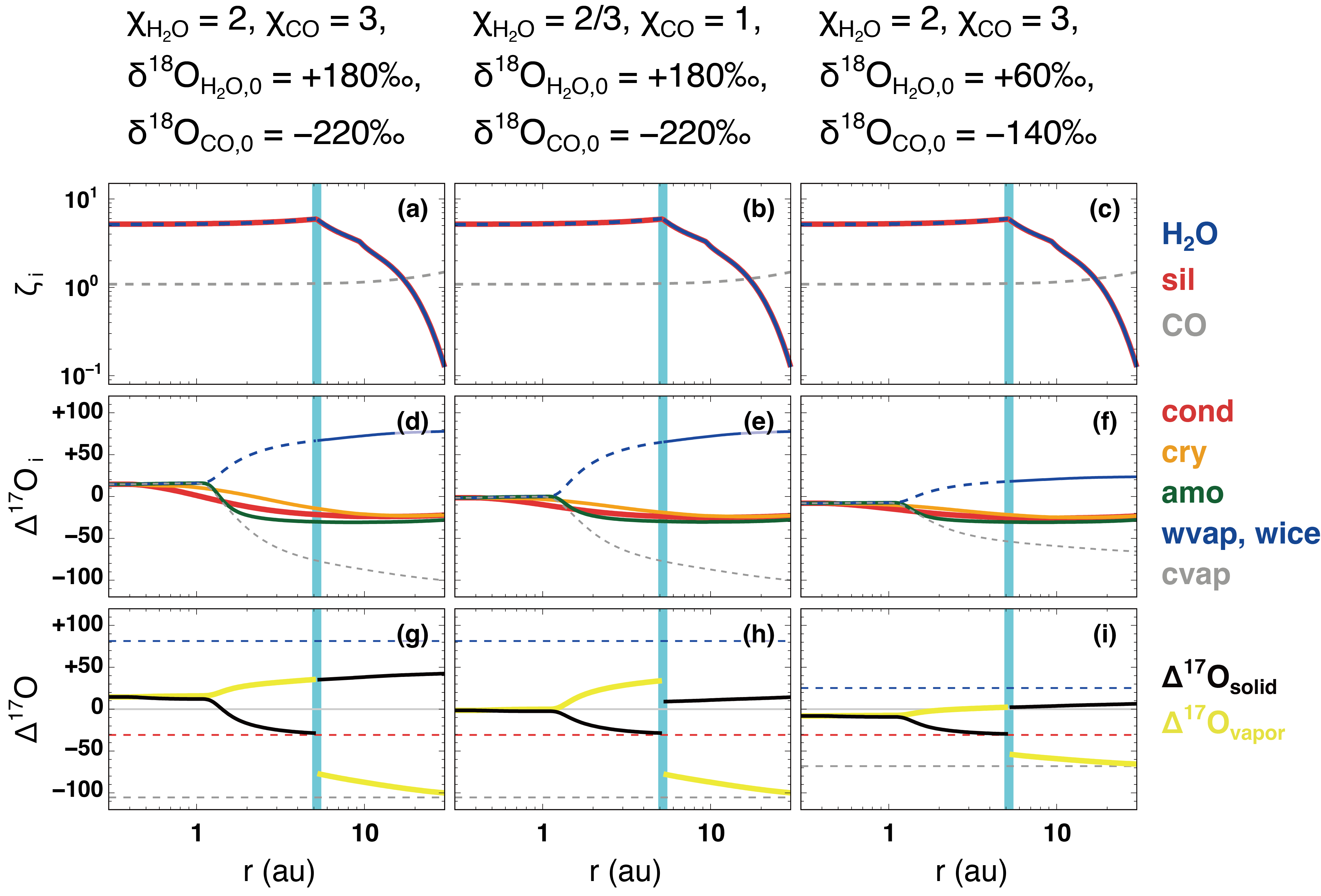}
\caption{
Radial distribution of each oxygen reservoir and their oxygen isotopic composition at $t = 1~{\rm Myr}$.
The left, middle, and right columns correspond to the fiducial case ($\chi_{\rm H_{2}O} = 2$, $\chi_{\rm CO} = 3$, $\updelta^{18}{\rm O}_{\rm H_{2}O, 0} = + 180\tcperthousand$ (i.e., $\Delta^{17}{\rm O}_{\rm H_{2}O, 0} = + 81.4\tcperthousand$), and $\updelta^{18}{\rm O}_{\rm CO, 0} = - 220\tcperthousand$ (i.e., $\Delta^{17}{\rm O}_{\rm CO, 0} = - 105.4\tcperthousand$)), the ice-depleted case ($\chi_{\rm H_{2}O} = 2/3$, $\chi_{\rm CO} = 1$, $\updelta^{18}{\rm O}_{\rm H_{2}O, 0} = + 180\tcperthousand$, and $\updelta^{18}{\rm O}_{\rm CO, 0} = - 220\tcperthousand$), and the weaker CO self-shielding case ($\chi_{\rm H_{2}O} = 2$, $\chi_{\rm CO} = 3$, $\updelta^{18}{\rm O}_{\rm H_{2}O, 0} = + 60\tcperthousand$ (i.e., $\Delta^{17}{\rm O}_{\rm H_{2}O, 0} = + 25.3\tcperthousand$), and $\updelta^{18}{\rm O}_{\rm CO, 0} = - 140\tcperthousand$ (i.e., $\Delta^{17}{\rm O}_{\rm CO, 0} = - 68.1\tcperthousand$)), respectively.
We set $a_{\rm agg, in} = 0.1~{\rm mm}$, $a_{\rm agg, out} = 10~{\rm mm}$, and $\omega_{\rm cd} = 3 \times 10^{-14}~{\rm s^{-1}}$ (i.e., $r_{\rm c, fin} \approx 100~{\rm au}$) in these calculations.
The vertical cyan lines represent the location of H$_{2}$O snow line.
Panels (a)--(c) show the enrichment factor of each oxygen reservoir, $\zeta_{i}$ ($i$ denotes {\sf sil}, {\sf H$_{2}$O}, or {\sf CO}).
Panels (d)--(f) show the oxygen isotopic composition of each species, $\Delta^{17}{\rm O}_{i}$ ($i$ denotes {\sf amo}, {\sf cry}, {\sf cond}, {\sf wice}, {\sf wvap}, or {\sf cvap}).
Panels (g)--(i) show the mean oxygen isotopic compositions of the solid and vapor phases, $\Delta^{17}{\rm O}_{\rm solid}$ and $\Delta^{17}{\rm O}_{\rm vapor}$.
The red, blue, and gray horizontal dashed lines in panels (g)--(i) indicate the oxygen isotopic compositions of the corresponding reservoirs in the parental cloud core.
}
\label{fig:7}
\end{figure*}

We demonstrate three models with different values of $\chi_{\rm H_{2}O}$ and $\updelta^{18}{\rm O}_{\rm H_{2}O, 0}$.
Figures \ref{fig:7}(a)--\ref{fig:7}(c) show the radial profiles of $\zeta_{i}$ at $t = 1~{\rm Myr}$, and Figures \ref{fig:7}(d)--\ref{fig:7}(f) show the radial profiles of $\Delta^{17}{\rm O}_{i}$ at the same time.
Figures \ref{fig:7}(g)--\ref{fig:7}(i) show the radial profiles of $\Delta^{17}{\rm O}_{\rm solid}$ and $\Delta^{17}{\rm O}_{\rm vapor}$.

Assuming the fiducial case of $\chi_{\rm H_{2}O} = 2$ and $\chi_{\rm CO} = 3$, we obtain $\Delta^{17}{\rm O}_{\rm solid} \sim +35$--$45\tcperthousand$ outside the snow line, $\Delta^{17}{\rm O}_{\rm solid} \sim -30\tcperthousand$ just interior to the snow line, and $\Delta^{17}{\rm O}_{\rm solid} \sim +10\tcperthousand$ at $r \lesssim 1~{\rm au}$ (Figure \ref{fig:7}(g)).
In contrast, for a more ice-depleted molecular cloud core composition with $\chi_{\rm H_{2}O} = 2/3$ and $\chi_{\rm CO} = 1$, we obtain $\Delta^{17}{\rm O}_{\rm solid} \sim +5$--$15\tcperthousand$ outside the snow line, $\Delta^{17}{\rm O}_{\rm solid} \sim -30\tcperthousand$ just interior to the snow line, and $\Delta^{17}{\rm O}_{\rm solid} \sim 0\tcperthousand$ at $r \lesssim 1~{\rm au}$ (Figure \ref{fig:7}(h)).
Moreover, if we assume weaker CO self-shielding in the parental molecular cloud, adopting $\updelta^{18}{\rm O}_{\rm H_{2}O, 0} = +60\tcperthousand$ and $\updelta^{18}{\rm O}_{\rm CO, 0} = -140\tcperthousand$, we obtain $\Delta^{17}{\rm O}_{\rm solid} \sim 0$--$10\tcperthousand$ outside the snow line, $\Delta^{17}{\rm O}_{\rm solid} \sim -30\tcperthousand$ just interior to the snow line, and $\Delta^{17}{\rm O}_{\rm solid} \sim -10\tcperthousand$ at $r \lesssim 1~{\rm au}$ (Figure \ref{fig:7}(i)).

If chondrules in carbonaceous chondrites formed in a solids-rich environment with $\Delta^{17}{\rm O}_{\rm solid} \sim 0\tcperthousand$ \citep[e.g.,][]{2018crpd.book..196T}, then either an ice-depleted molecular cloud core composition relative to the fiducial setting, or a scenario with weaker CO self-shielding, is favored for explaining their isotopic compositions.
Alternatively, moderate mixing of high-temperature materials in a disk formed with a moderate $\omega_{\rm cd}$ could also be plausible from the viewpoint of achieving $\Delta^{17}{\rm O}_{\rm solid} \sim 0\tcperthousand$ (Section \ref{sec:omega}).
We also emphasize that, in all cases, $\Delta^{17}{\rm O}_{\rm solid}$ becomes the solar-system average of $\approx -30\tcperthousand$ just interior to the snow line, which may not be suitable as a chondrule-forming environment (Section \ref{sec:inside}).

\section{Discussion}
\label{sec:discussion}

\subsection{Chondrule formation beyond the H$_{2}$O snow line}
\label{sec:beyond}

Here, we discuss a possible pathway for chondrule formation beyond the H$_{2}$O snow line after disk formation.
Previous studies \citep[e.g.,][]{2018crpd.book..196T} have reported a trend between the oxygen isotopic composition of chondrules in carbonaceous chondrites and their Mg\#.
Mg\# is a useful proxy for the redox state of the chondrule-forming environment and, in turn, for the dust-enrichment factor \citep[e.g.,][]{2018crpd.book..151E}.
In CM and CO carbonaceous chondrites, reduced chondrules with Mg\# $\approx 99$ show $\Delta^{17}{\rm O} \approx -5\tcperthousand$, whereas oxidized chondrules with Mg\# $\lesssim 98$ show $\Delta^{17}{\rm O} \approx -2\tcperthousand$ \citep[e.g.,][]{2018crpd.book..196T, 2021GeCoA.299..199C}.\footnote{
Much higher $\Delta^{17}{\rm O}$ values of $\approx +1\tcperthousand$ are observed in some oxidized chondrules from comet 81P/Wild 2, the Tagish Lake meteorite, and CR chondrites \citep[e.g.,][]{2021GeCoA.293..328U, 2024GeCoA.371..214Z}.
}

To explain the Mg\#--$\Delta^{17}{\rm O}$ trend, in which chondrules formed under more oxidizing conditions (lower Mg\#) exhibit higher $\Delta^{17}{\rm O}$ values, \citet{2015GeCoA.148..228T} and \citet{2018GeCoA.224..116H} proposed that chondrules formed in environments enriched in H$_{2}$O-ice-bearing silicate dust.
The H$_{2}$O-ice and silicate-dust abundances relative to H$_{2}$ gas are inferred to be enriched by a factor of typically $\sim 10^{2}$ for reduced chondrules and even more for oxidized chondrules, relative to the solar abundance ratio.
In this case, the oxygen isotope ratios of chondrules likely reflect the average oxygen isotopic composition of the solid components in the chondrule-forming region.

In Section \ref{sec:results}, we showed that disk formation via infall from the molecular cloud core can yield a mean isotopic composition of the solid phase beyond the snow line that is more $^{16}$O-poor than the solar-system average, i.e., $\Delta^{17}{\rm O}_{\rm solid} \sim 0\tcperthousand$.
However, in this case, silicates and H$_{2}$O ice can have markedly different isotopic compositions.
For example, in Figure \ref{fig:6}(h), silicates have $\Delta^{17}{\rm O}_{\rm amo} \approx \Delta^{17}{\rm O}_{\rm cry} \approx -30\tcperthousand$, whereas $\Delta^{17}{\rm O}_{\rm wice} \approx +10\tcperthousand$.
In contrast, chondrules in carbonaceous chondrites exhibit oxygen isotopic compositions closer to $\Delta^{17}{\rm O} \approx -2\tcperthousand$ and $-5\tcperthousand$ \citep[e.g.,][]{2018crpd.book..196T, 2021GeCoA.299..199C}.
This discrepancy in $\Delta^{17}{\rm O}$ may be interpreted as the result of efficient isotope exchange and equilibration between silicates and H$_{2}$O ice during silicate-melting flash-heating events that formed chondrules.
Since a large fraction of chondrules are inferred to have experienced multiple heating events \citep[e.g.,][]{1993Metic..28...14W, 2018crpd.book..196T}, such equilibration may also have occurred repeatedly in the solar protoplanetary disk.

Based on a mass-balance model \citep[e.g.,][]{2015GeCoA.148..228T, 2018GeCoA.224..116H}, the Mg\#--$\Delta^{17}{\rm O}$ trend of chondrules suggests that reduced chondrules with $\Delta^{17}{\rm O} \approx -5\tcperthousand$ and Mg\# $\approx 99$ formed in an environment with a lower H$_{2}$O-to-silicate mass ratio than that of the environment in which oxidized chondrules with $\Delta^{17}{\rm O} \approx -2\tcperthousand$ and Mg\# $\lesssim 98$ formed.
This difference in the H$_{2}$O-to-silicate mass ratio might have resulted from spatiotemporal variations in the accumulation of silicate and H$_{2}$O ice into the disk midplane \citep[e.g.,][]{2006Icar..181..178C}, where chondrules have formed.
However, here we discuss another possibility.

\begin{figure*}[]
\centering
\includegraphics[width = 0.8\textwidth]{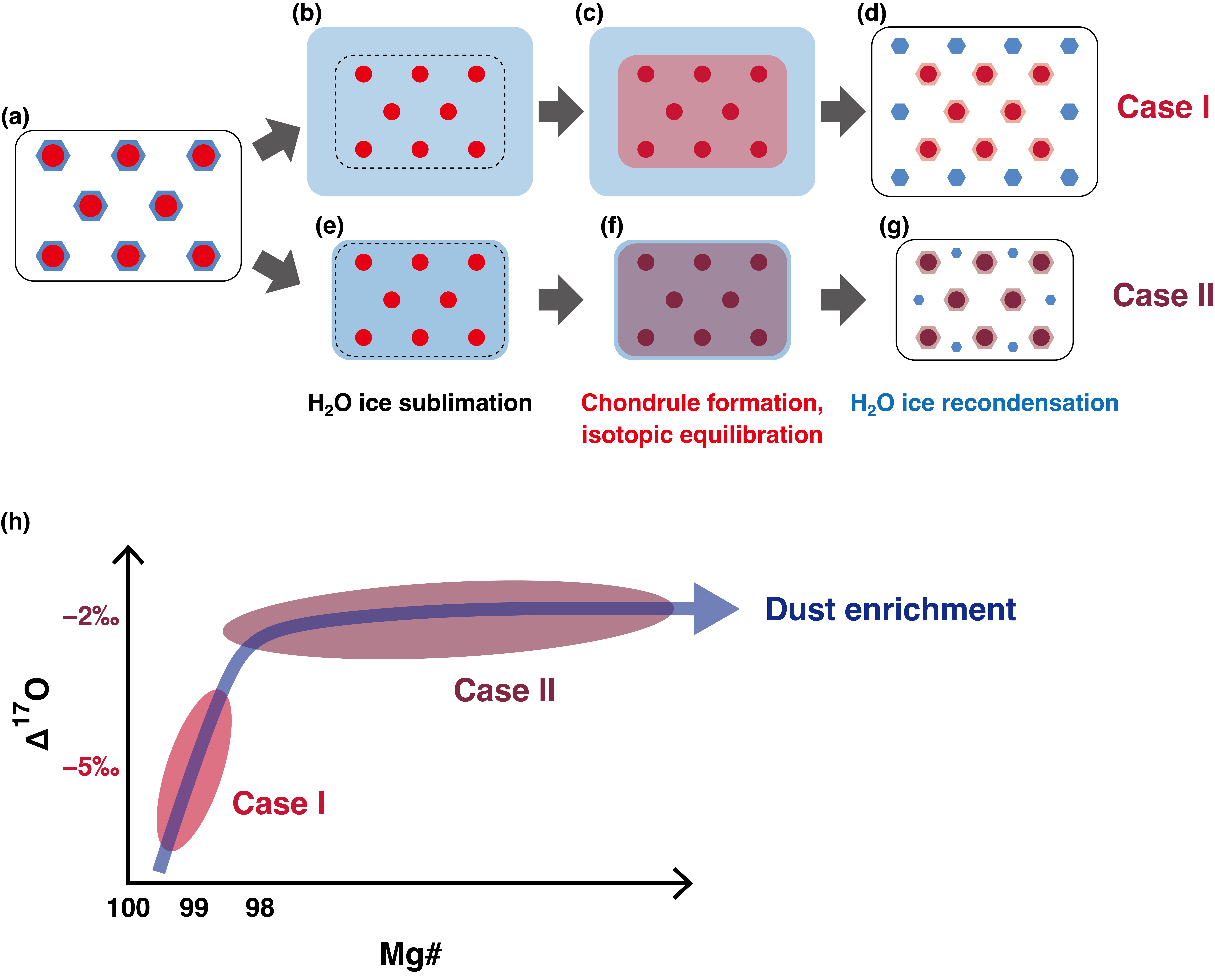}
\caption{
A schematic scenario that may explain the observed trend between the oxygen isotopic composition of chondrules and Mg\#.
(a) We consider an initial condition in which the mean isotopic composition of the solid phase in regions exterior to the snow line is $\Delta^{17}{\rm O}_{\rm solid} \approx -2\tcperthousand$.
Initially, silicates (red circles) have an isotopic composition close to the solar-system average ($\Delta^{17}{\rm O} \sim -30\tcperthousand$), whereas H$_{2}$O ice (blue hexagons) has a more $^{16}$O-poor composition.
(b) Localized flash heating beyond the snow line.
Part of the H$_{2}$O vapor may diffuse outward from the chondrule-forming heated region (enclosed by the dashed line).
(c) Chondrule formation.
During this stage, silicates exchange oxygen isotopes with the H$_{2}$O vapor present in the heated region.
(d) Subsequently, chondrules crystallize and H$_{2}$O vapor recondenses.
The H$_{2}$O that diffuses outward from the heated region avoids isotope exchange and retains its original composition.
Thus, isotope exchange between H$_{2}$O and silicates is incomplete, potentially explaining the formation of chondrules with $\Delta^{17}{\rm O} \sim -5\tcperthousand$ in regions enriched in icy dust with $\Delta^{17}{\rm O}_{\rm solid} \approx -2\tcperthousand$ (Case I).
(e--g) If the dust-enrichment factor in the heated region is sufficiently high, the dynamical back-reaction of silicates may inhibit vapor escape, potentially bringing the $\Delta^{17}{\rm O}$ value of chondrules closer to the $\Delta^{17}{\rm O}_{\rm solid}$ value (Case II).
(h) Schematic illustration of the $\Delta^{17}{\rm O}$--Mg\# systematics of chondrules in carbonaceous chondrites \citep[e.g.,][]{2012GeCoA..90..242U, 2018crpd.book..196T, 2021GeCoA.299..199C}.
The correlation between $\Delta^{17}{\rm O}$ and Mg\# may reflect differences in dust enrichment during chondrule formation.
}
\label{fig:8}
\end{figure*}

We suggest a scenario that may explain the observed trend between the oxygen isotopic composition of chondrules and Mg\# (Figure \ref{fig:8}).
First, disk formation via mass infall from the molecular cloud core, followed by viscous expansion, yields $\Delta^{17}{\rm O}_{\rm solid} \approx -2\tcperthousand$ in regions exterior to the snow line.
In this state, silicates (red circles) have an isotopic composition close to the solar-system average, whereas H$_{2}$O ice (blue hexagons) has a more $^{16}$O-poor composition (Figure \ref{fig:8}(a)).
After disk formation, we assume that localized flash-heating events occur in regions enriched in icy dust.
When icy aggregates are heated, H$_{2}$O ice sublimates, and part of the H$_{2}$O vapor is expected to diffuse outward from the chondrule-forming heated region (Figure \ref{fig:8}(b)).
As the temperature rises further and silicates melt, chondrules form.
During this stage, silicates exchange oxygen isotopes with the H$_{2}$O vapor present in the heated region (Figure \ref{fig:8}(c)).
Subsequently, as the temperature decreases, chondrules crystallize and H$_{2}$O vapor recondenses (Figure \ref{fig:8}(d)).
While H$_{2}$O within the heated region can exchange isotopes with the forming chondrules, the H$_{2}$O that diffuses outward from the heated region avoids isotope exchange and retains its original composition.
In other words, isotope exchange between H$_{2}$O and silicates is incomplete.
This incomplete isotopic exchange could explain the formation of chondrules with $\Delta^{17}{\rm O} \approx -5\tcperthousand$ in regions enriched in icy dust with $\Delta^{17}{\rm O}_{\rm solid} \approx -2\tcperthousand$ (Case I).

The resulting $\Delta^{17}{\rm O}$ value of chondrules after isotope exchange between H$_{2}$O and silicates depends on how much H$_{2}$O vapor escapes from the heated region.
If the dust-enrichment factor in the heated region is sufficiently high, dynamical back reaction from silicates may inhibit vapor escape \citep[e.g.,][]{1986Icar...67..375N}, potentially bringing the $\Delta^{17}{\rm O}$ value of chondrules closer to the $\Delta^{17}{\rm O}_{\rm solid}$ value (Case II; Figures \ref{fig:8}(e)--\ref{fig:8}(g)).
This scenario is consistent with the observed trend that chondrules inferred to have formed in more dust-enriched environments (Mg\# $\lesssim 98$) exhibit slightly higher $\Delta^{17}{\rm O}$ values ($\approx -2\tcperthousand$) than the other population (Mg\# $\sim 99$ and $\Delta^{17}{\rm O} \approx -5\tcperthousand$; Figure \ref{fig:8}(h)).

Some chondrules contain ``relict'' olivine grains that survived the last melting event \citep[e.g.,][]{2012GeCoA..90..242U}.
These grains are recognized by their distinct $\Delta^{17}{\rm O}$ values relative to the phenocrysts and mesostasis.
Importantly, relict grains in chondrules with $\Delta^{17}{\rm O} \approx -2\tcperthousand$ commonly show counterpart compositions of $\Delta^{17}{\rm O} \approx -5\tcperthousand$, and vice versa.
This observation indicates that (i) two major chondrule formation mechanisms or environments existed, and (ii) some chondrules underwent multiple heating events under different mechanisms or environments.

Note that oxygen isotopic equilibration between chondrule melts and ambient H$_{2}$O vapor cannot be achieved if the duration of chondrule-forming heating events is too short \citep[e.g.,][]{2021GeCoA.314..108Y}.
\citet{2023Icar..40515690A} evaluated the critical duration required for isotopic equilibration and found that, for an H$_{2}$O partial pressure of $\gtrsim 1~{\rm Pa}$, it is $\sim 10^{3}~{\rm s}$.
If the heating duration of chondrule-forming events is shorter than this value, isotope exchange is unlikely to proceed sufficiently.

\subsection{Chondrule formation inside the H$_{2}$O snow line}
\label{sec:inside}

Chondrules in ordinary chondrites are thought to have formed via flash-heating events in dust-enriched regions interior to the snow line, and trends in their oxygen isotopic compositions suggest that the chondrule-forming region had $\Delta^{17}{\rm O}_{\rm solid} \approx + 1\tcperthousand$ \citep[e.g.,][]{2018crpd.book..196T}.
In addition, the presence of primary troilite (FeS) within chondrules implies that the ambient temperature of the chondrule formation region was $\lesssim 650~{\rm K}$, and likely $\lesssim 500~{\rm K}$ \citep[e.g.,][]{1999GeCoA..63.2281R}.
However, in such a low-temperature environment, neither amorphous nor crystalline silicates undergo oxygen isotope exchange with CO and H$_{2}$O vapors \citep{2024GeCoA.374...93Y}.
Therefore, in our model, $\Delta^{17}{\rm O}_{\rm solid}$ remains close to the solar-system average ($- 30\tcperthousand$), which is inconsistent with the implications from cosmochemical analyses.
In other words, within a quasi-static, classical framework of disk formation and viscous evolution \citep[e.g.,][]{2005A&A...442..703H, 2012M&PS...47...99Y}, it is likely difficult to account for the oxygen isotope characteristics of ordinary-chondrite chondrules.
Thus, it may be necessary to consider more dynamic and time-variable disk processes, including outflow-driven circulation and/or decretion flows that transport inner-disk materials \citep[e.g.,][]{2021MNRAS.503..162H, 2021ApJ...920L..35T, 2024PASJ...76..616I, 2025ApJ...985...16T}.

\subsection{Calcium--aluminum-rich inclusions}
\label{sec:cai}

Calcium--aluminum-rich inclusions (CAIs) in chondrites are the oldest known refractory condensates in our solar system \citep[e.g.,][]{2010E&PSL.300..343A, 2012Sci...338..651C}.
As CAIs formed in the innermost region of the solar protoplanetary disk, measurements of their oxygen isotopic compositions may provide constraints on the oxygen isotopic composition of the innermost disk.
In our disk-formation model, the oxygen isotopic composition of condensates transported beyond the snow line remains identical to the solar-system average.
This is because the outward transport of condensates driven by viscous expansion occurs earlier than the increase in $\chi_{\rm H_{2}O}$ in the inner disk caused by the radial migration of icy aggregates (see Appendix \ref{app:zeta_delta}).

Our finding that condensates transported beyond the snow line had uniformly $^{16}{\rm O}$-rich isotopic compositions is basically consistent with previous observations of CAIs and AOAs in unmetamorphosed carbonaceous chondrites.
It is known that the oxygen isotopic compositions of primitive (least-altered) CAIs in type 3.0 carbonaceous chondrites show a concentrated distribution around $\Delta^{17}{\rm O} \approx -20\tcperthousand$ to $-25\tcperthousand$ \citep[e.g.,][]{2014E&PSL.401..327B, 2017GeCoA.201..103U, 2019M&PS...54.1647K}.\footnote{
We note that several previous studies \citep[e.g.,][]{2017GeCoA.201...83K, 2024M&PS...59.2388S} have reported that some CAIs exhibit variable oxygen isotopic compositions within a single CAI, which may be due to temporal variations in the oxygen isotopic composition of the innermost disk on a short timescale of $\lesssim 0.2~{\rm Myr}$.
However, our present disk evolution simulations cannot reproduce such short-timescale variability as long as dust aggregates remain no larger than $10~{\rm mm}$ in radius.}
Moreover, primitive AOAs, which are also regarded as high-temperature condensates, exhibit similarly $^{16}$O-rich oxygen isotopic compositions around $-25\tcperthousand$ \citep[e.g.,][]{2019PNAS..11623461M, 2021GeCoA.293..544F}.
Combined with our numerical results, these observations suggest that the oxygen isotopic compositions of CAIs/AOAs, the solar-system average, and the solar photosphere were all similar, with $\Delta^{17}{\rm O} \approx -25\tcperthousand$.

\subsection{Silicate crystallinity}

In the interstellar medium, silicate dust is thought to be amorphous, as indicated by broad and smooth absorption features \citep[e.g.,][]{2004ApJ...609..826K}.
In contrast, crystalline silicate features are often found toward disks around young stars \citep[e.g.,][]{2003ApJ...585L..59H}.
Comets and interplanetary dust particles in our solar system also contain crystalline silicates \citep[e.g.,][]{2007P&SS...55.1044O, 2023PSJ.....4..242H}.
These facts provide evidence that a fraction of the material delivered from the parental molecular cloud core experienced crystallization in the hot inner disk and was subsequently transported outward, possibly via viscous disk expansion \citep[e.g.,][]{2006ApJ...640L..67D}.
The presence of CAIs in comets, Ivuna-type (CI) chondrites, and returned samples from Ryugu and Bennu also supports the idea that outward transport of crystallized dust occurred in the solar protoplanetary disk \citep[e.g.,][]{2006Sci...314.1735Z, 2025ComEE...6..537K}.
As both $f_{\rm cry}$ and $f_{\rm cond}$ in the outer solar system are not negligible, a model with $r_{\rm c, fin} \gg 10~{\rm au}$ may not be appropriate for the solar protoplanetary disk (see Section \ref{sec:omega}).

\subsection{Future perspectives}

In this study, we investigated the spatiotemporal structure of the oxygen isotopic signature of the solar protoplanetary disk based on the classical picture of mass infall from a molecular cloud core and subsequent advection--diffusion transport in the disk \citep[e.g.,][]{2006ApJ...640L..67D, 2012M&PS...47...99Y, 2019ApJ...884...32J}.
In particular, we discussed the conditions under which the oxygen isotopic compositions of chondrules can be reproduced, such as the radial extent of mass infall from the cloud core (a function of the core angular velocity, $\omega_{\rm cd}$) and the dust-aggregate radius, $a_{\rm agg}$.
This work represents only a first step toward quantitatively describing the spatiotemporal evolution of the oxygen isotopic structure of the solar protoplanetary disk, and several important physical processes remain unaccounted for.

The mass-infall process from molecular cloud cores is being investigated in detail with magnetohydrodynamic simulations \citep[e.g.,][]{2023ASPC..534..317T}.
In particular, it has been pointed out that the mass-delivery process to disks can vary substantially depending on the relative orientation between the initial angular-momentum vector of the cloud core and the magnetic field direction \citep[e.g.,][]{2020ApJ...898..118H}.
Moreover, cloud cores are generally not in rigid-body rotation, and when magnetic braking is taken into account, the radial extent of mass deposition onto the disk does not necessarily coincide with the centrifugal radius $r_{\rm c}$ derived from a naive angular-momentum conservation argument \citep[e.g.,][]{2021A&A...648A.101L}.
Several studies have attempted to modify $r_{\rm c}$ to incorporate magnetic effects \citep[e.g.,][]{2022NatAs...6...72M}, but this issue remains under debate.
In addition, ''late accretion'' from the molecular cloud onto the disk might also be important for the mass growth of the protoplanetary disk \citep[e.g.,][]{2023ASPC..534..233P}.
Indeed, accretion flows known as streamers have been observed in several disks, and these streamers may contribute to isotopic heterogeneity in the solar system \citep[e.g.,][]{2023ApJ...947L..29A}.

The evolution of oxygen isotopic compositions is primarily controlled by the relative abundances of silicates, H$_{2}$O, and CO, as well as by the disk temperature structure.
Gas removal due to disk winds would be a plausible physical process that can modify the relative abundances of H$_{2}$O and CO in the region interior to the snow line.
Several theoretical studies have explored how photoevaporative and magnetically driven disk winds affect the structural evolution of protoplanetary disks \citep[e.g.,][]{2018ApJ...865...75N, 2017RSOS....470114E, 2020MNRAS.492.3849K}.
Assessing how disk winds influence the spatiotemporal evolution of isotopic compositions is an important topic for future study.

In addition, planetesimal formation effectively removes solid materials from the disk.
Solids can become concentrated through a variety of mechanisms in disks, leading to planetesimal formation \citep[e.g.,][]{2014prpl.conf..547J, 2025ApJ...983...15T}.
In the late stages of disk evolution, the relative abundances of H$_{2}$O and CO are expected to change as a result of planetesimal formation and gas dispersal.
How such changes affect the oxygen isotopic compositions of chondritic materials remains an important subject for future work.

In this study, we assumed that the temperature structure in the inner disk is set by accretion heating.
This assumption is widely adopted in classical disk evolution simulations \citep[e.g.,][]{2005A&A...442..703H, 2012M&PS...47...99Y}.
However, recent magnetohydrodynamic simulations of disks suggest that accretion heating may be inefficient, potentially leading to lower midplane temperatures $T$ than predicted by classical models \citep[e.g.,][]{2023ApJ...949..119K}.
As oxygen isotope exchange between amorphous silicates and H$_{2}$O/CO vapors requires $T \gtrsim 500~{\rm K}$, inefficient accretion heating would confine such regions to $r \ll 1~{\rm au}$.
Moreover, recent studies have suggested that the vertical structure of the accretion flow may be such that the gas advection velocity $v_{\rm g}$ is small near the midplane but becomes large in the disk surface layers \citep[e.g.,][]{2017ApJ...845...75B, 2024PASJ...76..616I}.
In such a disk, solids may remain in the disk for long times even if the disk gas is removed through accretion onto the central star \citep[e.g.,][]{2025PASJ...77..162O}.
As a result, the disk could become progressively depleted in CO gas over long timescales \citep[e.g.,][]{2016ApJ...828...46A}.

For simplicity, we considered only three oxygen reservoirs: silicates, H$_{2}$O, and CO.
However, additional oxygen-bearing reservoirs have been reported in molecular clouds and protoplanetary disks.
For example, CO$_{2}$ ice is observed in molecular clouds \citep[e.g.,][]{2025NatAs...9..883S} and is also predicted to be present in chemical evolution models of protoplanetary disks \citep[e.g.,][]{2020ApJ...899..134K, 2022ApJ...938...29F}.
Moreover, carbonates have been identified in carbonaceous chondrites \citep[e.g.,][]{2019NatAs...3..910F} as well as in returned samples from C-type asteroids Ryugu and Bennu \citep[e.g.,][]{2023Sci...379.8671N, 2024M&PS...59.2453L}, which further supports the presence of CO$_{2}$ ice in the solar protoplanetary disk.
As CO$_{2}$ forms via chemical reactions involving H$_{2}$O and CO, it may carry an oxygen isotopic composition distinct from those of H$_{2}$O and CO.
Extending our model to include CO$_{2}$ will therefore be essential in future work.
We note, however, that the isotope-exchange kinetics between CO$_{2}$ and silicates are currently poorly constrained, and laboratory measurements will be crucial.

\section{Summary}

Chondrites and their components preserve a record of physicochemical processes in the solar protoplanetary disk.
Chondrules are thought to have formed through transient flash-heating events in dust-enriched regions of the disk; thus, their oxygen isotopic compositions are expected to reflect the mean isotopic composition of the disk's solid component.
Chondrules in carbonaceous chondrites are thought to have formed in the cold outer disk, where H$_{2}$O existed as ice.
In contrast, chondrules in ordinary chondrites are thought to have formed in the hot inner regions, where H$_{2}$O existed as vapor.

The three major oxygen reservoirs in the disk and its parental molecular cloud core are H$_{2}$O, CO, and silicates.
These reservoirs have distinct isotopic compositions in the parental cloud core.
A fraction of the material delivered from the cloud core falls into the hot inner disk, where these reservoirs can exchange oxygen isotopes.
The processed material is subsequently transported outward by radial expansion during the disk-formation stage.
By comparing our numerical modeling with the observed oxygen isotopic characteristics of chondrules, we can discuss plausible conditions for the formation and evolution of the solar protoplanetary disk.

The key findings of this study are summarized as follows:
\begin{itemize}

\item
The radial extent of mass infall, controlled by the cloud-core angular velocity $\omega_{\rm cd}$ (and effectively by $r_{\rm c, fin}$), primarily determines whether solids delivered to the outer disk retain their parental cloud-core isotopic signature or are overprinted by high-temperature processing in the inner disk.

\item
For small aggregates ($a_{\rm agg, in} = a_{\rm agg, out} = 0.1~{\rm mm}$), the enrichment factors of H$_{2}$O, CO, and silicates remain nearly constant ($\zeta_{i} \approx 1$), indicating tight gas--dust coupling and weak differential transport.
In this tightly coupled setting, the mean solid isotopic composition interior to the snow line remains close to the solar-system average ($\Delta^{17}{\rm O}_{\rm solid} \approx -30\tcperthousand$) with little dependence on $\omega_{\rm cd}$, whereas beyond the snow line $\Delta^{17}{\rm O}_{\rm solid}$ can become substantially more $^{16}$O-poor depending on $\omega_{\rm cd}$ (Section \ref{sec:omega}).

\item
When aggregates are large beyond the snow line ($a_{\rm agg, out} = 10~{\rm mm}$), differential radial drift enhances H$_{2}$O vapor abundances interior to the snow line ($\zeta_{\rm H_{2}O} > 1$), enabling $^{16}$O-poor silicates to form through isotope exchange with ambient vapors in regions where $T \gtrsim 500~{\rm K}$.
Since isotope-exchange kinetics differ strongly between amorphous and crystalline silicates, the inner-disk $\Delta^{17}{\rm O}_{\rm solid}$ can exhibit a two-step transition: amorphous silicates become $^{16}$O-poor at $T \gtrsim 500$--$600~{\rm K}$, while crystalline/condensate silicates require much higher temperatures ($T \gtrsim 1300~{\rm K}$) to exchange efficiently (Section \ref{sec:a_agg}).

\item
Matching the inference that $\Delta^{17}{\rm O}_{\rm solid} \sim 0\tcperthousand$ in the chondrule-forming region(s) favors either (i) moderately extended infall (moderate $\omega_{\rm cd}$) with moderate mixing of high-temperature materials or (ii) a parental cloud core that is more ice-depleted and/or experiences weaker CO self-shielding than the commonly assumed fiducial values (Section \ref{sec:chi_and_C}).

\item
In all explored models, $\Delta^{17}{\rm O}_{\rm solid}$ tends to remain near the solar-system average of $-30\tcperthousand$ interior to the snow line where $T \lesssim 500~{\rm K}$, suggesting that classical quasi-steady disk formation and viscous evolution alone may be insufficient to satisfy some constraints implied by chondrules in ordinary chondrites, and that more dynamic, time-variable processes may be required.

\item
We suggest a scenario for the origin of chondrules in carbonaceous chondrites that may explain the observed trend between their oxygen isotopic compositions and Mg\#, which serves as a proxy for the redox state of the chondrule-forming environment.
Chondrules in carbonaceous chondrites may have formed through flash-heating events in icy-dust-enriched environments after disk formation.
When icy aggregates are heated, H$_{2}$O ice sublimates, and a fraction of the resulting H$_{2}$O vapor is expected to diffuse outward from the chondrule-forming heated region.
This partial escape of H$_{2}$O vapor leads to incomplete isotope exchange between H$_{2}$O and silicates, potentially producing the observed trend in which chondrules formed in more dust-enriched environments exhibit slightly more $^{16}$O-poor compositions than the dominant population (Section \ref{sec:beyond}).

\end{itemize}

We acknowledge that our numerical model is based on the classical picture of disk formation and evolution.
With recent advances in numerical simulations and astronomical observations, the emerging picture of the disk formation process is undergoing substantial revision.
More comprehensive modeling should incorporate additional physics and chemistry that can modify reservoir abundances and temperatures (e.g., disk winds, planetesimal formation, and additional oxygen reservoirs such as CO$_{2}$).
Moreover, recent progress is making it possible to measure the oxygen isotopic compositions of reservoirs such as H$_{2}$O not only in the solar system but also in extrasolar protoplanetary disks \citep[e.g.,][]{2026ApJ...999...92S}.
In the future, by comparing astronomical observations with isotopic analyses of meteorites, we should be able to discuss whether our solar system is a typical or atypical planetary system \citep[e.g.,][]{2025SciA...11x7892S}.

\section*{}
The authors thank Lily Ishizaki for fruitful discussions.
SA was supported by JSPS KAKENHI Grants JP24K17118, JP24KK0072, JP25K00025, and JP25H00678; RTT by JP24KJ1041; and TU by JP22K18741, JP24H00259, and JP25K07382.

%







\appendix

\restartappendixnumbering

\section{Effective viscosity}
\label{app:nu}

The radial advection velocity of gas, $v_{\rm g}$, is given by \citep[e.g.,][]{1974MNRAS.168..603L}
\begin{equation}
v_{\rm g} = - \frac{3}{2} \frac{\nu}{r} {\left[ 1 + 2 \frac{\partial \ln{\left( \Sigma_{\rm g} \nu \right)}}{\partial \ln{r}} \right]}.
\end{equation}
The effective gas viscosity, $\nu$, is given by the classical $\alpha$-viscosity formalism \citep{1973A&A....24..337S}: $\nu = \alpha c_{\rm s} h_{\rm g}$, where $h_{\rm g} = c_{\rm s} / \Omega_{\rm K}$ is the gas scale height, $\Omega_{\rm K}$ is the Keplerian angular velocity, and $\alpha$ is a dimensionless viscosity parameter.

The $\alpha$ value depends on the gravitational stability of the disk.
The stability of the disk is measured by Toomre's $Q$ value, defined as $Q = { c_{\rm s} \Omega_{\rm K} } / {( \pi \mathcal{G} \Sigma_{\rm g} )}$ \citep{1964ApJ...139.1217T}.
Regions of the disk that satisfy $Q \ge 2$ are gravitationally stable, and we set $\alpha = \alpha_{0}$.
Here, $\alpha_{0} = 10^{-3}$ is the $\alpha$ value in the stable region.
In contrast, regions with $Q < 2$ are gravitationally unstable, where angular momentum transport is enhanced, corresponding to a larger $\alpha$.
In this study, we adopt the following equation \citep{2001MNRAS.324..705A}:
\begin{equation}
\alpha = \alpha_{0} + \alpha_{\rm GI} {\left[ {\left( \frac{2}{\min{\left( Q, 2 \right)}} \right)}^{2} - 1 \right]},
\end{equation}
where $\alpha_{\rm GI} = 10^{-2}$.

\section{Disk temperature}
\label{app:T}

The disk midplane temperature, $T$, is given by Equation \eqref{eq:T} (Section \ref{sec:disk_evolution}).
The accretion heating term, $T_{\rm acc}$, dominates in the inner disk, whereas the irradiation heating term dominates in the outer disk.
Assuming that accretion energy is released at the disk midplane, $T_{\rm acc}$ is given by \citep[e.g.,][]{1994ApJ...421..640N}
\begin{equation}
T_{\rm acc} = {\left( \frac{27 \alpha {c_{\rm s}}^{2} \kappa_{\rm sil} \Sigma_{\rm sil} \Sigma_{\rm g} \Omega_{\rm K}}{128 \sigma_{\rm SB}} \right)}^{1/4},
\end{equation}
where $\kappa_{\rm sil}$ is the dust opacity and $\sigma_{\rm SB}$ is the Stefan--Boltzmann constant.
The dust opacity depends on $T$, and we adopt the following prescription \citep{2018ApJ...868..118H}:
\begin{equation}
\kappa_{\rm sil} = 100 \times \min{\left[ {\left( \frac{T}{170~{\rm K}} \right)}^{2}, 1 \right]}~{\rm cm}^{2}~{\rm g}^{-1}.
\end{equation}
The irradiation heating term is given by \citep[e.g.,][]{1997ApJ...490..368C}:
\begin{equation}
T_{\rm irr} = 150 {\left( \frac{r}{1~{\rm au}} \right)}^{- 3/7}~{\rm K}.
\end{equation}

\section{Exchange reaction timescale}
\label{app:exchange}

In this study, we consider oxygen isotope exchange between silicate aggregates and ambient H$_{2}$O/CO vapors.
\citet{2024ApJ...974..199A} showed that the isotope exchange timescale for an aggregate is the same as that for its constituent grains as long as the aggregate radius is smaller than a few centimeters.

For amorphous silicate grains, the isotope exchange timescale with an ambient species $i$ ($i$ denotes {\sf wvap} or {\sf cvap}) is given as follows:
\begin{equation}
t_{{\rm amo} \leftrightarrow i} = t_{{\rm amo} \leftrightarrow i}^{\rm (diff)} + t_{{\rm amo} \leftrightarrow i}^{\rm (sup)},
\end{equation}
where the intra-grain diffusion timescale, $t_{{\rm amo} \leftrightarrow i}^{\rm (diff)}$, and the supply timescale, $t_{{\rm amo} \leftrightarrow i}^{\rm (sup)}$, are given by \citep[e.g.,][]{2018ApJ...865...98Y, 2020M&PS...55.1281Y}
\begin{align}
t_{{\rm amo} \leftrightarrow i}^{\rm (diff)} & = \frac{{a_{\bullet}}^{2}}{\pi^{2} D_{{\rm amo} \leftrightarrow i}} \exp{\left( \frac{E_{{\rm amo} \leftrightarrow i}}{R_{\rm gas} T} \right)}, \\
t_{{\rm amo} \leftrightarrow i}^{\rm (sup)}  & = \frac{\gamma_{\rm Fo} N_{\rm A} r_{\bullet}}{3 \Omega_{\rm Fo} \beta_{i} J_{i}}.
\end{align}
Here, $a_{\bullet} = 0.1~{\rm \upmu m}$ is the constituent grain radius, $\gamma_{\rm Fo} = 4$ is the number of oxygen atoms in the formula unit of Mg$_{2}$SiO$_{4}$, and $\Omega_{\rm Fo} = 50~{\rm cm}^{3}~{\rm mol}^{-1}$ is the molar volume of Mg$_{2}$SiO$_{4}$.

The diffusive isotope exchange coefficient, $D_{{\rm amo} \leftrightarrow i}$, and the activation energy for the exchange reaction, $E_{{\rm amo} \leftrightarrow i}$ are listed in Table \ref{table:coeff}.
The isotopic exchange efficiency at the grain surface, $\beta_{i}$, also depends on the ambient species $i$.
For {\sf wvap}, we set $\beta_{\rm wvap} = 7.4 \times 10^{-6}$ and assume that it is independent of $T$ \citep{2018ApJ...865...98Y}.
In contrast, for {\sf cvap} we adopt a temperature-dependent form:
\begin{equation}
\beta_{\rm cvap} = \beta_{\rm cvap, 0} \exp{\left( - \frac{E_{\beta, {\rm cvap}}}{R_{\rm gas} T} \right)},
\end{equation}
where $\beta_{\rm cvap, 0} = 9.8 \times 10^{-5}$ is a prefactor and $E_{\beta, {\rm cvap}} = 43.1~{\rm kJ}~{\rm mol}^{-1}$ is the activation energy for the grain surface exchange reaction \citep{2024GeCoA.374...93Y}.

The supply flux of oxygen atoms from species $i$, $J_{i}$, is given by $J_{i} = P_{i} \gamma_{i} / \sqrt{2 \pi m_{i} k_{\rm B} T}$, where $\gamma_{i}$ is the number of oxygen atoms in the chemical formula of $i$; we set $\gamma_{i} = 1$ for both {\sf wvap} and {\sf cvap}.
The partial pressure of species $i$ at the disk midplane, $P_{i}$, is given by
\begin{equation}
P_{i} = \frac{\Sigma_{i}}{\sqrt{2 \pi} h_{\rm g} m_{i}}  k_{\rm B} T,
\end{equation}
where $m_{i}$ is the molecular mass of species $i$.

\begin{table}[h]
\caption{Diffusive isotope exchange coefficients, $D_{j \leftrightarrow i}$, and activation energy for exchange reaction, $E_{j \leftrightarrow i}$ ($i$ denotes denotes {\sf wvap} or {\sf cvap}, whereas $j$ denotes denotes {\sf amo} or {\sf cry}).}
\label{table:coeff}
\centering
\begin{tabular}{lll}
\hline
 {\bf Symbol}                                  & {\bf Value}                                        & {\bf Reference}    \\ \hline
$D_{{\rm amo} \leftrightarrow {\rm wvap}}$     & $1.4 \times 10^{-10}~{\rm cm}^{2}~{\rm s}^{-1}$    & \citet{2024GeCoA.374...93Y} \\
$D_{{\rm amo} \leftrightarrow {\rm cvap}}$     & $2.0 \times 10^{-17}~{\rm cm}^{2}~{\rm s}^{-1}$    & \citet{2024GeCoA.374...93Y} \\
$D_{{\rm cry} \leftrightarrow {\rm wvap}}$     & $5.4 \times 10^{-9}~{\rm cm}^{2}~{\rm s}^{-1}$     & \citet{1980EPSL..47..391J}  \\
$D_{{\rm cry} \leftrightarrow {\rm cvap}}$     & $5.4 \times 10^{-9}~{\rm cm}^{2}~{\rm s}^{-1}$     & \citet{1980EPSL..47..391J}  \\
\\
$E_{{\rm amo} \leftrightarrow {\rm wvap}}$     & $122.7~{\rm kJ}~{\rm mol}^{-1}$                    & \citet{2024GeCoA.374...93Y} \\
$E_{{\rm amo} \leftrightarrow {\rm cvap}}$     & $41.7~{\rm kJ}~{\rm mol}^{-1}$                     & \citet{2024GeCoA.374...93Y} \\
$E_{{\rm cry} \leftrightarrow {\rm wvap}}$     & $320~{\rm kJ}~{\rm mol}^{-1}$                      & \citet{1980EPSL..47..391J}  \\
$E_{{\rm cry} \leftrightarrow {\rm cvap}}$     & $320~{\rm kJ}~{\rm mol}^{-1}$                      & \citet{1980EPSL..47..391J}  \\
\hline
\end{tabular}
\end{table}

For crystalline silicate grains, the isotopic exchange timescale is limited by intra-grain diffusion and is given by
\begin{equation}
t_{{\rm cry} \leftrightarrow i} = \frac{{a_{\bullet}}^{2}}{\pi^{2} D_{{\rm cry} \leftrightarrow i}} \exp{\left( \frac{E_{{\rm cry} \leftrightarrow i}}{R_{\rm gas} T} \right)}.
\end{equation}
The diffusive isotope exchange coefficient, $D_{{\rm cry} \leftrightarrow i}$, and the activation energy for the exchange reaction, $E_{{\rm cry} \leftrightarrow i}$ are listed in Table \ref{table:coeff}.
We also assume that the isotopic exchange timescale for condensate silicates is identical to that for crystalline silicates: $t_{{\rm cond} \leftrightarrow i} = t_{{\rm cry} \leftrightarrow i}$.

We also consider oxygen isotope exchange between H$_{2}$O and CO via gas-phase chemical reactions.
In gas-phase chemical reaction networks for protoplanetary disks, radicals such as OH play a key role, and oxygen isotope exchange between H$_{2}$O and CO is rate-limited by the reaction ${\rm CO} + {\rm OH} \to {\rm CO}_2 + {\rm H}$ \citep[e.g.,][]{2004GeCoA..68.3943A}.
As the equilibrium partial pressure of OH depends on $P_{\rm g}$ and $T$, the isotope exchange timescale, $t_{\rm wvap \leftrightarrow cvap}$, can be approximated by the following expression \citep[e.g.,][]{2004GeCoA..68.3943A, 2009GeCoA..73.4998L}:
\begin{equation}
t_{\rm wvap \leftrightarrow cvap} = t_{\rm ref} {\left( \frac{P_{\rm g}}{P_{\rm ref}} \right)}^{- 1/2} \exp{\left( \frac{T_{\rm act}}{T} \right)},
\end{equation}
where $t_{\rm ref} = 6 \times 10^{-8}~{\rm s}$, $P_{\rm ref} = 10^{5}~{\rm Pa}$, and $T_{\rm act} = 3.5 \times 10^{4}~{\rm K}$.

Oxygen isotope exchange between H$_{2}$O and CO via gas-phase chemical reactions needs to be considered in the innermost hot regions of the disk, where $T \gtrsim 1000~{\rm K}$.
However, \citet{2024GeCoA.374...93Y} noted that, in the presence of amorphous silicates, H$_{2}$O and CO can exchange oxygen isotopes and reach equilibrium via interactions with amorphous silicates even at low temperatures with $T \ll 1000~{\rm K}$.
We also confirm this behavior in our numerical simulations (see Figures \ref{fig:5}--\ref{fig:7}).

\section{Enrichment factors and oxygen isotopic compositions at the end of mass infall from the molecular cloud core}
\label{app:zeta_delta}

We show the radial profiles of $\zeta_{i}$ ($i$ denotes {\sf sil}, {\sf H$_{2}$O}, or {\sf CO}) and $\Delta^{17}{\rm O}_{i}$ ($i$ denotes {\sf amo}, {\sf cry}, {\sf cond}, {\sf svap}, {\sf wice}, {\sf wvap}, or {\sf cvap}) at the end of mass infall from the molecular cloud core ($t = 0.1~{\rm Myr}$).
At this time, $\zeta_{i} \sim 1$ throughout the disk for all three species in all cases, indicating that the radial drift timescales of dust aggregates are longer than $0.1~{\rm Myr}$ and that the aggregates remain dynamically coupled to the gas during the early stage of disk formation, when viscous expansion drives the outward transport of disk material (see Section \ref{sec:cai}).

\begin{figure*}[]
\centering
\includegraphics[width = \textwidth]{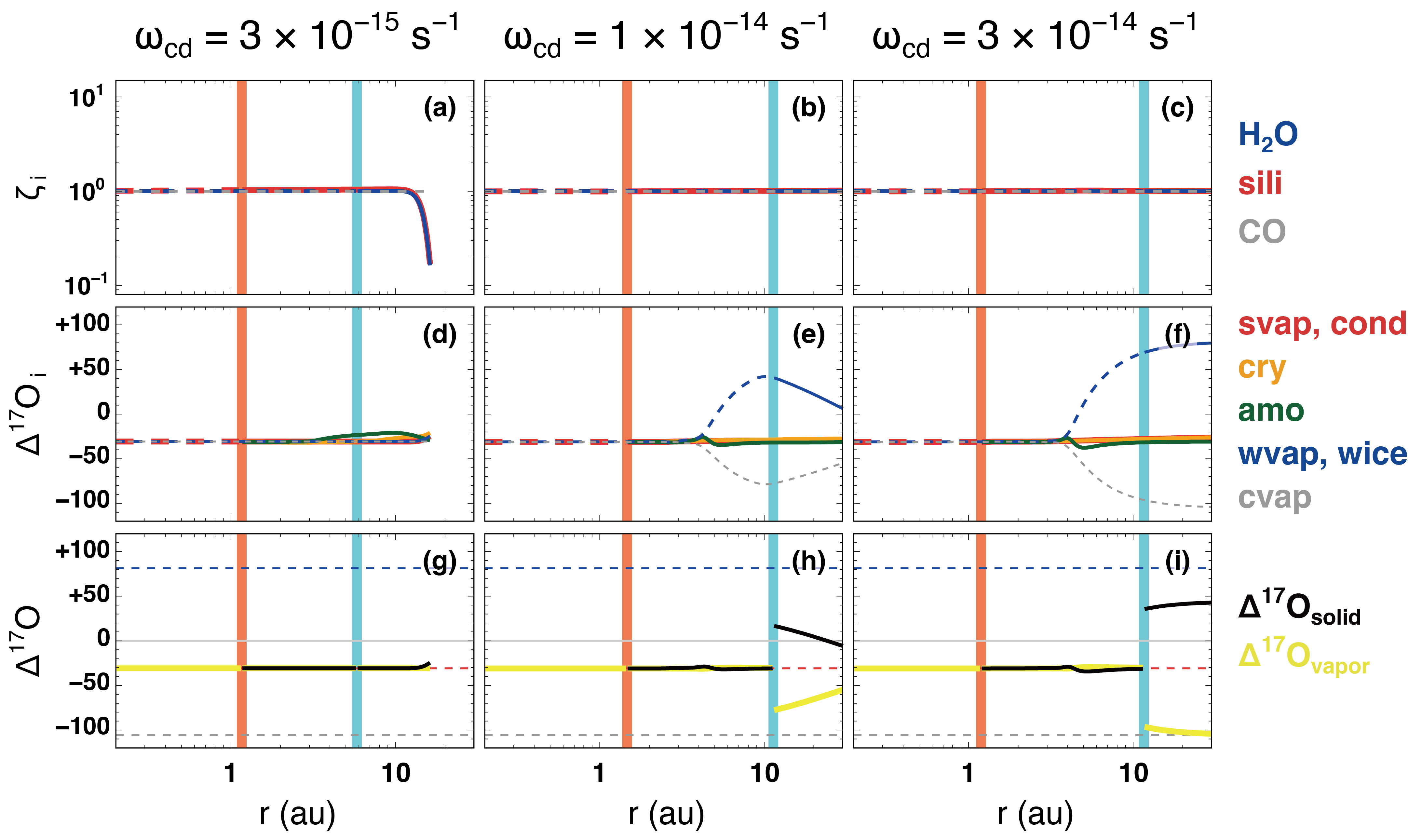}
\caption{
Radial distribution of each oxygen reservoir and their oxygen isotopic composition at $t = 0.1~{\rm Myr}$, instead of $t = 1~{\rm Myr}$ (Figure \ref{fig:5}).
The left, middle, and right columns correspond to $\omega_{\rm cd} = 3 \times 10^{-15}~{\rm s}^{-1}$ (i.e., $r_{\rm c, fin} \approx 1~{\rm au}$), $\omega_{\rm cd} = 1 \times 10^{-14}~{\rm s}^{-1}$ (i.e., $r_{\rm c, fin} \approx 10~{\rm au}$), and $\omega_{\rm cd} = 3 \times 10^{-14}~{\rm s}^{-1}$ (i.e., $r_{\rm c, fin} \approx 100~{\rm au}$), respectively.
We set $a_{\rm agg, in} = a_{\rm agg, out} = 0.1~{\rm mm}$ in these calculations.
The vertical cyan lines represent the location of H$_{2}$O snow line, whereas the vertical orange lines represent the location of the silicate sublimation line.
Panels (a)--(c) show the enrichment factor of each oxygen reservoir, $\zeta_{i}$ ($i$ denotes {\sf sil}, {\sf H$_{2}$O}, or {\sf CO}).
Panels (d)--(f) show the oxygen isotopic composition of each species, $\Delta^{17}{\rm O}_{i}$ ($i$ denotes {\sf amo}, {\sf cry}, {\sf cond}, {\sf svap}, {\sf wice}, {\sf wvap}, or {\sf cvap}).
Panels (g)--(i) show the mean oxygen isotopic compositions of the solid and vapor phases, $\Delta^{17}{\rm O}_{\rm solid}$ and $\Delta^{17}{\rm O}_{\rm vapor}$.
The red, blue, and gray horizontal dashed lines in panels (g)--(i) indicate the oxygen isotopic compositions of the corresponding reservoirs in the parental cloud core.
}
\label{fig:A1}
\end{figure*}

Figure \ref{fig:A1} shows the radial distribution of each oxygen reservoir and their oxygen isotopic composition at $t = 0.1~{\rm Myr}$.
The left, middle, and right columns correspond to $\omega_{\rm cd} = 3 \times 10^{-15}~{\rm s}^{-1}$ (i.e., $r_{\rm c, fin} \approx 1~{\rm au}$), $\omega_{\rm cd} = 1 \times 10^{-14}~{\rm s}^{-1}$ (i.e., $r_{\rm c, fin} \approx 10~{\rm au}$), and $\omega_{\rm cd} = 3 \times 10^{-14}~{\rm s}^{-1}$ (i.e., $r_{\rm c, fin} \approx 100~{\rm au}$), respectively.
We set $a_{\rm agg, in} = a_{\rm agg, out} = 0.1~{\rm mm}$ in these calculations.
These settings are the same as those adopted for the calculation shown in Figure \ref{fig:5}.
At $t = 0.1~{\rm Myr}$, the radial extent of the silicate sublimation region defined by $T = 2000~{\rm K}$ exceeds $1~{\rm au}$, and the H$_{2}$O snow line is located at $5$--$10~{\rm au}$.

We acknowledge that a deviation of $\zeta_{i}$ from $1$ is visible at $r \gtrsim 10~{\rm au}$ in Figure \ref{fig:A1}(a); however, this does not affect our main conclusions.
For $\omega_{\rm cd} = 3 \times 10^{-15}~{\rm s}^{-1}$, both $\Sigma_{\rm g}$ and $\Sigma_{\rm sil}$ drop sharply at $r \gtrsim 10~{\rm au}$, and $\Sigma_{\rm sil} < 10^{-3}~{\rm g}~{\rm cm}^{-2}$ at $r > 16~{\rm au}$ (Figure \ref{fig:4}(a)).
The tight dynamical coupling between dust aggregates and gas breaks down at $r \gtrsim 10~{\rm au}$ because of the low gas density.
Such a region, where the tight coupling breaks down, is effectively outside the disk in the sense that the surface densities are already extremely low, and discussion of compositional evolution there is not meaningful.
By contrast, in the region at $r \lesssim 10~{\rm au}$ where $\Sigma_{\rm sil} \gg 10^{-3}~{\rm g}~{\rm cm}^{-2}$, we obtain $\zeta_{i} \sim 1$ at $t = 0.1~{\rm Myr}$, as in the other cases.

\begin{figure*}[]
\centering
\includegraphics[width = \textwidth]{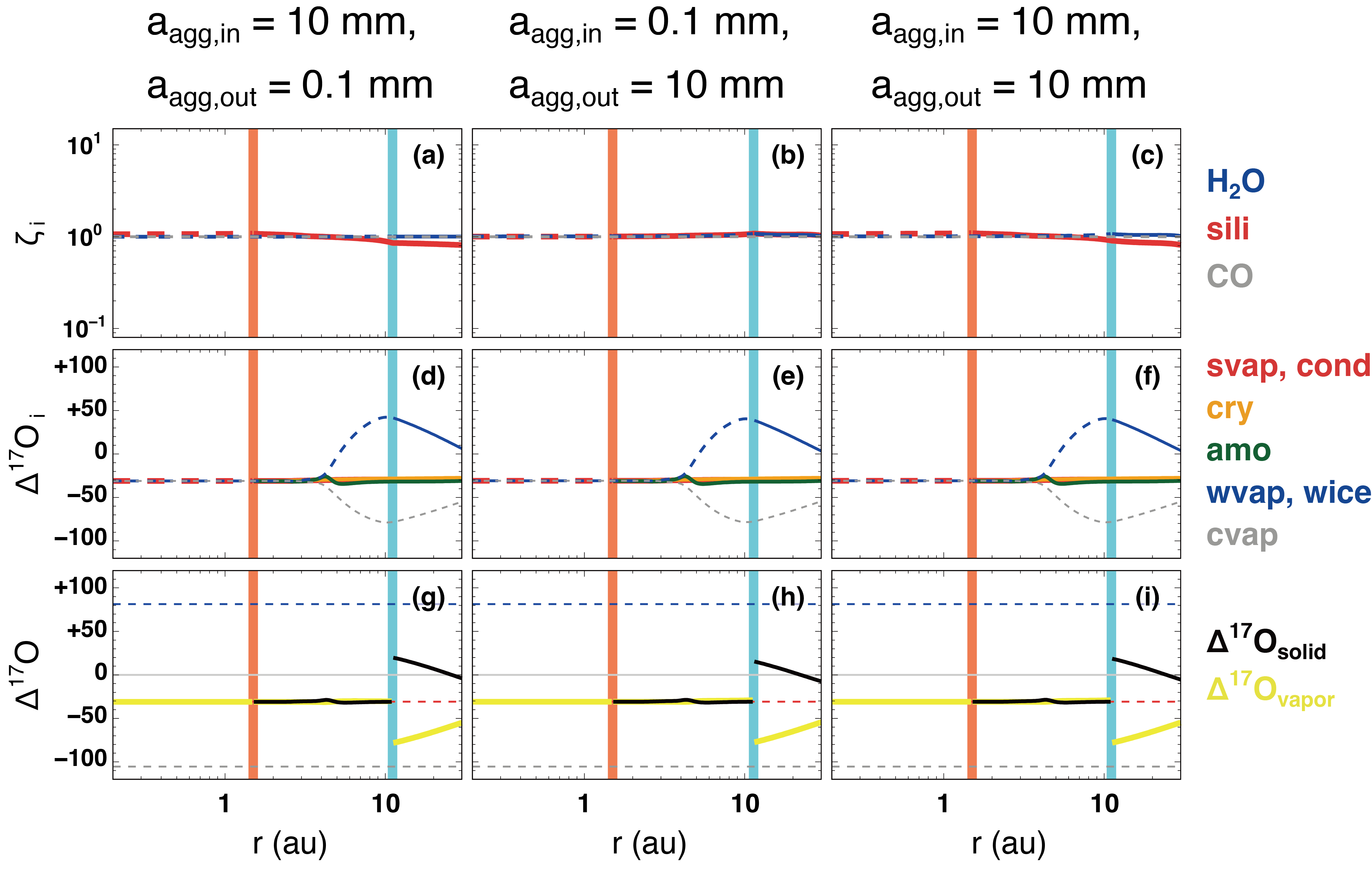}
\caption{
Radial distribution of each oxygen reservoir and their oxygen isotopic composition at $t = 0.1~{\rm Myr}$, instead of $t = 1~{\rm Myr}$ (Figure \ref{fig:6}).
The left, middle, and right columns correspond to case (i) ($a_{\rm agg, in} = 10~{\rm mm}$ and $a_{\rm agg, out} = 0.1~{\rm mm}$), case (ii) ($a_{\rm agg, in} = 0.1~{\rm mm}$ and $a_{\rm agg, out} = 10~{\rm mm}$), and case (iii) ($a_{\rm agg, in} = 10~{\rm mm}$ and $a_{\rm agg, out} = 10~{\rm mm}$), respectively.
We set $\omega_{\rm cd} = 1 \times 10^{-14}~{\rm s^{-1}}$ (i.e., $r_{\rm c, fin} \approx 10~{\rm au}$) in these calculations.
The vertical cyan lines represent the location of H$_{2}$O snow line, whereas the vertical orange lines represent the location of the silicate sublimation line.
Panels (a)--(c) show the enrichment factor of each oxygen reservoir, $\zeta_{i}$ ($i$ denotes {\sf sil}, {\sf H$_{2}$O}, or {\sf CO}).
Panels (d)--(f) show the oxygen isotopic composition of each species, $\Delta^{17}{\rm O}_{i}$ ($i$ denotes {\sf amo}, {\sf cry}, {\sf cond}, {\sf svap}, {\sf wice}, {\sf wvap}, or {\sf cvap}).
Panels (g)--(i) show the mean oxygen isotopic compositions of the solid and vapor phases, $\Delta^{17}{\rm O}_{\rm solid}$ and $\Delta^{17}{\rm O}_{\rm vapor}$.
The red, blue, and gray horizontal dashed lines in panels (g)--(i) indicate the oxygen isotopic compositions of the corresponding reservoirs in the parental cloud core.
}
\label{fig:A2}
\end{figure*}

Figure \ref{fig:A2} shows the radial distribution of each oxygen reservoir and their oxygen isotopic composition at $t = 0.1~{\rm Myr}$.
The left, middle, and right columns correspond to case (i) ($a_{\rm agg, in} = 10~{\rm mm}$ and $a_{\rm agg, out} = 0.1~{\rm mm}$), case (ii) ($a_{\rm agg, in} = 0.1~{\rm mm}$ and $a_{\rm agg, out} = 10~{\rm mm}$), and case (iii) ($a_{\rm agg, in} = 10~{\rm mm}$ and $a_{\rm agg, out} = 10~{\rm mm}$), respectively.
We set $\omega_{\rm cd} = 1 \times 10^{-14}~{\rm s^{-1}}$ (i.e., $r_{\rm c, fin} \approx 10~{\rm au}$) in these calculations.
These settings are the same as those adopted for the calculation shown in Figure \ref{fig:6}.

\bibliography{sample631}{}

@ARTICLE{2005A&A...442..703H,
       author = {{Hueso}, R. and {Guillot}, T.},
        title = "{Evolution of protoplanetary disks: constraints from DM Tauri and GM Aurigae}",
      journal = {\aap},
         year = 2005,
        month = nov,
       volume = {442},
       number = {2},
        pages = {703-725},
          doi = {10.1051/0004-6361:20041905},
       adsurl = {https://ui.adsabs.harvard.edu/abs/2005A&A...442..703H}
}

@INCOLLECTION{2018crpd.book..196T,
       author = {{Tenner}, Travis J. and {Ushikubo}, Takayuki and {Nakashima}, Daisuke and {Schrader}, Devin L. and {Weisberg}, Michael K. and {Kimura}, Makoto and {Kita}, Noriko T.},
        title = "{Oxygen Isotope Characteristics of Chondrules from Recent Studies by Secondary Ion Mass Spectrometry}",
    booktitle = {Chondrules: Records of Protoplanetary Disk Processes},
         year = 2018,
       editor = {{Russell}, Sara S. and {Connolly}, Jr., Harold C. and {Krot}, Alexander N.},
        pages = {196-246},
          doi = {10.1017/9781108284073.008},
       adsurl = {https://ui.adsabs.harvard.edu/abs/2018crpd.book..196T}
}

@ARTICLE{2018E&PSL.496..132M,
       author = {{Marrocchi}, Yves and {Villeneuve}, Johan and {Batanova}, Valentina and {Piani}, Laurette and {Jacquet}, Emmanuel},
        title = "{Oxygen isotopic diversity of chondrule precursors and the nebular origin of chondrules}",
      journal = {Earth and Planetary Science Letters},
         year = 2018,
        month = aug,
       volume = {496},
        pages = {132-141},
          doi = {10.1016/j.epsl.2018.05.042},
       adsurl = {https://ui.adsabs.harvard.edu/abs/2018E&PSL.496..132M}
}

@ARTICLE{2017GeCoA.201...83K,
       author = {{Kawasaki}, Noriyuki and {Itoh}, Shoichi and {Sakamoto}, Naoya and {Yurimoto}, Hisayoshi},
        title = "{Chronological study of oxygen isotope composition for the solar protoplanetary disk recorded in a fluffy Type A CAI from Vigarano}",
      journal = {\gca},
         year = 2017,
        month = mar,
       volume = {201},
        pages = {83-102},
          doi = {10.1016/j.gca.2015.12.031},
       adsurl = {https://ui.adsabs.harvard.edu/abs/2017GeCoA.201...83K}
}

@ARTICLE{2012GeCoA..90..242U,
       author = {{Ushikubo}, Takayuki and {Kimura}, Makoto and {Kita}, Noriko T. and {Valley}, John W.},
        title = "{Primordial oxygen isotope reservoirs of the solar nebula recorded in chondrules in Acfer 094 carbonaceous chondrite}",
      journal = {\gca},
         year = 2012,
        month = aug,
       volume = {90},
        pages = {242-264},
          doi = {10.1016/j.gca.2012.05.010},
       adsurl = {https://ui.adsabs.harvard.edu/abs/2012GeCoA..90..242U}
}

@ARTICLE{1977E&PSL..34..209C,
       author = {{Clayton}, R.~N. and {Onuma}, N. and {Grossman}, L. and {Mayeda}, T.~K.},
        title = "{Distribution of the pre-solar component in Allende and other carbonaceous chondrites}",
      journal = {Earth and Planetary Science Letters},
         year = 1977,
        month = mar,
       volume = {34},
       number = {2},
        pages = {209-224},
          doi = {10.1016/0012-821X(77)90005-X},
       adsurl = {https://ui.adsabs.harvard.edu/abs/1977E&PSL..34..209C}
}

@ARTICLE{1998Sci...282..452Y,
       author = {{Young}, Edward D. and {Russell}, Sara S.},
        title = "{Oxygen Reservoirs in the Early Solar Nebula Inferred from an Allende CAI}",
      journal = {Science},
         year = 1998,
        month = oct,
       volume = {282},
        pages = {452},
          doi = {10.1126/science.282.5388.452},
       adsurl = {https://ui.adsabs.harvard.edu/abs/1998Sci...282..452Y}
}

@ARTICLE{2004Sci...305.1763Y,
       author = {{Yurimoto}, Hisayoshi and {Kuramoto}, Kiyoshi},
        title = "{Molecular Cloud Origin for the Oxygen Isotope Heterogeneity in the Solar System}",
      journal = {Science},
         year = 2004,
        month = sep,
       volume = {305},
       number = {5691},
        pages = {1763-1766},
          doi = {10.1126/science.1100989},
       adsurl = {https://ui.adsabs.harvard.edu/abs/2004Sci...305.1763Y}
}

@ARTICLE{2020SciA....6.2724K,
       author = {{Krot}, Alexander N. and {Nagashima}, Kazuhide and {Lyons}, James R. and {Lee}, Jeong-Eun and {Bizzarro}, Martin},
        title = "{Oxygen isotopic heterogeneity in the early Solar System inherited from the protosolar molecular cloud}",
      journal = {Science Advances},
         year = 2020,
        month = oct,
       volume = {6},
       number = {42},
        pages = {eaay2724},
          doi = {10.1126/sciadv.aay2724},
       adsurl = {https://ui.adsabs.harvard.edu/abs/2020SciA....6.2724K}
}

@ARTICLE{2019ARA&A..57..113A,
       author = {{Altwegg}, Kathrin and {Balsiger}, Hans and {Fuselier}, Stephen A.},
        title = "{Cometary Chemistry and the Origin of Icy Solar System Bodies: The View After Rosetta}",
      journal = {\araa},
         year = 2019,
        month = aug,
       volume = {57},
        pages = {113-155},
          doi = {10.1146/annurev-astro-091918-104409},
       adsurl = {https://ui.adsabs.harvard.edu/abs/2019ARA&A..57..113A}
}

@ARTICLE{2007Sci...317..231S,
       author = {{Sakamoto}, Naoya and {Seto}, Yusuke and {Itoh}, Shoichi and {Kuramoto}, Kiyoshi and {Fujino}, Kiyoshi and {Nagashima}, Kazuhide and {Krot}, Alexander N. and {Yurimoto}, Hisayoshi},
        title = "{Remnants of the Early Solar System Water Enriched in Heavy Oxygen Isotopes}",
      journal = {Science},
         year = 2007,
        month = jul,
       volume = {317},
       number = {5835},
        pages = {231},
          doi = {10.1126/science.1142021},
       adsurl = {https://ui.adsabs.harvard.edu/abs/2007Sci...317..231S}
}

@ARTICLE{2004ApJ...614..490C,
       author = {{Cuzzi}, Jeffrey N. and {Zahnle}, Kevin J.},
        title = "{Material Enhancement in Protoplanetary Nebulae by Particle Drift through Evaporation Fronts}",
      journal = {\apj},
         year = 2004,
        month = oct,
       volume = {614},
       number = {1},
        pages = {490-496},
          doi = {10.1086/423611},
       adsurl = {https://ui.adsabs.harvard.edu/abs/2004ApJ...614..490C}
}

@ARTICLE{2024GeCoA.374...93Y,
       author = {{Yamamoto}, Daiki and {Kawasaki}, Noriyuki and {Tachibana}, Shogo and {Ishizaki}, Lily and {Sakurai}, Ryosuke and {Yurimoto}, Hisayoshi},
        title = "{An experimental simulation of oxygen isotope exchange reaction between amorphous silicate dust and carbon monoxide gas in the early Solar System}",
      journal = {\gca},
         year = 2024,
        month = jun,
       volume = {374},
        pages = {93-105},
          doi = {10.1016/j.gca.2024.04.014},
       adsurl = {https://ui.adsabs.harvard.edu/abs/2024GeCoA.374...93Y}
}

@ARTICLE{2018ApJ...865...98Y,
       author = {{Yamamoto}, Daiki and {Kuroda}, Minami and {Tachibana}, Shogo and {Sakamoto}, Naoya and {Yurimoto}, Hisayoshi},
        title = "{Oxygen Isotopic Exchange between Amorphous Silicate and Water Vapor and Its Implications for Oxygen Isotopic Evolution in the Early Solar System}",
      journal = {\apj},
         year = 2018,
        month = oct,
       volume = {865},
       number = {2},
          eid = {98},
        pages = {98},
          doi = {10.3847/1538-4357/aadcee},
       adsurl = {https://ui.adsabs.harvard.edu/abs/2018ApJ...865...98Y}
}

@ARTICLE{2004ApJ...609..826K,
       author = {{Kemper}, F. and {Vriend}, W.~J. and {Tielens}, A.~G.~G.~M.},
        title = "{The Absence of Crystalline Silicates in the Diffuse Interstellar Medium}",
      journal = {\apj},
         year = 2004,
        month = jul,
       volume = {609},
       number = {2},
        pages = {826-837},
          doi = {10.1086/421339},
       adsurl = {https://ui.adsabs.harvard.edu/abs/2004ApJ...609..826K}
}

@ARTICLE{2012M&PS...47...99Y,
       author = {{Yang}, Le and {Ciesla}, Fred J.},
        title = "{The effects of disk building on the distributions of refractory materials in the solar nebula}",
      journal = {\maps},
         year = 2012,
        month = jan,
       volume = {47},
       number = {1},
        pages = {99-119},
          doi = {10.1111/j.1945-5100.2011.01315.x},
       adsurl = {https://ui.adsabs.harvard.edu/abs/2012M&PS...47...99Y}
}

@ARTICLE{2006ApJ...640L..67D,
       author = {{Dullemond}, C.~P. and {Apai}, D. and {Walch}, S.},
        title = "{Crystalline Silicates as a Probe of Disk Formation History}",
      journal = {\apjl},
         year = 2006,
        month = mar,
       volume = {640},
       number = {1},
        pages = {L67-L70},
          doi = {10.1086/503100},
       adsurl = {https://ui.adsabs.harvard.edu/abs/2006ApJ...640L..67D}
}

@ARTICLE{1977MNRAS.180...57W,
       author = {{Weidenschilling}, S.~J.},
        title = "{Aerodynamics of solid bodies in the solar nebula.}",
      journal = {\mnras},
         year = 1977,
        month = jul,
       volume = {180},
        pages = {57-70},
          doi = {10.1093/mnras/180.2.57},
       adsurl = {https://ui.adsabs.harvard.edu/abs/1977MNRAS.180...57W}
}

@ARTICLE{2024A&A...687A..65W,
       author = {{Woitke}, P. and {Dr{\k{a}}{\.z}kowska}, J. and {Lammer}, H. and {Kadam}, K. and {Marigo}, P.},
        title = "{CAI formation in the early Solar System}",
      journal = {\aap},
         year = 2024,
        month = jul,
       volume = {687},
          eid = {A65},
        pages = {A65},
          doi = {10.1051/0004-6361/202450289},
       adsurl = {https://ui.adsabs.harvard.edu/abs/2024A&A...687A..65W}
}

@ARTICLE{2021ApJ...920...27A,
       author = {{Arakawa}, Sota and {Matsumoto}, Yuji and {Honda}, Mitsuhiko},
        title = "{On the Crystallinity of Silicate Dust in Evolving Protoplanetary Disks due to Magnetically Driven Disk Winds}",
      journal = {\apj},
         year = 2021,
        month = oct,
       volume = {920},
       number = {1},
          eid = {27},
        pages = {27},
          doi = {10.3847/1538-4357/ac157e},
       adsurl = {https://ui.adsabs.harvard.edu/abs/2021ApJ...920...27A}
}

@ARTICLE{2024ApJ...974..199A,
       author = {{Arakawa}, Sota and {Yamamoto}, Daiki and {Ishizaki}, Lily and {Okamoto}, Tamami and {Kawasaki}, Noriyuki},
        title = "{Oxygen Isotope Exchange between Dust Aggregates and Ambient Nebular Gas}",
      journal = {\apj},
         year = 2024,
        month = oct,
       volume = {974},
       number = {2},
          eid = {199},
        pages = {199},
          doi = {10.3847/1538-4357/ad7795},
       adsurl = {https://ui.adsabs.harvard.edu/abs/2024ApJ...974..199A}
}

@ARTICLE{2023Icar..40515690A,
       author = {{Arakawa}, Sota and {Yamamoto}, Daiki and {Ushikubo}, Takayuki and {Kaneko}, Hiroaki and {Tanaka}, Hidekazu and {Hirose}, Shigenobu and {Nakamoto}, Taishi},
        title = "{Oxygen isotope exchange between molten silicate spherules and ambient water vapor with nonzero relative velocity: Implication for chondrule formation environment}",
      journal = {\icarus},
         year = 2023,
        month = nov,
       volume = {405},
          eid = {115690},
        pages = {115690},
          doi = {10.1016/j.icarus.2023.115690},
       adsurl = {https://ui.adsabs.harvard.edu/abs/2023Icar..40515690A}
}

@ARTICLE{2023ApJ...957...47I,
       author = {{Ishizaki}, Lily and {Tachibana}, Shogo and {Okamoto}, Tamami and {Yamamoto}, Daiki and {Ida}, Shigeru},
        title = "{Effective Reaction Temperatures of Irreversible Dust Chemical Reactions in a Protoplanetary Disk}",
      journal = {\apj},
         year = 2023,
        month = nov,
       volume = {957},
       number = {1},
          eid = {47},
        pages = {47},
          doi = {10.3847/1538-4357/acf310},
       adsurl = {https://ui.adsabs.harvard.edu/abs/2023ApJ...957...47I}
}

@ARTICLE{2024A&A...691A.147M,
       author = {{Morbidelli}, Alessandro and {Marrocchi}, Yves and {Ahmad}, Adnan Ali and {Bhandare}, Asmita and {Charnoz}, S{\'e}bastien and {Commer{\c{c}}on}, Beno{\^\i}t and {Dullemond}, Cornelis P. and {Guillot}, Tristan and {Hennebelle}, Patrick and {Lee}, Yueh-Ning and {Lovascio}, Francesco and {Marschall}, Raphael and {Marty}, Bernard and {Maury}, Ana{\"e}lle and {Tamami}, Okamoto},
        title = "{Formation and evolution of a protoplanetary disk: Combining observations, simulations, and cosmochemical constraints}",
      journal = {\aap},
         year = 2024,
        month = nov,
       volume = {691},
          eid = {A147},
        pages = {A147},
          doi = {10.1051/0004-6361/202451388},
       adsurl = {https://ui.adsabs.harvard.edu/abs/2024A&A...691A.147M}
}

@ARTICLE{2022NatAs...6...72M,
       author = {{Morbidelli}, A. and {Bailli{\'e}}, K. and {Batygin}, K. and {Charnoz}, S. and {Guillot}, T. and {Rubie}, D.~C. and {Kleine}, T.},
        title = "{Contemporary formation of early Solar System planetesimals at two distinct radial locations}",
      journal = {Nature Astronomy},
         year = 2022,
        month = jan,
       volume = {6},
        pages = {72-79},
          doi = {10.1038/s41550-021-01517-7},
       adsurl = {https://ui.adsabs.harvard.edu/abs/2022NatAs...6...72M}
}

@ARTICLE{2024A&A...687A.158B,
       author = {{Bhandare}, Asmita and {Commer{\c{c}}on}, Beno{\^\i}t and {Laibe}, Guillaume and {Flock}, Mario and {Kuiper}, Rolf and {Henning}, Thomas and {Mignone}, Andrea and {Marleau}, Gabriel-Dominique},
        title = "{Mixing is easy: New insights for cosmochemical evolution from pre-stellar core collapse}",
      journal = {\aap},
         year = 2024,
        month = jul,
       volume = {687},
          eid = {A158},
        pages = {A158},
          doi = {10.1051/0004-6361/202449594},
       adsurl = {https://ui.adsabs.harvard.edu/abs/2024A&A...687A.158B}
}

@INCOLLECTION{2018crpd.book..151E,
       author = {{Ebel}, Denton S. and {Alexander}, Conel M. O'D. and {Libourel}, Guy},
        title = "{Vapor{\textendash}Melt Exchange}",
    booktitle = {Chondrules: Records of Protoplanetary Disk Processes},
         year = 2018,
       editor = {{Russell}, Sara S. and {Connolly}, Jr., Harold C. and {Krot}, Alexander N.},
        pages = {151-174},
          doi = {10.1017/9781108284073.006},
       adsurl = {https://ui.adsabs.harvard.edu/abs/2018crpd.book..151E}
}

@INCOLLECTION{2018crpd.book...57J,
       author = {{Jones}, Rhian H. and {Villeneuve}, Johan and {Libourel}, Guy},
        title = "{Thermal Histories of Chondrules}",
    booktitle = {Chondrules: Records of Protoplanetary Disk Processes},
         year = 2018,
       editor = {{Russell}, Sara S. and {Connolly}, Jr., Harold C. and {Krot}, Alexander N.},
        pages = {57-90},
          doi = {10.1017/9781108284073.003},
       adsurl = {https://ui.adsabs.harvard.edu/abs/2018crpd.book...57J}
}

@ARTICLE{1998A&A...330..375G,
       author = {{Greenberg}, J. Mayo},
        title = "{Making a comet nucleus}",
      journal = {\aap},
         year = 1998,
        month = feb,
       volume = {330},
        pages = {375-380},
       adsurl = {https://ui.adsabs.harvard.edu/abs/1998A&A...330..375G}
}

@ARTICLE{2025PASJ...77..162O,
       author = {{Okuzumi}, Satoshi},
        title = "{Surface accretion as a dust retention mechanism in protoplanetary disks. I. Formulation and proof-of-concept simulations}",
      journal = {\pasj},
         year = 2025,
        month = feb,
       volume = {77},
       number = {1},
        pages = {162-177},
          doi = {10.1093/pasj/psae107},
       adsurl = {https://ui.adsabs.harvard.edu/abs/2025PASJ...77..162O}
}

@ARTICLE{2011Sci...332.1528M,
       author = {{McKeegan}, K.~D. and {Kallio}, A.~P.~A. and {Heber}, V.~S. and {Jarzebinski}, G. and {Mao}, P.~H. and {Coath}, C.~D. and {Kunihiro}, T. and {Wiens}, R.~C. and {Nordholt}, J.~E. and {Moses}, R.~W. and {Reisenfeld}, D.~B. and {Jurewicz}, A.~J.~G. and {Burnett}, D.~S.},
        title = "{The Oxygen Isotopic Composition of the Sun Inferred from Captured Solar Wind}",
      journal = {Science},
         year = 2011,
        month = jun,
       volume = {332},
       number = {6037},
        pages = {1528},
          doi = {10.1126/science.1204636},
       adsurl = {https://ui.adsabs.harvard.edu/abs/2011Sci...332.1528M}
}

@ARTICLE{1999GeCoA..63.2281R,
       author = {{Rubin}, Alan E. and {Sailer}, Alan L. and {Wasson}, John T.},
        title = "{Troilite in the chondrules of type-3 ordinary chondrites: Implications for chondrule formation}",
      journal = {\gca},
         year = 1999,
        month = aug,
       volume = {63},
       number = {15},
        pages = {2281-2298},
          doi = {10.1016/S0016-7037(99)00119-2},
       adsurl = {https://ui.adsabs.harvard.edu/abs/1999GeCoA..63.2281R}
}

@ARTICLE{1976E&PSL..31..341B,
       author = {{Baertschi}, P.},
        title = "{Absolute $^{18}$O content of standard mean ocean water}",
      journal = {Earth and Planetary Science Letters},
         year = 1976,
        month = aug,
       volume = {31},
       number = {3},
        pages = {341-344},
          doi = {10.1016/0012-821X(76)90115-1},
       adsurl = {https://ui.adsabs.harvard.edu/abs/1976E&PSL..31..341B}
}

@ARTICLE{2013ApJ...770...71T,
       author = {{Takahashi}, Sanemichi Z. and {Inutsuka}, Shu-ichiro and {Machida}, Masahiro N.},
        title = "{A Semi-analytical Description for the Formation and Gravitational Evolution of Protoplanetary Disks}",
      journal = {\apj},
         year = 2013,
        month = jun,
       volume = {770},
       number = {1},
          eid = {71},
        pages = {71},
          doi = {10.1088/0004-637X/770/1/71},
       adsurl = {https://ui.adsabs.harvard.edu/abs/2013ApJ...770...71T}
}

@ARTICLE{2025ApJ...985..106K,
       author = {{Kawasaki}, Yoshihiro and {Machida}, Masahiro N.},
        title = "{Impact of Magnetohydrodynamic Disk Wind on Early Evolutionary Stage of Protoplanetary Disk and Dust Growth}",
      journal = {\apj},
         year = 2025,
        month = may,
       volume = {985},
       number = {1},
          eid = {106},
        pages = {106},
          doi = {10.3847/1538-4357/adc67b},
       adsurl = {https://ui.adsabs.harvard.edu/abs/2025ApJ...985..106K}
}

@ARTICLE{1994ApJ...421..640N,
       author = {{Nakamoto}, Taishi and {Nakagawa}, Yoshitsugo},
        title = "{Formation, Early Evolution, and Gravitational Stability of Protoplanetary Disks}",
      journal = {\apj},
         year = 1994,
        month = feb,
       volume = {421},
        pages = {640},
          doi = {10.1086/173678},
       adsurl = {https://ui.adsabs.harvard.edu/abs/1994ApJ...421..640N}
}

@ARTICLE{2019ApJ...884...32J,
       author = {{Jacquet}, Emmanuel and {Pignatale}, Francesco C. and {Chaussidon}, Marc and {Charnoz}, S{\'e}bastien},
        title = "{Fingerprints of the Protosolar Cloud Collapse in the Solar System. II. Nucleosynthetic Anomalies in Meteorites}",
      journal = {\apj},
         year = 2019,
        month = oct,
       volume = {884},
       number = {1},
          eid = {32},
        pages = {32},
          doi = {10.3847/1538-4357/ab38c1},
       adsurl = {https://ui.adsabs.harvard.edu/abs/2019ApJ...884...32J}
}

@ARTICLE{2024PASJ...76..881H,
       author = {{Homma}, Kazuaki A. and {Okuzumi}, Satoshi and {Arakawa}, Sota and {Fukai}, Ryota},
        title = "{Isotopic variation of non-carbonaceous meteorites caused by dust leakage across the Jovian gap in the solar nebula}",
      journal = {\pasj},
         year = 2024,
        month = oct,
       volume = {76},
       number = {5},
        pages = {881-894},
          doi = {10.1093/pasj/psae052},
       adsurl = {https://ui.adsabs.harvard.edu/abs/2024PASJ...76..881H}
}

@ARTICLE{2017A&A...600A.140S,
       author = {{Stammler}, Sebastian Markus and {Birnstiel}, Tilman and {Pani{\'c}}, Olja and {Dullemond}, Cornelis Petrus and {Dominik}, Carsten},
        title = "{Redistribution of CO at the location of the CO ice line in evolving gas and dust disks}",
      journal = {\aap},
         year = 2017,
        month = apr,
       volume = {600},
          eid = {A140},
        pages = {A140},
          doi = {10.1051/0004-6361/201629041},
       adsurl = {https://ui.adsabs.harvard.edu/abs/2017A&A...600A.140S}
}

@ARTICLE{2018ESC.....2..778Y,
       author = {{Yamamoto}, Daiki and {Tachibana}, Shogo},
        title = "{Water Vapor Pressure Dependence of Crystallization Kinetics of Amorphous Forsterite}",
      journal = {ACS Earth and Space Chemistry},
         year = 2018,
        month = aug,
       volume = {2},
       number = {8},
        pages = {778-786},
          doi = {10.1021/acsearthspacechem.8b00047},
       adsurl = {https://ui.adsabs.harvard.edu/abs/2018ESC.....2..778Y}
}

@ARTICLE{2004GeCoA..68.3943A,
       author = {{Alexander}, C.~M.~O. 'D.},
        title = "{Chemical equilibrium and kinetic constraints for chondrule and CAI formation conditions}",
      journal = {\gca},
         year = 2004,
        month = oct,
       volume = {68},
       number = {19},
        pages = {3943-3969},
          doi = {10.1016/j.gca.2004.03.030},
       adsurl = {https://ui.adsabs.harvard.edu/abs/2004GeCoA..68.3943A}
}

@ARTICLE{2009GeCoA..73.4998L,
       author = {{Lyons}, J.~R. and {Bergin}, E.~A. and {Ciesla}, F.~J. and {Davis}, A.~M. and {Desch}, S.~J. and {Hashizume}, K. and {Lee}, J.-E.},
        title = "{Timescales for the evolution of oxygen isotope compositions in the solar nebula}",
      journal = {\gca},
         year = 2009,
        month = sep,
       volume = {73},
       number = {17},
        pages = {4998-5017},
          doi = {10.1016/j.gca.2009.01.041},
       adsurl = {https://ui.adsabs.harvard.edu/abs/2009GeCoA..73.4998L}
}

@ARTICLE{1973A&A....24..337S,
       author = {{Shakura}, N.~I. and {Sunyaev}, R.~A.},
        title = "{Black holes in binary systems. Observational appearance.}",
      journal = {\aap},
         year = 1973,
        month = jan,
       volume = {24},
        pages = {337-355},
       adsurl = {https://ui.adsabs.harvard.edu/abs/1973A&A....24..337S}
}

@ARTICLE{1964ApJ...139.1217T,
       author = {{Toomre}, A.},
        title = "{On the gravitational stability of a disk of stars.}",
      journal = {\apj},
         year = 1964,
        month = may,
       volume = {139},
        pages = {1217-1238},
          doi = {10.1086/147861},
       adsurl = {https://ui.adsabs.harvard.edu/abs/1964ApJ...139.1217T}
}

@ARTICLE{2001MNRAS.324..705A,
       author = {{Armitage}, Philip J. and {Livio}, Mario and {Pringle}, J.~E.},
        title = "{Episodic accretion in magnetically layered protoplanetary discs}",
      journal = {\mnras},
         year = 2001,
        month = jun,
       volume = {324},
       number = {3},
        pages = {705-711},
          doi = {10.1046/j.1365-8711.2001.04356.x},
       adsurl = {https://ui.adsabs.harvard.edu/abs/2001MNRAS.324..705A}
}

@ARTICLE{2018ApJ...868..118H,
       author = {{Homma}, Kenji and {Nakamoto}, Taishi},
        title = "{Collisional Growth of Icy Dust Aggregates in the Disk Formation Stage: Difficulties for Planetesimal Formation via Direct Collisional Growth outside the Snowline}",
      journal = {\apj},
         year = 2018,
        month = dec,
       volume = {868},
       number = {2},
          eid = {118},
        pages = {118},
          doi = {10.3847/1538-4357/aae0fb},
       adsurl = {https://ui.adsabs.harvard.edu/abs/2018ApJ...868..118H}
}

@ARTICLE{1997ApJ...490..368C,
       author = {{Chiang}, E.~I. and {Goldreich}, P.},
        title = "{Spectral Energy Distributions of T Tauri Stars with Passive Circumstellar Disks}",
      journal = {\apj},
         year = 1997,
        month = nov,
       volume = {490},
       number = {1},
        pages = {368-376},
          doi = {10.1086/304869},
       adsurl = {https://ui.adsabs.harvard.edu/abs/1997ApJ...490..368C}
}

@ARTICLE{1974MNRAS.168..603L,
       author = {{Lynden-Bell}, D. and {Pringle}, J.~E.},
        title = "{The evolution of viscous discs and the origin of the nebular variables.}",
      journal = {\mnras},
         year = 1974,
        month = sep,
       volume = {168},
        pages = {603-637},
          doi = {10.1093/mnras/168.3.603},
       adsurl = {https://ui.adsabs.harvard.edu/abs/1974MNRAS.168..603L}
}

@ARTICLE{1980EPSL..47..391J,
       author = {{Jaoul}, O. and {Froidevaux}, C. and {Durham}, W.~B. and {Michaut}, M.},
        title = "{Oxygen self-diffusion in forsterite: Implications for the high-temperature creep mechanism}",
      journal = {Earth and Planetary Science Letters},
         year = 1980,
        month = may,
       volume = {47},
       number = {3},
        pages = {391-397},
          doi = {10.1016/0012-821X(80)90026-6},
       adsurl = {https://ui.adsabs.harvard.edu/abs/1980E&PSL..47..391J}
}

@ARTICLE{2020M&PS...55.1281Y,
       author = {{Yamamoto}, Daiki and {Tachibana}, Shogo and {Kawasaki}, Noriyuki and {Yurimoto}, Hisayoshi},
        title = "{Survivability of presolar oxygen isotopic signature of amorphous silicate dust in the protosolar disk}",
      journal = {\maps},
         year = 2020,
        month = jun,
       volume = {55},
       number = {6},
        pages = {1281-1292},
          doi = {10.1111/maps.13365},
       adsurl = {https://ui.adsabs.harvard.edu/abs/2020M&PS...55.1281Y}
}

@ARTICLE{2007Icar..192..588Y,
       author = {{Youdin}, Andrew N. and {Lithwick}, Yoram},
        title = "{Particle stirring in turbulent gas disks: Including orbital oscillations}",
      journal = {\icarus},
         year = 2007,
        month = dec,
       volume = {192},
       number = {2},
        pages = {588-604},
          doi = {10.1016/j.icarus.2007.07.012},
       adsurl = {https://ui.adsabs.harvard.edu/abs/2007Icar..192..588Y}
}

@ARTICLE{2013ChRv..113.9043V,
       author = {{van Dishoeck}, Ewine F. and {Herbst}, Eric and {Neufeld}, David A.},
        title = "{Interstellar Water Chemistry: From Laboratory to Observations}",
      journal = {Chemical Reviews},
         year = 2013,
        month = dec,
       volume = {113},
       number = {12},
        pages = {9043-9085},
          doi = {10.1021/cr4003177},
       adsurl = {https://ui.adsabs.harvard.edu/abs/2013ChRv..113.9043V}
}

@ARTICLE{2010ApJ...710.1009W,
       author = {{Whittet}, D.~C.~B.},
        title = "{Oxygen Depletion in the Interstellar Medium: Implications for Grain Models and the Distribution of Elemental Oxygen}",
      journal = {\apj},
         year = 2010,
        month = feb,
       volume = {710},
       number = {2},
        pages = {1009-1016},
          doi = {10.1088/0004-637X/710/2/1009},
       adsurl = {https://ui.adsabs.harvard.edu/abs/2010ApJ...710.1009W}
}

@ARTICLE{2025NatAs...9..883S,
       author = {{Smith}, Z.~L. and {Dickinson}, H.~J. and {Fraser}, H.~J. and {McClure}, M.~K. and {Noble}, J.~A. and {Boogert}, A.~C.~A. and {Sun}, F. and {Egami}, E. and {Dartois}, E. and {Erkal}, J. and {Shimonishi}, T. and {Beck}, T.~L. and {Bergner}, J.~B. and {Caselli}, P. and {Charnley}, S.~B. and {Chu}, L. and {Drozdovskaya}, M.~N. and {Garrod}, R. and {Harsono}, D. and {Ioppolo}, S. and {Jimenez-Serra}, I. and {J{\o}rgensen}, J.~K. and {Melnick}, G.~J. and {{\~A}-berg}, K.~I. and {Palumbo}, M.~E. and {Pendleton}, Y.~J. and {Perotti}, G. and {Pontoppidan}, K.~M. and {Qasim}, D. and {Rocha}, W.~R.~M. and {Sturm}, J.~A. and {Taillard}, A. and {Urso}, R.~G. and {van Dishoeck}, E.~F.},
        title = "{Cospatial ice mapping of H$_{2}$O with CO$_{2}$ and CO across a molecular cloud with JWST/NIRCam}",
      journal = {Nature Astronomy},
         year = 2025,
        month = jun,
       volume = {9},
        pages = {883-894},
          doi = {10.1038/s41550-025-02511-z},
       adsurl = {https://ui.adsabs.harvard.edu/abs/2025NatAs...9..883S}
}

@ARTICLE{2024NatAs...8.1148U,
       author = {{Ueda}, Takahiro and {Tazaki}, Ryo and {Okuzumi}, Satoshi and {Flock}, Mario and {Sudarshan}, Prakruti},
        title = "{Support for fragile porous dust in a gravitationally self-regulated disk around IM Lup}",
      journal = {Nature Astronomy},
         year = 2024,
        month = sep,
       volume = {8},
        pages = {1148-1158},
          doi = {10.1038/s41550-024-02308-6},
       adsurl = {https://ui.adsabs.harvard.edu/abs/2024NatAs...8.1148U}
}

@ARTICLE{2024A&A...682A.144D,
       author = {{Dominik}, C. and {Dullemond}, C.~P.},
        title = "{The bouncing barrier revisited: Impact on key planet formation processes and observational signatures}",
      journal = {\aap},
         year = 2024,
        month = feb,
       volume = {682},
          eid = {A144},
        pages = {A144},
          doi = {10.1051/0004-6361/202347716},
       adsurl = {https://ui.adsabs.harvard.edu/abs/2024A&A...682A.144D}
}

@ARTICLE{2022A&A...664A.137G,
       author = {{Guidi}, G. and {Isella}, A. and {Testi}, L. and {Chandler}, C.~J. and {Liu}, H.~B. and {Schmid}, H.~M. and {Rosotti}, G. and {Meng}, C. and {Jennings}, J. and {Williams}, J.~P. and {Carpenter}, J.~M. and {de Gregorio-Monsalvo}, I. and {Li}, H. and {Liu}, S.~F. and {Ortolani}, S. and {Quanz}, S.~P. and {Ricci}, L. and {Tazzari}, M.},
        title = "{Distribution of solids in the rings of the HD 163296 disk: a multiwavelength study}",
      journal = {\aap},
         year = 2022,
        month = aug,
       volume = {664},
          eid = {A137},
        pages = {A137},
          doi = {10.1051/0004-6361/202142303},
       adsurl = {https://ui.adsabs.harvard.edu/abs/2022A&A...664A.137G}
}

@ARTICLE{2023ApJ...957...11D,
       author = {{Doi}, Kiyoaki and {Kataoka}, Akimasa},
        title = "{Constraints on the Dust Size Distributions in the HD 163296 Disk from the Difference of the Apparent Dust Ring Widths between Two ALMA Bands}",
      journal = {\apj},
         year = 2023,
        month = nov,
       volume = {957},
       number = {1},
          eid = {11},
        pages = {11},
          doi = {10.3847/1538-4357/acf5df},
       adsurl = {https://ui.adsabs.harvard.edu/abs/2023ApJ...957...11D}
}

@ARTICLE{2021ApJ...920L..35T,
       author = {{Tsukamoto}, Yusuke and {Machida}, Masahiro N. and {Inutsuka}, Shu-ichiro},
        title = "{``Ashfall'' Induced by Molecular Outflow in Protostar Evolution}",
      journal = {\apjl},
         year = 2021,
        month = oct,
       volume = {920},
       number = {2},
          eid = {L35},
        pages = {L35},
          doi = {10.3847/2041-8213/ac2b2f},
       adsurl = {https://ui.adsabs.harvard.edu/abs/2021ApJ...920L..35T}
}

@ARTICLE{2025ApJ...985...16T,
       author = {{Takasao}, Shinsuke and {Hosokawa}, Takashi and {Tomida}, Kengo and {Iwasaki}, Kazunari},
        title = "{Connecting a Magnetized Disk to a Convective Low-mass Protostar: A Global 3D Model of Boundary Layer Accretion}",
      journal = {\apj},
         year = 2025,
        month = may,
       volume = {985},
       number = {1},
          eid = {16},
        pages = {16},
          doi = {10.3847/1538-4357/adc37b},
       adsurl = {https://ui.adsabs.harvard.edu/abs/2025ApJ...985...16T}
}

@ARTICLE{1993Metic..28...14W,
       author = {{Wasson}, J.~T.},
        title = "{Constraints on chondrule origins.}",
      journal = {Meteoritics},
         year = 1993,
        month = mar,
       volume = {28},
       number = {1},
        pages = {14-28},
          doi = {10.1111/j.1945-5100.1993.tb00244.x},
       adsurl = {https://ui.adsabs.harvard.edu/abs/1993Metic..28...14W}
}

@ARTICLE{2021GeCoA.314..108Y,
       author = {{Yamamoto}, Daiki and {Kawasaki}, Noriyuki and {Tachibana}, Shogo and {Kamibayashi}, Michiru and {Yurimoto}, Hisayoshi},
        title = "{An experimental study on oxygen isotope exchange reaction between CAI melt and low-pressure water vapor under simulated Solar nebular conditions}",
      journal = {\gca},
         year = 2021,
        month = dec,
       volume = {314},
        pages = {108-120},
          doi = {10.1016/j.gca.2021.09.016},
       adsurl = {https://ui.adsabs.harvard.edu/abs/2021GeCoA.314..108Y}
}

@INPROCEEDINGS{2023ASPC..534..317T,
       author = {{Tsukamoto}, Y. and {Maury}, A. and {Commercon}, B. and {Alves}, F.~O. and {Cox}, E.~G. and {Sakai}, N. and {Ray}, T. and {Zhao}, B. and {Machida}, M.~N.},
        title = "{The Role of Magnetic Fields in the Formation of Protostars, Disks, and Outflows}",
    booktitle = {Protostars and Planets VII},
         year = 2023,
       editor = {{Inutsuka}, S. and {Aikawa}, Y. and {Muto}, T. and {Tomida}, K. and {Tamura}, M.},
       series = {Astronomical Society of the Pacific Conference Series},
       volume = {534},
        month = jul,
        pages = {317},
          doi = {10.48550/arXiv.2209.13765},
       adsurl = {https://ui.adsabs.harvard.edu/abs/2023ASPC..534..317T}
}

@ARTICLE{2020ApJ...898..118H,
       author = {{Hirano}, Shingo and {Tsukamoto}, Yusuke and {Basu}, Shantanu and {Machida}, Masahiro N.},
        title = "{The Effect of Misalignment between the Rotation Axis and Magnetic Field on the Circumstellar Disk}",
      journal = {\apj},
         year = 2020,
        month = aug,
       volume = {898},
       number = {2},
          eid = {118},
        pages = {118},
          doi = {10.3847/1538-4357/ab9f9d},
       adsurl = {https://ui.adsabs.harvard.edu/abs/2020ApJ...898..118H}
}

@ARTICLE{2023ApJ...947L..29A,
       author = {{Arzoumanian}, Doris and {Arakawa}, Sota and {Kobayashi}, Masato I.~N. and {Iwasaki}, Kazunari and {Fukuda}, Kohei and {Mori}, Shoji and {Hirai}, Yutaka and {Kunitomo}, Masanobu and {Kumar}, M.~S. Nanda and {Kokubo}, Eiichiro},
        title = "{Insights on the Sun Birth Environment in the Context of Star Cluster Formation in Hub-Filament Systems}",
      journal = {\apjl},
         year = 2023,
        month = apr,
       volume = {947},
       number = {2},
          eid = {L29},
        pages = {L29},
          doi = {10.3847/2041-8213/acc849},
       adsurl = {https://ui.adsabs.harvard.edu/abs/2023ApJ...947L..29A}
}

@ARTICLE{2021A&A...648A.101L,
       author = {{Lee}, Yueh-Ning and {Charnoz}, S{\'e}bastien and {Hennebelle}, Patrick},
        title = "{Protoplanetary disk formation from the collapse of a prestellar core}",
      journal = {\aap},
         year = 2021,
        month = apr,
       volume = {648},
          eid = {A101},
        pages = {A101},
          doi = {10.1051/0004-6361/202038105},
       adsurl = {https://ui.adsabs.harvard.edu/abs/2021A&A...648A.101L}
}

@INPROCEEDINGS{2023ASPC..534..233P,
       author = {{Pineda}, J.~E. and {Arzoumanian}, D. and {Andre}, P. and {Friesen}, R.~K. and {Zavagno}, A. and {Clarke}, S.~D. and {Inoue}, T. and {Chen}, C. and {Lee}, Y. and {Soler}, J.~D. and {Kuffmeier}, M.},
        title = "{From Bubbles and Filaments to Cores and Disks: Gas Gathering and Growth of Structure Leading to the Formation of Stellar Systems}",
    booktitle = {Protostars and Planets VII},
         year = 2023,
       editor = {{Inutsuka}, S. and {Aikawa}, Y. and {Muto}, T. and {Tomida}, K. and {Tamura}, M.},
       series = {Astronomical Society of the Pacific Conference Series},
       volume = {534},
        month = jul,
        pages = {233},
          doi = {10.48550/arXiv.2205.03935},
       adsurl = {https://ui.adsabs.harvard.edu/abs/2023ASPC..534..233P}
}

@ARTICLE{2020MNRAS.492.3849K,
       author = {{Kunitomo}, Masanobu and {Suzuki}, Takeru K. and {Inutsuka}, Shu-ichiro},
        title = "{Dispersal of protoplanetary discs by the combination of magnetically driven and photoevaporative winds}",
      journal = {\mnras},
         year = 2020,
        month = mar,
       volume = {492},
       number = {3},
        pages = {3849-3858},
          doi = {10.1093/mnras/staa087},
       adsurl = {https://ui.adsabs.harvard.edu/abs/2020MNRAS.492.3849K}
}

@ARTICLE{2018ApJ...865...75N,
       author = {{Nakatani}, Riouhei and {Hosokawa}, Takashi and {Yoshida}, Naoki and {Nomura}, Hideko and {Kuiper}, Rolf},
        title = "{Radiation Hydrodynamics Simulations of Photoevaporation of Protoplanetary Disks. II. Metallicity Dependence of UV and X-Ray Photoevaporation}",
      journal = {\apj},
         year = 2018,
        month = sep,
       volume = {865},
       number = {1},
          eid = {75},
        pages = {75},
          doi = {10.3847/1538-4357/aad9fd},
       adsurl = {https://ui.adsabs.harvard.edu/abs/2018ApJ...865...75N}
}

@ARTICLE{2025ApJ...983...15T,
       author = {{Tominaga}, Ryosuke T. and {Tanaka}, Hidekazu},
        title = "{Dust Coagulation Assisted by Streaming Instability in Protoplanetary Disks}",
      journal = {\apj},
         year = 2025,
        month = apr,
       volume = {983},
       number = {1},
          eid = {15},
        pages = {15},
          doi = {10.3847/1538-4357/adbbca},
       adsurl = {https://ui.adsabs.harvard.edu/abs/2025ApJ...983...15T}
}

@ARTICLE{2023ApJ...949..119K,
       author = {{Kondo}, Katsushi and {Okuzumi}, Satoshi and {Mori}, Shoji},
        title = "{The Roles of Dust Growth in the Temperature Evolution and Snow Line Migration in Magnetically Accreting Protoplanetary Disks}",
      journal = {\apj},
         year = 2023,
        month = jun,
       volume = {949},
       number = {2},
          eid = {119},
        pages = {119},
          doi = {10.3847/1538-4357/acc840},
archivePrefix = {arXiv},
       eprint = {2205.13511},
 primaryClass = {astro-ph.EP},
       adsurl = {https://ui.adsabs.harvard.edu/abs/2023ApJ...949..119K}
}

@ARTICLE{2024PASJ...76..616I,
       author = {{Iwasaki}, Kazunari and {Tomida}, Kengo and {Takasao}, Shinsuke and {Okuzumi}, Satoshi and {Suzuki}, Takeru K.},
        title = "{Dynamics near the inner dead-zone edges in a proprotoplanetary disk}",
      journal = {\pasj},
         year = 2024,
        month = aug,
       volume = {76},
       number = {4},
        pages = {616-652},
          doi = {10.1093/pasj/psae036},
       adsurl = {https://ui.adsabs.harvard.edu/abs/2024PASJ...76..616I}
}

@ARTICLE{2017ApJ...845...75B,
       author = {{Bai}, Xue-Ning},
        title = "{Global Simulations of the Inner Regions of Protoplanetary Disks with Comprehensive Disk Microphysics}",
      journal = {\apj},
         year = 2017,
        month = aug,
       volume = {845},
       number = {1},
          eid = {75},
        pages = {75},
          doi = {10.3847/1538-4357/aa7dda},
       adsurl = {https://ui.adsabs.harvard.edu/abs/2017ApJ...845...75B}
}

@ARTICLE{2020ApJ...899..134K,
       author = {{Krijt}, Sebastiaan and {Bosman}, Arthur D. and {Zhang}, Ke and {Schwarz}, Kamber R. and {Ciesla}, Fred J. and {Bergin}, Edwin A.},
        title = "{CO Depletion in Protoplanetary Disks: A Unified Picture Combining Physical Sequestration and Chemical Processing}",
      journal = {\apj},
         year = 2020,
        month = aug,
       volume = {899},
       number = {2},
          eid = {134},
        pages = {134},
          doi = {10.3847/1538-4357/aba75d},
       adsurl = {https://ui.adsabs.harvard.edu/abs/2020ApJ...899..134K}
}

@ARTICLE{2022ApJ...938...29F,
       author = {{Furuya}, Kenji and {Lee}, Seokho and {Nomura}, Hideko},
        title = "{Different Degrees of Nitrogen and Carbon Depletion in the Warm Molecular Layers of Protoplanetary Disks}",
      journal = {\apj},
         year = 2022,
        month = oct,
       volume = {938},
       number = {1},
          eid = {29},
        pages = {29},
          doi = {10.3847/1538-4357/ac9233},
       adsurl = {https://ui.adsabs.harvard.edu/abs/2022ApJ...938...29F}
}

@ARTICLE{2019NatAs...3..910F,
       author = {{Fujiya}, W. and {Hoppe}, P. and {Ushikubo}, T. and {Fukuda}, K. and {Lindgren}, P. and {Lee}, M.~R. and {Koike}, M. and {Shirai}, K. and {Sano}, Y.},
        title = "{Migration of D-type asteroids from the outer Solar System inferred from carbonate in meteorites}",
      journal = {Nature Astronomy},
         year = 2019,
        month = jul,
       volume = {3},
        pages = {910-915},
          doi = {10.1038/s41550-019-0801-4},
       adsurl = {https://ui.adsabs.harvard.edu/abs/2019NatAs...3..910F}
}

@ARTICLE{2023Sci...379.8671N,
       author = {{Nakamura}, T. and {Matsumoto}, M. and {Amano}, K. and {Enokido}, Y. and {Zolensky}, M.~E. and {Mikouchi}, T. and {Genda}, H. and {Tanaka}, S. and {Zolotov}, M.~Y. and {Kurosawa}, K. and {Wakita}, S. and {Hyodo}, R. and {Nagano}, H. and {Nakashima}, D. and {Takahashi}, Y. and {Fujioka}, Y. and {Kikuiri}, M. and {Kagawa}, E. and {Matsuoka}, M. and {Brearley}, A.~J. and {Tsuchiyama}, A. and {Uesugi}, M. and {Matsuno}, J. and {Kimura}, Y. and {Sato}, M. and {Milliken}, R.~E. and {Tatsumi}, E. and {Sugita}, S. and {Hiroi}, T. and {Kitazato}, K. and {Brownlee}, D. and {Joswiak}, D.~J. and {Takahashi}, M. and {Ninomiya}, K. and {Takahashi}, T. and {Osawa}, T. and {Terada}, K. and {Brenker}, F.~E. and {Tkalcec}, B.~J. and {Vincze}, L. and {Brunetto}, R. and {Al{\'e}on-Toppani}, A. and {Chan}, Q.~H.~S. and {Roskosz}, M. and {Viennet}, J.-C. and {Beck}, P. and {Alp}, E.~E. and {Michikami}, T. and {Nagaashi}, Y. and {Tsuji}, T. and {Ino}, Y. and {Martinez}, J. and {Han}, J. and {Dolocan}, A. and {Bodnar}, R.~J. and {Tanaka}, M. and {Yoshida}, H. and {Sugiyama}, K. and {King}, A.~J. and {Fukushi}, K. and {Suga}, H. and {Yamashita}, S. and {Kawai}, T. and {Inoue}, K. and {Nakato}, A. and {Noguchi}, T. and {Vilas}, F. and {Hendrix}, A.~R. and {Jaramillo-Correa}, C. and {Domingue}, D.~L. and {Dominguez}, G. and {Gainsforth}, Z. and {Engrand}, C. and {Duprat}, J. and {Russell}, S.~S. and {Bonato}, E. and {Ma}, C. and {Kawamoto}, T. and {Wada}, T. and {Watanabe}, S. and {Endo}, R. and {Enju}, S. and {Riu}, L. and {Rubino}, S. and {Tack}, P. and {Takeshita}, S. and {Takeichi}, Y. and {Takeuchi}, A. and {Takigawa}, A. and {Takir}, D. and {Tanigaki}, T. and {Taniguchi}, A. and {Tsukamoto}, K. and {Yagi}, T. and {Yamada}, S. and {Yamamoto}, K. and {Yamashita}, Y. and {Yasutake}, M. and {Uesugi}, K. and {Umegaki}, I. and {Chiu}, I. and {Ishizaki}, T. and {Okumura}, S. and {Palomba}, E. and {Pilorget}, C. and {Potin}, S.~M. and {Alasli}, A. and {Anada}, S. and {Araki}, Y. and {Sakatani}, N. and {Schultz}, C. and {Sekizawa}, O. and {Sitzman}, S.~D. and {Sugiura}, K. and {Sun}, M. and {Dartois}, E. and {De Pauw}, E. and {Dionnet}, Z. and {Djouadi}, Z. and {Falkenberg}, G. and {Fujita}, R. and {Fukuma}, T. and {Gearba}, I.~R. and {Hagiya}, K. and {Hu}, M.~Y. and {Kato}, T. and {Kawamura}, T. and {Kimura}, M. and {Kubo}, M.~K. and {Langenhorst}, F. and {Lantz}, C. and {Lavina}, B. and {Lindner}, M. and {Zhao}, J. and {Vekemans}, B. and {Baklouti}, D. and {Bazi}, B. and {Borondics}, F. and {Nagasawa}, S. and {Nishiyama}, G. and {Nitta}, K. and {Mathurin}, J. and {Matsumoto}, T. and {Mitsukawa}, I. and {Miura}, H. and {Miyake}, A. and {Miyake}, Y. and {Yurimoto}, H. and {Okazaki}, R. and {Yabuta}, H. and {Naraoka}, H. and {Sakamoto}, K. and {Tachibana}, S. and {Connolly}, H.~C. and {Lauretta}, D.~S. and {Yoshitake}, M. and {Yoshikawa}, M. and {Yoshikawa}, K. and {Yoshihara}, K. and {Yokota}, Y. and {Yogata}, K. and {Yano}, H. and {Yamamoto}, Y. and {Yamamoto}, D. and {Yamada}, M. and {Yamada}, T. and {Yada}, T. and {Wada}, K. and {Usui}, T. and {Tsukizaki}, R. and {Terui}, F. and {Takeuchi}, H. and {Takei}, Y. and {Iwamae}, A. and {Soejima}, H. and {Shirai}, K. and {Shimaki}, Y. and {Senshu}, H. and {Sawada}, H. and {Saiki}, T. and {Ozaki}, M. and {Ono}, G. and {Okada}, T. and {Ogawa}, N. and {Ogawa}, K. and {Noguchi}, R. and {Noda}, H. and {Nishimura}, M. and {Namiki}, N. and {Nakazawa}, S. and {Morota}, T. and {Miyazaki}, A. and {Miura}, A. and {Mimasu}, Y. and {Matsumoto}, K. and {Kumagai}, K. and {Kouyama}, T. and {Kikuchi}, S. and {Kawahara}, K. and {Kameda}, S.},
        title = "{Formation and evolution of carbonaceous asteroid Ryugu: Direct evidence from returned samples}",
      journal = {Science},
         year = 2023,
        month = mar,
       volume = {379},
       number = {6634},
          eid = {abn8671},
        pages = {abn8671},
          doi = {10.1126/science.abn8671},
       adsurl = {https://ui.adsabs.harvard.edu/abs/2023Sci...379.8671N}
}

@ARTICLE{2024M&PS...59.2453L,
       author = {{Lauretta}, Dante S. and {Connolly}, Harold C. and {Aebersold}, Joseph E. and {Alexander}, Conel M. O'D. and {Ballouz}, Ronald-L. and {Barnes}, Jessica J. and {Bates}, Helena C. and {Bennett}, Carina A. and {Blanche}, Laurinne and {Blumenfeld}, Erika H. and {Clemett}, Simon J. and {Cody}, George D. and {DellaGiustina}, Daniella N. and {Dworkin}, Jason P. and {Eckley}, Scott A. and {Foustoukos}, Dionysis I. and {Franchi}, Ian A. and {Glavin}, Daniel P. and {Greenwood}, Richard C. and {Haenecour}, Pierre and {Hamilton}, Victoria E. and {Hill}, Dolores H. and {Hiroi}, Takahiro and {Ishimaru}, Kana and {Jourdan}, Fred and {Kaplan}, Hannah H. and {Keller}, Lindsay P. and {King}, Ashley J. and {Koefoed}, Piers and {Kontogiannis}, Melissa K. and {Le}, Loan and {Macke}, Robert J. and {McCoy}, Timothy J. and {Milliken}, Ralph E. and {Najorka}, Jens and {Nguyen}, Ann N. and {Pajola}, Maurizio and {Polit}, Anjani T. and {Righter}, Kevin and {Roper}, Heather L. and {Russell}, Sara S. and {Ryan}, Andrew J. and {Sandford}, Scott A. and {Schofield}, Paul F. and {Schultz}, Cody D. and {Seifert}, Laura B. and {Tachibana}, Shogo and {Thomas-Keprta}, Kathie L. and {Thompson}, Michelle S. and {Tu}, Valerie and {Tusberti}, Filippo and {Wang}, Kun and {Zega}, Thomas J. and {Wolner}, C.~W.~V.},
        title = "{Asteroid (101955) Bennu in the laboratory: Properties of the sample collected by OSIRIS-REx}",
      journal = {\maps},
         year = 2024,
        month = sep,
       volume = {59},
       number = {9},
        pages = {2453-2486},
          doi = {10.1111/maps.14227},
       adsurl = {https://ui.adsabs.harvard.edu/abs/2024M&PS...59.2453L}
}

@ARTICLE{2019M&PS...54.1647K,
       author = {{Krot}, Alexander N.},
        title = "{Refractory inclusions in carbonaceous chondrites: Records of early solar system processes}",
      journal = {\maps},
         year = 2019,
        month = aug,
       volume = {54},
       number = {8},
        pages = {1647-1691},
          doi = {10.1111/maps.13350},
       adsurl = {https://ui.adsabs.harvard.edu/abs/2019M&PS...54.1647K}
}

@ARTICLE{2010E&PSL.300..343A,
       author = {{Amelin}, Yuri and {Kaltenbach}, Angela and {Iizuka}, Tsuyoshi and {Stirling}, Claudine H. and {Ireland}, Trevor R. and {Petaev}, Michail and {Jacobsen}, Stein B.},
        title = "{U-Pb chronology of the Solar System's oldest solids with variable $^{238}$U/ $^{235}$U}",
      journal = {Earth and Planetary Science Letters},
         year = 2010,
        month = dec,
       volume = {300},
       number = {3-4},
        pages = {343-350},
          doi = {10.1016/j.epsl.2010.10.015},
       adsurl = {https://ui.adsabs.harvard.edu/abs/2010E&PSL.300..343A}
}

@ARTICLE{2012Sci...338..651C,
       author = {{Connelly}, James N. and {Bizzarro}, Martin and {Krot}, Alexander N. and {Nordlund}, {\r{A}}ke and {Wielandt}, Daniel and {Ivanova}, Marina A.},
        title = "{The Absolute Chronology and Thermal Processing of Solids in the Solar Protoplanetary Disk}",
      journal = {Science},
         year = 2012,
        month = nov,
       volume = {338},
       number = {6107},
        pages = {651},
          doi = {10.1126/science.1226919},
       adsurl = {https://ui.adsabs.harvard.edu/abs/2012Sci...338..651C}
}

@ARTICLE{2021GeCoA.293..544F,
       author = {{Fukuda}, Kohei and {Brownlee}, Donald E. and {Joswiak}, David J. and {Tenner}, Travis J. and {Kimura}, Makoto and {Kita}, Noriko T.},
        title = "{Correlated isotopic and chemical evidence for condensation origins of olivine in comet 81P/Wild 2 and in AOAs from CV and CO chondrites}",
      journal = {\gca},
         year = 2021,
        month = jan,
       volume = {293},
        pages = {544-574},
          doi = {10.1016/j.gca.2020.09.036},
       adsurl = {https://ui.adsabs.harvard.edu/abs/2021GeCoA.293..544F}
}

@ARTICLE{2021GeCoA.299..199C,
       author = {{Chaumard}, No{\"e}l and {Defouilloy}, C{\'e}line and {Hertwig}, Andreas T. and {Kita}, Noriko T.},
        title = "{Oxygen isotope systematics of chondrules in the Paris CM2 chondrite: Indication for a single large formation region across snow line}",
      journal = {\gca},
         year = 2021,
        month = apr,
       volume = {299},
        pages = {199-218},
          doi = {10.1016/j.gca.2021.02.012},
       adsurl = {https://ui.adsabs.harvard.edu/abs/2021GeCoA.299..199C}
}

@ARTICLE{2014E&PSL.401..327B,
       author = {{Bod{\'e}nan}, Jean-David and {Starkey}, Natalie A. and {Russell}, Sara S. and {Wright}, Ian P. and {Franchi}, Ian A.},
        title = "{An oxygen isotope study of Wark-Lovering rims on type A CAIs in primitive carbonaceous chondrites}",
      journal = {Earth and Planetary Science Letters},
         year = 2014,
        month = sep,
       volume = {401},
        pages = {327-336},
          doi = {10.1016/j.epsl.2014.05.035},
       adsurl = {https://ui.adsabs.harvard.edu/abs/2014E&PSL.401..327B}
}

@ARTICLE{2003ApJ...585L..59H,
       author = {{Honda}, Mitsuhiko and {Kataza}, Hirokazu and {Okamoto}, Yoshiko K. and {Miyata}, Takashi and {Yamashita}, Takuya and {Sako}, Shigeyuki and {Takubo}, Shinya and {Onaka}, Takashi},
        title = "{Detection of Crystalline Silicates around the T Tauri Star Hen 3-600A}",
      journal = {\apjl},
         year = 2003,
        month = mar,
       volume = {585},
       number = {1},
        pages = {L59-L63},
          doi = {10.1086/374034},
       adsurl = {https://ui.adsabs.harvard.edu/abs/2003ApJ...585L..59H}
}

@ARTICLE{2007P&SS...55.1044O,
       author = {{Ootsubo}, Takafumi and {Watanabe}, Jun-ichi and {Kawakita}, Hideyo and {Honda}, Mitsuhiko and {Furusho}, Reiko},
        title = "{Grain properties of Oort cloud comets: Modeling the mineralogical composition of cometary dust from mid-infrared emission features}",
      journal = {\planss},
         year = 2007,
        month = jun,
       volume = {55},
       number = {9},
        pages = {1044-1049},
          doi = {10.1016/j.pss.2006.11.012},
       adsurl = {https://ui.adsabs.harvard.edu/abs/2007P&SS...55.1044O}
}

@ARTICLE{2023PSJ.....4..242H,
       author = {{Harker}, David E. and {Wooden}, Diane H. and {Kelley}, Michael S.~P. and {Woodward}, Charles E.},
        title = "{Dust Properties of Comets Observed by Spitzer}",
      journal = {\psj},
         year = 2023,
        month = dec,
       volume = {4},
       number = {12},
          eid = {242},
        pages = {242},
          doi = {10.3847/PSJ/ad0382},
       adsurl = {https://ui.adsabs.harvard.edu/abs/2023PSJ.....4..242H}
}

@ARTICLE{2025ComEE...6..537K,
       author = {{Kawasaki}, Noriyuki and {Arakawa}, Sota and {Miyamoto}, Yushi and {Sakamoto}, Naoya and {Yamamoto}, Daiki and {Russell}, Sara S. and {Yurimoto}, Hisayoshi},
        title = "{Solar System's earliest solids as tracers of the accretion region of Ryugu and Ivuna-type carbonaceous chondrites}",
      journal = {Communications Earth and Environment},
         year = 2025,
        month = jul,
       volume = {6},
       number = {1},
          eid = {537},
        pages = {537},
          doi = {10.1038/s43247-025-02511-x},
       adsurl = {https://ui.adsabs.harvard.edu/abs/2025ComEE...6..537K}
}

@ARTICLE{2006Sci...314.1735Z,
       author = {{Zolensky}, Michael E. and {Zega}, Thomas J. and {Yano}, Hajime and {Wirick}, Sue and {Westphal}, Andrew J. and {Weisberg}, Mike K. and {Weber}, Iris and {Warren}, Jack L. and {Velbel}, Michael A. and {Tsuchiyama}, Akira and {Tsou}, Peter and {Toppani}, Alice and {Tomioka}, Naotaka and {Tomeoka}, Kazushige and {Teslich}, Nick and {Taheri}, Mitra and {Susini}, Jean and {Stroud}, Rhonda and {Stephan}, Thomas and {Stadermann}, Frank J. and {Snead}, Christopher J. and {Simon}, Steven B. and {Simionovici}, Alexandre and {See}, Thomas H. and {Robert}, Fran{\c{c}}ois and {Rietmeijer}, Frans J.~M. and {Rao}, William and {Perronnet}, Murielle C. and {Papanastassiou}, Dimitri A. and {Okudaira}, Kyoko and {Ohsumi}, Kazumasa and {Ohnishi}, Ichiro and {Nakamura-Messenger}, Keiko and {Nakamura}, Tomoki and {Mostefaoui}, Smail and {Mikouchi}, Takashi and {Meibom}, Anders and {Matrajt}, Graciela and {Marcus}, Matthew A. and {Leroux}, Hugues and {Lemelle}, Laurence and {Le}, Loan and {Lanzirotti}, Antonio and {Langenhorst}, Falko and {Krot}, Alexander N. and {Keller}, Lindsay P. and {Kearsley}, Anton T. and {Joswiak}, David and {Jacob}, Damien and {Ishii}, Hope and {Harvey}, Ralph and {Hagiya}, Kenji and {Grossman}, Lawrence and {Grossman}, Jeffrey N. and {Graham}, Giles A. and {Gounelle}, Matthieu and {Gillet}, Philippe and {Genge}, Matthew J. and {Flynn}, George and {Ferroir}, Tristan and {Fallon}, Stewart and {Ebel}, Denton S. and {Dai}, Zu Rong and {Cordier}, Patrick and {Clark}, Benton and {Chi}, Miaofang and {Butterworth}, Anna L. and {Brownlee}, Donald E. and {Bridges}, John C. and {Brennan}, Sean and {Brearley}, Adrian and {Bradley}, John P. and {Bleuet}, Pierre and {Bland}, Phil A. and {Bastien}, Ron},
        title = "{Mineralogy and Petrology of Comet 81P/Wild 2 Nucleus Samples}",
      journal = {Science},
         year = 2006,
        month = dec,
       volume = {314},
       number = {5806},
        pages = {1735},
          doi = {10.1126/science.1135842},
       adsurl = {https://ui.adsabs.harvard.edu/abs/2006Sci...314.1735Z}
}

@ARTICLE{2002Natur.415..860C,
       author = {{Clayton}, Robert N.},
        title = "{Solar System: Self-shielding in the solar nebula}",
      journal = {\nat},
         year = 2002,
        month = feb,
       volume = {415},
       number = {6874},
        pages = {860-861},
          doi = {10.1038/415860b},
       adsurl = {https://ui.adsabs.harvard.edu/abs/2002Natur.415..860C}
}

@ARTICLE{2005Natur.435..317L,
       author = {{Lyons}, J.~R. and {Young}, E.~D.},
        title = "{CO self-shielding as the origin of oxygen isotope anomalies in the early solar nebula}",
      journal = {\nat},
         year = 2005,
        month = may,
       volume = {435},
       number = {7040},
        pages = {317-320},
          doi = {10.1038/nature03557},
       adsurl = {https://ui.adsabs.harvard.edu/abs/2005Natur.435..317L}
}

@ARTICLE{1981Icar...48..353C,
       author = {{Cassen}, P. and {Moosman}, A.},
        title = "{On the formation of protostellar disks}",
      journal = {\icarus},
         year = 1981,
        month = dec,
       volume = {48},
       number = {3},
        pages = {353-376},
          doi = {10.1016/0019-1035(81)90051-8},
       adsurl = {https://ui.adsabs.harvard.edu/abs/1981Icar...48..353C}
}

@ARTICLE{2020SSRv..216...55K,
       author = {{Kleine}, T. and {Budde}, G. and {Burkhardt}, C. and {Kruijer}, T.~S. and {Worsham}, E.~A. and {Morbidelli}, A. and {Nimmo}, F.},
        title = "{The Non-carbonaceous-Carbonaceous Meteorite Dichotomy}",
      journal = {\ssr},
         year = 2020,
        month = may,
       volume = {216},
       number = {4},
          eid = {55},
        pages = {55},
          doi = {10.1007/s11214-020-00675-w},
       adsurl = {https://ui.adsabs.harvard.edu/abs/2020SSRv..216...55K}
}

@ARTICLE{2018ApJ...854..164S,
       author = {{Scott}, Edward R.~D. and {Krot}, Alexander N. and {Sanders}, Ian S.},
        title = "{Isotopic Dichotomy among Meteorites and Its Bearing on the Protoplanetary Disk}",
      journal = {\apj},
         year = 2018,
        month = feb,
       volume = {854},
       number = {2},
          eid = {164},
        pages = {164},
          doi = {10.3847/1538-4357/aaa5a5},
       adsurl = {https://ui.adsabs.harvard.edu/abs/2018ApJ...854..164S}
}

@ARTICLE{1988ApJ...334..771V,
       author = {{van Dishoeck}, Ewine F. and {Black}, John H.},
        title = "{The Photodissociation and Chemistry of Interstellar CO}",
      journal = {\apj},
         year = 1988,
        month = nov,
       volume = {334},
        pages = {771},
          doi = {10.1086/166877},
       adsurl = {https://ui.adsabs.harvard.edu/abs/1988ApJ...334..771V}
}

@ARTICLE{2009A&A...503..323V,
       author = {{Visser}, R. and {van Dishoeck}, E.~F. and {Black}, J.~H.},
        title = "{The photodissociation and chemistry of CO isotopologues: applications to interstellar clouds and circumstellar disks}",
      journal = {\aap},
         year = 2009,
        month = aug,
       volume = {503},
       number = {2},
        pages = {323-343},
          doi = {10.1051/0004-6361/200912129},
       adsurl = {https://ui.adsabs.harvard.edu/abs/2009A&A...503..323V}
}

@ARTICLE{2021A&A...653A.141A,
       author = {{Asplund}, M. and {Amarsi}, A.~M. and {Grevesse}, N.},
        title = "{The chemical make-up of the Sun: A 2020 vision}",
      journal = {\aap},
         year = 2021,
        month = sep,
       volume = {653},
          eid = {A141},
        pages = {A141},
          doi = {10.1051/0004-6361/202140445},
       adsurl = {https://ui.adsabs.harvard.edu/abs/2021A&A...653A.141A}
}

@ARTICLE{1981PThPS..70...35H,
       author = {{Hayashi}, C.},
        title = "{Structure of the Solar Nebula, Growth and Decay of Magnetic Fields and Effects of Magnetic and Turbulent Viscosities on the Nebula}",
      journal = {Progress of Theoretical Physics Supplement},
         year = 1981,
        month = jan,
       volume = {70},
        pages = {35-53},
          doi = {10.1143/PTPS.70.35},
       adsurl = {https://ui.adsabs.harvard.edu/abs/1981PThPS..70...35H}
}

@ARTICLE{2016ApJ...821...82O,
       author = {{Okuzumi}, Satoshi and {Momose}, Munetake and {Sirono}, Sin-iti and {Kobayashi}, Hiroshi and {Tanaka}, Hidekazu},
        title = "{Sintering-induced Dust Ring Formation in Protoplanetary Disks: Application to the HL Tau Disk}",
      journal = {\apj},
         year = 2016,
        month = apr,
       volume = {821},
       number = {2},
          eid = {82},
        pages = {82},
          doi = {10.3847/0004-637X/821/2/82},
       adsurl = {https://ui.adsabs.harvard.edu/abs/2016ApJ...821...82O}
}

@ARTICLE{2010ARA&A..48...21H,
       author = {{Henning}, Thomas},
        title = "{Cosmic Silicates}",
      journal = {\araa},
         year = 2010,
        month = sep,
       volume = {48},
        pages = {21-46},
          doi = {10.1146/annurev-astro-081309-130815},
       adsurl = {https://ui.adsabs.harvard.edu/abs/2010ARA&A..48...21H}
}

@ARTICLE{1955ZA.....37..217E,
       author = {{Ebert}, R.},
        title = "{{\"U}ber die Verdichtung von H I-Gebieten. Mit 5 Textabbildungen}",
      journal = {\zap},
         year = 1955,
        month = jan,
       volume = {37},
        pages = {217},
       adsurl = {https://ui.adsabs.harvard.edu/abs/1955ZA.....37..217E}
}

@ARTICLE{1956MNRAS.116..351B,
       author = {{Bonnor}, W.~B.},
        title = "{Boyle's Law and gravitational instability}",
      journal = {\mnras},
         year = 1956,
        month = jan,
       volume = {116},
        pages = {351},
          doi = {10.1093/mnras/116.3.351},
       adsurl = {https://ui.adsabs.harvard.edu/abs/1956MNRAS.116..351B}
}

@INPROCEEDINGS{2014prpl.conf...27A,
       author = {{Andr{\'e}}, P. and {Di Francesco}, J. and {Ward-Thompson}, D. and {Inutsuka}, S.-I. and {Pudritz}, R.~E. and {Pineda}, J.~E.},
        title = "{From Filamentary Networks to Dense Cores in Molecular Clouds: Toward a New Paradigm for Star Formation}",
    booktitle = {Protostars and Planets VI},
         year = 2014,
       editor = {{Beuther}, Henrik and {Klessen}, Ralf S. and {Dullemond}, Cornelis P. and {Henning}, Thomas},
        month = jan,
        pages = {27-51},
          doi = {10.2458/azu_uapress_9780816531240-ch002},
       adsurl = {https://ui.adsabs.harvard.edu/abs/2014prpl.conf...27A}
}

@ARTICLE{2018ApJ...865..102T,
       author = {{Takahashi}, Sanemichi Z. and {Muto}, Takayuki},
        title = "{Structure Formation in a Young Protoplanetary Disk by a Magnetic Disk Wind}",
      journal = {\apj},
         year = 2018,
        month = oct,
       volume = {865},
       number = {2},
          eid = {102},
        pages = {102},
          doi = {10.3847/1538-4357/aadda0},
       adsurl = {https://ui.adsabs.harvard.edu/abs/2018ApJ...865..102T}
}

@ARTICLE{2010ARA&A..48..205D,
       author = {{Dullemond}, C.~P. and {Monnier}, J.~D.},
        title = "{The Inner Regions of Protoplanetary Disks}",
      journal = {\araa},
         year = 2010,
        month = sep,
       volume = {48},
        pages = {205-239},
          doi = {10.1146/annurev-astro-081309-130932},
       adsurl = {https://ui.adsabs.harvard.edu/abs/2010ARA&A..48..205D}
}

@INPROCEEDINGS{2014prpl.conf..339T,
       author = {{Testi}, L. and {Birnstiel}, T. and {Ricci}, L. and {Andrews}, S. and {Blum}, J. and {Carpenter}, J. and {Dominik}, C. and {Isella}, A. and {Natta}, A. and {Williams}, J.~P. and {Wilner}, D.~J.},
        title = "{Dust Evolution in Protoplanetary Disks}",
    booktitle = {Protostars and Planets VI},
         year = 2014,
       editor = {{Beuther}, Henrik and {Klessen}, Ralf S. and {Dullemond}, Cornelis P. and {Henning}, Thomas},
        month = jan,
        pages = {339-361},
          doi = {10.2458/azu_uapress_9780816531240-ch015},
       adsurl = {https://ui.adsabs.harvard.edu/abs/2014prpl.conf..339T}
}

@ARTICLE{1986Icar...67..375N,
       author = {{Nakagawa}, Y. and {Sekiya}, M. and {Hayashi}, C.},
        title = "{Settling and growth of dust particles in a laminar phase of a low-mass solar nebula}",
      journal = {\icarus},
         year = 1986,
        month = sep,
       volume = {67},
       number = {3},
        pages = {375-390},
          doi = {10.1016/0019-1035(86)90121-1},
       adsurl = {https://ui.adsabs.harvard.edu/abs/1986Icar...67..375N}
}

@ARTICLE{2000ApJ...535..247H,
       author = {{Hallenbeck}, Susan L. and {Nuth}, III, Joseph A. and {Nelson}, Robert N.},
        title = "{Evolving Optical Properties of Annealing Silicate Grains: From Amorphous Condensate to Crystalline Mineral}",
      journal = {\apj},
         year = 2000,
        month = may,
       volume = {535},
       number = {1},
        pages = {247-255},
          doi = {10.1086/308810},
       adsurl = {https://ui.adsabs.harvard.edu/abs/2000ApJ...535..247H}
}

@ARTICLE{2018SSRv..214...52B,
       author = {{Blum}, J{\"u}rgen},
        title = "{Dust Evolution in Protoplanetary Discs and the Formation of Planetesimals. What Have We Learned from Laboratory Experiments?}",
      journal = {\ssr},
         year = 2018,
        month = mar,
       volume = {214},
       number = {2},
          eid = {52},
        pages = {52},
          doi = {10.1007/s11214-018-0486-5},
       adsurl = {https://ui.adsabs.harvard.edu/abs/2018SSRv..214...52B}
}

@ARTICLE{2015ARA&A..53..541B,
       author = {{Boogert}, A.~C. Adwin and {Gerakines}, Perry A. and {Whittet}, Douglas C.~B.},
        title = "{Observations of the icy universe.}",
      journal = {\araa},
         year = 2015,
        month = aug,
       volume = {53},
        pages = {541-581},
          doi = {10.1146/annurev-astro-082214-122348},
       adsurl = {https://ui.adsabs.harvard.edu/abs/2015ARA&A..53..541B}
}

@ARTICLE{2022Eleme..18..175V,
       author = {{Vacher}, Lionel G. and {Fujiya}, Wataru},
        title = "{Recent Advances in our Understanding of Water and Aqueous Activity in Chondrites}",
      journal = {Elements},
         year = 2022,
        month = jun,
       volume = {18},
       number = {3},
        pages = {175-180},
          doi = {10.2138/gselements.18.3.175},
       adsurl = {https://ui.adsabs.harvard.edu/abs/2022Eleme..18..175V}
}

@ARTICLE{2017M&PS...52.1797A,
       author = {{Alexander}, Conel M. O'd. and {Nittler}, Larry R. and {Davidson}, Jemma and {Ciesla}, Fred J.},
        title = "{Measuring the level of interstellar inheritance in the solar protoplanetary disk}",
      journal = {\maps},
         year = 2017,
        month = sep,
       volume = {52},
       number = {9},
        pages = {1797-1821},
          doi = {10.1111/maps.12891},
       adsurl = {https://ui.adsabs.harvard.edu/abs/2017M&PS...52.1797A}
}

@ARTICLE{2017RSOS....470114E,
       author = {{Ercolano}, Barbara and {Pascucci}, Ilaria},
        title = "{The dispersal of planet-forming discs: theory confronts observations}",
      journal = {Royal Society Open Science},
         year = 2017,
        month = apr,
       volume = {4},
       number = {4},
          eid = {170114},
        pages = {170114},
          doi = {10.1098/rsos.170114},
       adsurl = {https://ui.adsabs.harvard.edu/abs/2017RSOS....470114E}
}

@INPROCEEDINGS{2014prpl.conf..547J,
       author = {{Johansen}, A. and {Blum}, J. and {Tanaka}, H. and {Ormel}, C. and {Bizzarro}, M. and {Rickman}, H.},
        title = "{The Multifaceted Planetesimal Formation Process}",
    booktitle = {Protostars and Planets VI},
         year = 2014,
       editor = {{Beuther}, Henrik and {Klessen}, Ralf S. and {Dullemond}, Cornelis P. and {Henning}, Thomas},
        month = jan,
        pages = {547-570},
          doi = {10.2458/azu_uapress_9780816531240-ch024},
       adsurl = {https://ui.adsabs.harvard.edu/abs/2014prpl.conf..547J}
}

@ARTICLE{2016ApJ...828...46A,
       author = {{Ansdell}, M. and {Williams}, J.~P. and {van der Marel}, N. and {Carpenter}, J.~M. and {Guidi}, G. and {Hogerheijde}, M. and {Mathews}, G.~S. and {Manara}, C.~F. and {Miotello}, A. and {Natta}, A. and {Oliveira}, I. and {Tazzari}, M. and {Testi}, L. and {van Dishoeck}, E.~F. and {van Terwisga}, S.~E.},
        title = "{ALMA Survey of Lupus Protoplanetary Disks. I. Dust and Gas Masses}",
      journal = {\apj},
         year = 2016,
        month = sep,
       volume = {828},
       number = {1},
          eid = {46},
        pages = {46},
          doi = {10.3847/0004-637X/828/1/46},
       adsurl = {https://ui.adsabs.harvard.edu/abs/2016ApJ...828...46A}
}

@ARTICLE{2010GeCoA..74.6610K,
       author = {{Kita}, Noriko T. and {Nagahara}, Hiroko and {Tachibana}, Shogo and {Tomomura}, Shin and {Spicuzza}, Michael J. and {Fournelle}, John H. and {Valley}, John W.},
        title = "{High precision SIMS oxygen three isotope study of chondrules in LL3 chondrites: Role of ambient gas during chondrule formation}",
      journal = {\gca},
         year = 2010,
        month = nov,
       volume = {74},
       number = {22},
        pages = {6610-6635},
          doi = {10.1016/j.gca.2010.08.011},
       adsurl = {https://ui.adsabs.harvard.edu/abs/2010GeCoA..74.6610K}
}

@ARTICLE{2021GeCoA.293..328U,
       author = {{Ushikubo}, Takayuki and {Kimura}, Makoto},
        title = "{Oxygen-isotope systematics of chondrules and olivine fragments from Tagish Lake C2 chondrite: Implications of chondrule-forming regions in protoplanetary disk}",
      journal = {\gca},
         year = 2021,
        month = jan,
       volume = {293},
        pages = {328-343},
          doi = {10.1016/j.gca.2020.11.003},
       adsurl = {https://ui.adsabs.harvard.edu/abs/2021GeCoA.293..328U}
}

@ARTICLE{2017GeCoA.201..103U,
       author = {{Ushikubo}, Takayuki and {Tenner}, Travis J. and {Hiyagon}, Hajime and {Kita}, Noriko T.},
        title = "{A long duration of the $^{16}$O-rich reservoir in the solar nebula, as recorded in fine-grained refractory inclusions from the least metamorphosed carbonaceous chondrites}",
      journal = {\gca},
         year = 2017,
        month = mar,
       volume = {201},
        pages = {103-122},
          doi = {10.1016/j.gca.2016.08.032},
       adsurl = {https://ui.adsabs.harvard.edu/abs/2017GeCoA.201..103U}
}

@ARTICLE{2018E&PSL.481..264F,
       author = {{Fujiya}, Wataru},
        title = "{Oxygen isotopic ratios of primordial water in carbonaceous chondrites}",
      journal = {Earth and Planetary Science Letters},
         year = 2018,
        month = jan,
       volume = {481},
        pages = {264-272},
          doi = {10.1016/j.epsl.2017.10.046},
       adsurl = {https://ui.adsabs.harvard.edu/abs/2018E&PSL.481..264F}
}

@ARTICLE{2015E&PSL.430..308M,
       author = {{Marrocchi}, Yves and {Chaussidon}, Marc},
        title = "{A systematic for oxygen isotopic variation in meteoritic chondrules}",
      journal = {Earth and Planetary Science Letters},
         year = 2015,
        month = nov,
       volume = {430},
        pages = {308-315},
          doi = {10.1016/j.epsl.2015.08.032},
       adsurl = {https://ui.adsabs.harvard.edu/abs/2015E&PSL.430..308M}
}

@ARTICLE{2019PNAS..11623461M,
       author = {{Marrocchi}, Yves and {Villeneuve}, Johan and {Jacquet}, Emmanuel and {Piralla}, Maxime and {Chaussidon}, Marc},
        title = "{Rapid condensation of the first Solar System solids}",
      journal = {Proceedings of the National Academy of Science},
         year = 2019,
        month = nov,
       volume = {116},
       number = {47},
        pages = {23461-23466},
          doi = {10.1073/pnas.1912479116},
       adsurl = {https://ui.adsabs.harvard.edu/abs/2019PNAS..11623461M}
}

@ARTICLE{2008Sci...320.1617A,
       author = {{Alexander}, C.~M.~O. 'D. and {Grossman}, J.~N. and {Ebel}, D.~S. and {Ciesla}, F.~J.},
        title = "{The Formation Conditions of Chondrules and Chondrites}",
      journal = {Science},
         year = 2008,
        month = jun,
       volume = {320},
       number = {5883},
        pages = {1617},
          doi = {10.1126/science.1156561},
       adsurl = {https://ui.adsabs.harvard.edu/abs/2008Sci...320.1617A}
}

@ARTICLE{2025Natur.643..649M,
       author = {{McClure}, M.~K. and {van't Hoff}, Merel and {Francis}, Logan and {Bergin}, Edwin and {Rocha}, Will R.~M. and {Sturm}, J.~A. and {Harsono}, Daniel and {van Dishoeck}, Ewine F. and {Black}, John H. and {Noble}, J.~A. and {Qasim}, D. and {Dartois}, E.},
        title = "{Refractory solid condensation detected in an embedded protoplanetary disk}",
      journal = {\nat},
         year = 2025,
        month = jul,
       volume = {643},
       number = {8072},
        pages = {649-653},
          doi = {10.1038/s41586-025-09163-z},
       adsurl = {https://ui.adsabs.harvard.edu/abs/2025Natur.643..649M}
}

@ARTICLE{2025SSRv..221...85T,
       author = {{Tissot}, Fran{\c{c}}ois L.~H. and {Burkhardt}, Christoph and {Kuznetsova}, Aleksandra and {Pack}, Andreas and {Schiller}, Martin and {Spitzer}, Fridolin and {Van Kooten}, Elishevah M.~M.~E. and {Yap}, Teng Ee},
        title = "{Infall and Disk Processes {\textendash} the Message from Meteorites}",
      journal = {\ssr},
         year = 2025,
        month = sep,
       volume = {221},
       number = {7},
          eid = {85},
        pages = {85},
          doi = {10.1007/s11214-025-01207-0},
       adsurl = {https://ui.adsabs.harvard.edu/abs/2025SSRv..221...85T}
}

@ARTICLE{2025SciA...11x7892S,
       author = {{Sawada}, Ryo and {Kurokawa}, Hiroyuki and {Suwa}, Yudai and {Taki}, Tetsuo and {Lee}, Shiu-Hang and {Tanikawa}, Ataru},
        title = "{Cosmic-ray bath in a past supernova gives birth to Earth-like planets}",
      journal = {Science Advances},
         year = 2025,
        month = dec,
       volume = {11},
        pages = {24.7892},
          doi = {10.1126/sciadv.adx7892},
       adsurl = {https://ui.adsabs.harvard.edu/abs/2025SciA...11x7892S}
}

@ARTICLE{2024SSRv..220...69M,
       author = {{Marrocchi}, Yves and {Jones}, Rhian H. and {Russell}, Sara S. and {Hezel}, Dominik C. and {Barosch}, Jens and {Kuznetsova}, Aleksandra},
        title = "{Chondrule Properties and Formation Conditions}",
      journal = {\ssr},
         year = 2024,
        month = sep,
       volume = {220},
       number = {6},
          eid = {69},
        pages = {69},
          doi = {10.1007/s11214-024-01102-0},
       adsurl = {https://ui.adsabs.harvard.edu/abs/2024SSRv..220...69M}
}

@ARTICLE{2018NatCo...9..908L,
       author = {{Lyons}, James R. and {Gharib-Nezhad}, Ehsan and {Ayres}, Thomas R.},
        title = "{A light carbon isotope composition for the Sun}",
      journal = {Nature Communications},
         year = 2018,
        month = mar,
       volume = {9},
          eid = {908},
        pages = {908},
          doi = {10.1038/s41467-018-03093-3},
       adsurl = {https://ui.adsabs.harvard.edu/abs/2018NatCo...9..908L}
}

@ARTICLE{2023SSRv..219...32H,
       author = {{Hoppe}, Peter and {Rubin}, Martin and {Altwegg}, Kathrin},
        title = "{A Comparison of Presolar Isotopic Signatures in Laboratory-Studied Primitive Solar System Materials and Comet 67P/Churyumov-Gerasimenko: New Insights from Light Elements, Halogens, and Noble Gases}",
      journal = {\ssr},
         year = 2023,
        month = jun,
       volume = {219},
       number = {4},
          eid = {32},
        pages = {32},
          doi = {10.1007/s11214-023-00977-9},
       adsurl = {https://ui.adsabs.harvard.edu/abs/2023SSRv..219...32H}
}

@ARTICLE{2015GeCoA.148..228T,
       author = {{Tenner}, Travis J. and {Nakashima}, Daisuke and {Ushikubo}, Takayuki and {Kita}, Noriko T. and {Weisberg}, Michael K.},
        title = "{Oxygen isotope ratios of FeO-poor chondrules in CR3 chondrites: Influence of dust enrichment and H$_{2}$O during chondrule formation}",
      journal = {\gca},
         year = 2015,
        month = jan,
       volume = {148},
        pages = {228-250},
          doi = {10.1016/j.gca.2014.09.025},
       adsurl = {https://ui.adsabs.harvard.edu/abs/2015GeCoA.148..228T}
}

@ARTICLE{2020GeCoA.282..133S,
       author = {{Schrader}, Devin L. and {Nagashima}, Kazuhide and {Davidson}, Jemma and {McCoy}, Timothy J. and {Ogliore}, Ryan C. and {Fu}, Roger R.},
        title = "{Outward migration of chondrule fragments in the early Solar System: O-isotopic evidence for rocky material crossing the Jupiter Gap?}",
      journal = {\gca},
         year = 2020,
        month = aug,
       volume = {282},
        pages = {133-155},
          doi = {10.1016/j.gca.2020.05.014},
       adsurl = {https://ui.adsabs.harvard.edu/abs/2020GeCoA.282..133S}
}

@ARTICLE{2011GeCoA..75.6556W,
       author = {{Weisberg}, Michael K. and {Ebel}, Denton S. and {Connolly}, Harold C. and {Kita}, Noriko T. and {Ushikubo}, Takayuki},
        title = "{Petrology and oxygen isotope compositions of chondrules in E3 chondrites}",
      journal = {\gca},
         year = 2011,
        month = nov,
       volume = {75},
       number = {21},
        pages = {6556-6569},
          doi = {10.1016/j.gca.2011.08.040},
       adsurl = {https://ui.adsabs.harvard.edu/abs/2011GeCoA..75.6556W}
}

@ARTICLE{2017GeCoA.209...24M,
       author = {{Miller}, K.~E. and {Lauretta}, D.~S. and {Connolly}, H.~C. and {Berger}, E.~L. and {Nagashima}, K. and {Domanik}, K.},
        title = "{Formation of unequilibrated R chondrite chondrules and opaque phases}",
      journal = {\gca},
         year = 2017,
        month = jul,
       volume = {209},
        pages = {24-50},
          doi = {10.1016/j.gca.2017.04.009},
       adsurl = {https://ui.adsabs.harvard.edu/abs/2017GeCoA.209...24M}
}

@ARTICLE{2025ApJ...979L..29I,
       author = {{Iizuka}, Tsuyoshi and {Hibiya}, Yuki and {Yoshihara}, Satoshi and {Hayakawa}, Takehito},
        title = "{Timescales of Solar System Formation Based on Al{\textendash}Ti Isotope Correlation by Supernova Ejecta}",
      journal = {\apjl},
         year = 2025,
        month = feb,
       volume = {979},
       number = {2},
          eid = {L29},
        pages = {L29},
          doi = {10.3847/2041-8213/ada554},
archivePrefix = {arXiv},
       eprint = {2412.20022},
 primaryClass = {astro-ph.EP},
       adsurl = {https://ui.adsabs.harvard.edu/abs/2025ApJ...979L..29I}
}

@ARTICLE{2022GeCoA.322..194F,
       author = {{Fukuda}, Kohei and {Tenner}, Travis J. and {Kimura}, Makoto and {Tomioka}, Naotaka and {Siron}, Guillaume and {Ushikubo}, Takayuki and {Chaumard}, No{\"e}l and {Hertwig}, Andreas T. and {Kita}, Noriko T.},
        title = "{A temporal shift of chondrule generation from the inner to outer Solar System inferred from oxygen isotopes and Al-Mg chronology of chondrules from primitive CM and CO chondrites}",
      journal = {\gca},
         year = 2022,
        month = apr,
       volume = {322},
        pages = {194-226},
          doi = {10.1016/j.gca.2021.12.027},
       adsurl = {https://ui.adsabs.harvard.edu/abs/2022GeCoA.322..194F}
}

@ARTICLE{2019GeCoA.253..111H,
       author = {{Hertwig}, Andreas T. and {Kimura}, Makoto and {Ushikubo}, Takayuki and {Defouilloy}, C{\'e}line and {Kita}, Noriko T.},
        title = "{The $^{26}$Al-$^{26}$Mg systematics of FeO-rich chondrules from Acfer 094: Two chondrule generations distinct in age and oxygen isotope ratios}",
      journal = {\gca},
         year = 2019,
        month = may,
       volume = {253},
        pages = {111-126},
          doi = {10.1016/j.gca.2019.02.020},
       adsurl = {https://ui.adsabs.harvard.edu/abs/2019GeCoA.253..111H}
}

@ARTICLE{2019GeCoA.260..133T,
       author = {{Tenner}, Travis J. and {Nakashima}, Daisuke and {Ushikubo}, Takayuki and {Tomioka}, Naotaka and {Kimura}, Makoto and {Weisberg}, Michael K. and {Kita}, Noriko T.},
        title = "{Extended chondrule formation intervals in distinct physicochemical environments: Evidence from Al-Mg isotope systematics of CR chondrite chondrules with unaltered plagioclase}",
      journal = {\gca},
         year = 2019,
        month = sep,
       volume = {260},
        pages = {133-160},
          doi = {10.1016/j.gca.2019.06.023},
       adsurl = {https://ui.adsabs.harvard.edu/abs/2019GeCoA.260..133T}
}

@ARTICLE{2021GeCoA.293..103S,
       author = {{Siron}, Guillaume and {Fukuda}, Kohei and {Kimura}, Makoto and {Kita}, Noriko T.},
        title = "{New constraints from $^{26}$Al-$^{26}$Mg chronology of anorthite bearing chondrules in unequilibrated ordinary chondrites}",
      journal = {\gca},
         year = 2021,
        month = jan,
       volume = {293},
        pages = {103-126},
          doi = {10.1016/j.gca.2020.10.025},
       adsurl = {https://ui.adsabs.harvard.edu/abs/2021GeCoA.293..103S}
}

@ARTICLE{2019GeCoA.244..416P,
       author = {{Pape}, J. and {Mezger}, K. and {Bouvier}, A.-S. and {Baumgartner}, L.~P.},
        title = "{Time and duration of chondrule formation: Constraints from $^{26}$Al-$^{26}$Mg ages of individual chondrules}",
      journal = {\gca},
         year = 2019,
        month = jan,
       volume = {244},
        pages = {416-436},
          doi = {10.1016/j.gca.2018.10.017},
       adsurl = {https://ui.adsabs.harvard.edu/abs/2019GeCoA.244..416P}
}

@ARTICLE{2012M&PS...47.1108K,
       author = {{Kita}, Noriko T. and {Ushikubo}, Takayuki},
        title = "{Evolution of protoplanetary disk inferred from $^{26}$Al chronology of individual chondrules}",
      journal = {\maps},
         year = 2012,
        month = jul,
       volume = {47},
       number = {7},
        pages = {1108-1119},
          doi = {10.1111/j.1945-5100.2011.01264.x},
       adsurl = {https://ui.adsabs.harvard.edu/abs/2012M&PS...47.1108K}
}

@ARTICLE{2009Sci...325..985V,
       author = {{Villeneuve}, Johan and {Chaussidon}, Marc and {Libourel}, Guy},
        title = "{Homogeneous Distribution of $^{26}$Al in the Solar System from the Mg Isotopic Composition of Chondrules}",
      journal = {Science},
         year = 2009,
        month = aug,
       volume = {325},
       number = {5943},
        pages = {985},
          doi = {10.1126/science.1173907},
       adsurl = {https://ui.adsabs.harvard.edu/abs/2009Sci...325..985V}
}

@ARTICLE{2023Icar..40215607D,
       author = {{Desch}, Steven J. and {Dunlap}, Daniel R. and {Dunham}, Emilie T. and {Williams}, Curtis D. and {Mane}, Prajkta},
        title = "{Statistical chronometry of meteorites. I. A Test of $^{26}$Al homogeneity and the Pb-Pb age of the solar system's t = 0}",
      journal = {\icarus},
         year = 2023,
        month = sep,
       volume = {402},
          eid = {115607},
        pages = {115607},
          doi = {10.1016/j.icarus.2023.115607},
archivePrefix = {arXiv},
       eprint = {2212.00390},
 primaryClass = {astro-ph.EP},
       adsurl = {https://ui.adsabs.harvard.edu/abs/2023Icar..40215607D}
}

@ARTICLE{2017PNAS..114.6712K,
       author = {{Kruijer}, Thomas S. and {Burkhardt}, Christoph and {Budde}, Gerrit and {Kleine}, Thorsten},
        title = "{Age of Jupiter inferred from the distinct genetics and formation times of meteorites}",
      journal = {Proceedings of the National Academy of Science},
         year = 2017,
        month = jun,
       volume = {114},
       number = {26},
        pages = {6712-6716},
          doi = {10.1073/pnas.1704461114},
       adsurl = {https://ui.adsabs.harvard.edu/abs/2017PNAS..114.6712K}
}

@ARTICLE{2020NatAs...4...32K,
       author = {{Kruijer}, Thomas S. and {Kleine}, Thorsten and {Borg}, Lars E.},
        title = "{The great isotopic dichotomy of the early Solar System}",
      journal = {Nature Astronomy},
         year = 2020,
        month = jan,
       volume = {4},
        pages = {32-40},
          doi = {10.1038/s41550-019-0959-9},
       adsurl = {https://ui.adsabs.harvard.edu/abs/2020NatAs...4...32K}
}

@ARTICLE{2023Icar..39415427P,
       author = {{Piralla}, Maxime and {Villeneuve}, Johan and {Schnuriger}, Nicolas and {Bekaert}, David V. and {Marrocchi}, Yves},
        title = "{A unified chronology of dust formation in the early solar system}",
      journal = {\icarus},
         year = 2023,
        month = apr,
       volume = {394},
          eid = {115427},
        pages = {115427},
          doi = {10.1016/j.icarus.2023.115427},
       adsurl = {https://ui.adsabs.harvard.edu/abs/2023Icar..39415427P}
}

@ARTICLE{2013GeCoA.102..226T,
       author = {{Tenner}, Travis J. and {Ushikubo}, Takayuki and {Kurahashi}, Erika and {Kita}, Noriko T. and {Nagahara}, Hiroko},
        title = "{Oxygen isotope systematics of chondrule phenocrysts from the CO3.0 chondrite Yamato 81020: Evidence for two distinct oxygen isotope reservoirs}",
      journal = {\gca},
         year = 2013,
        month = feb,
       volume = {102},
        pages = {226-245},
          doi = {10.1016/j.gca.2012.10.034},
       adsurl = {https://ui.adsabs.harvard.edu/abs/2013GeCoA.102..226T}
}

@ARTICLE{2008A&A...477....9H,
       author = {{Hennebelle}, P. and {Fromang}, S.},
        title = "{Magnetic processes in a collapsing dense core. I. Accretion and ejection}",
      journal = {\aap},
         year = 2008,
        month = jan,
       volume = {477},
       number = {1},
        pages = {9-24},
          doi = {10.1051/0004-6361:20078309},
archivePrefix = {arXiv},
       eprint = {0709.2886},
 primaryClass = {astro-ph},
       adsurl = {https://ui.adsabs.harvard.edu/abs/2008A&A...477....9H}
}

@ARTICLE{2009ApJ...698..922M,
       author = {{Mellon}, Richard R. and {Li}, Zhi-Yun},
        title = "{Magnetic Braking and Protostellar Disk Formation: Ambipolar Diffusion}",
      journal = {\apj},
         year = 2009,
        month = jun,
       volume = {698},
       number = {1},
        pages = {922-927},
          doi = {10.1088/0004-637X/698/1/922},
archivePrefix = {arXiv},
       eprint = {0809.3593},
 primaryClass = {astro-ph},
       adsurl = {https://ui.adsabs.harvard.edu/abs/2009ApJ...698..922M}
}

@ARTICLE{2011PASJ...63..555M,
       author = {{Machida}, Masahiro N. and {Inutsuka}, Shu-ichiro and {Matsumoto}, Tomoaki},
        title = "{Effect of Magnetic Braking on Circumstellar Disk Formation in a Strongly Magnetized Cloud}",
      journal = {\pasj},
         year = 2011,
        month = jun,
       volume = {63},
       number = {3},
        pages = {555-573},
          doi = {10.1093/pasj/63.3.555},
archivePrefix = {arXiv},
       eprint = {1009.2140},
 primaryClass = {astro-ph.SR},
       adsurl = {https://ui.adsabs.harvard.edu/abs/2011PASJ...63..555M}
}

@ARTICLE{2016MNRAS.457.1037W,
       author = {{Wurster}, James and {Price}, Daniel J. and {Bate}, Matthew R.},
        title = "{Can non-ideal magnetohydrodynamics solve the magnetic braking catastrophe?}",
      journal = {\mnras},
         year = 2016,
        month = mar,
       volume = {457},
       number = {1},
        pages = {1037-1061},
          doi = {10.1093/mnras/stw013},
archivePrefix = {arXiv},
       eprint = {1512.01597},
 primaryClass = {astro-ph.SR},
       adsurl = {https://ui.adsabs.harvard.edu/abs/2016MNRAS.457.1037W}
}

@ARTICLE{2021MNRAS.503..162H,
       author = {{Hu}, Zitao and {Bai}, Xue-Ning},
        title = "{Dust transport in protoplanetary discs with wind-driven accretion}",
      journal = {\mnras},
         year = 2021,
        month = may,
       volume = {503},
       number = {1},
        pages = {162-175},
          doi = {10.1093/mnras/stab542},
archivePrefix = {arXiv},
       eprint = {2102.01110},
 primaryClass = {astro-ph.EP},
       adsurl = {https://ui.adsabs.harvard.edu/abs/2021MNRAS.503..162H}
}

@ARTICLE{2026ApJ...999...92S,
       author = {{Salyk}, Colette and {Pontoppidan}, Klaus M. and {Zhang}, Ke and {Heinzen}, Sophie and {Calahan}, Jenny K. and {Banzatti}, Andrea and {Dickson-Vandervelde}, D. Annie and {Bergin}, Edwin A. and {Blake}, Geoffrey A. and {Arulanantham}, Nicole and {Krijt}, Sebastiaan and {Carr}, John and {Najita}, Joan and {Green}, Joel and {Romero-Mirza}, Carlos},
        title = "{H218O in the Terrestrial Planet-forming Regions of Protoplanetary Disks}",
      journal = {\apj},
         year = 2026,
        month = mar,
       volume = {999},
       number = {1},
          eid = {92},
        pages = {92},
          doi = {10.3847/1538-4357/ae3db1},
archivePrefix = {arXiv},
       eprint = {2601.20096},
 primaryClass = {astro-ph.EP},
       adsurl = {https://ui.adsabs.harvard.edu/abs/2026ApJ...999...92S}
}

@ARTICLE{2020GeCoA.279....1K,
       author = {{Kawasaki}, Noriyuki and {Wada}, Sohei and {Park}, Changkun and {Sakamoto}, Naoya and {Yurimoto}, Hisayoshi},
        title = "{Variations in initial $^{26}$Al/$^{27}$Al ratios among fine-grained Ca-Al-rich inclusions from reduced CV chondrites}",
      journal = {\gca},
         year = 2020,
        month = jun,
       volume = {279},
        pages = {1-15},
          doi = {10.1016/j.gca.2020.03.045},
       adsurl = {https://ui.adsabs.harvard.edu/abs/2020GeCoA.279....1K}
}

@ARTICLE{2024GeCoA.371..214Z,
       author = {{Zhang}, Mingming and {Chaumard}, No{\"e}l and {Defouilloy}, C{\'e}line and {Nachlas}, William O. and {Brownlee}, Donald E. and {Joswiak}, David J. and {Westphal}, Andrew J. and {Gainsforth}, Zack and {Kitajima}, Kouki and {Kita}, Noriko T.},
        title = "{Comet 81P/Wild 2 dust impactors of Stardust turnip-like tracks analogous to cluster IDPs}",
      journal = {\gca},
         year = 2024,
        month = apr,
       volume = {371},
        pages = {214-227},
          doi = {10.1016/j.gca.2024.02.013},
       adsurl = {https://ui.adsabs.harvard.edu/abs/2024GeCoA.371..214Z}
}

@ARTICLE{2018GeCoA.224..116H,
       author = {{Hertwig}, Andreas T. and {Defouilloy}, C{\'e}line and {Kita}, Noriko T.},
        title = "{Formation of chondrules in a moderately high dust enriched disk: Evidence from oxygen isotopes of chondrules from the Kaba CV3 chondrite}",
      journal = {\gca},
         year = 2018,
        month = mar,
       volume = {224},
        pages = {116-131},
          doi = {10.1016/j.gca.2017.12.013},
       adsurl = {https://ui.adsabs.harvard.edu/abs/2018GeCoA.224..116H}
}

@ARTICLE{2024M&PS...59.2388S,
       author = {{Suzumura}, Akimasa and {Kawasaki}, Noriyuki and {Yurimoto}, Hisayoshi and {Itoh}, Shoichi},
        title = "{Condensation of refractory minerals on igneous compact type A Ca-Al-rich inclusion from Northwest Africa 7865 CV chondrite}",
      journal = {\maps},
         year = 2024,
        month = sep,
       volume = {59},
       number = {9},
        pages = {2388-2402},
          doi = {10.1111/maps.14222},
       adsurl = {https://ui.adsabs.harvard.edu/abs/2024M&PS...59.2388S}
}

@ARTICLE{2006Icar..181..178C,
       author = {{Ciesla}, Fred J. and {Cuzzi}, Jeffrey N.},
        title = "{The evolution of the water distribution in a viscous protoplanetary disk}",
      journal = {\icarus},
         year = 2006,
        month = mar,
       volume = {181},
       number = {1},
        pages = {178-204},
          doi = {10.1016/j.icarus.2005.11.009},
archivePrefix = {arXiv},
       eprint = {astro-ph/0511372},
 primaryClass = {astro-ph},
       adsurl = {https://ui.adsabs.harvard.edu/abs/2006Icar..181..178C}
}
\bibliographystyle{aasjournal}



\end{document}